\documentclass[letterpaper,twocolumn,english,aps,showpacs,showkeys,preprintnumbers,prd,floatfix,nofootinbib,superscriptaddress]{revtex4-1}

\usepackage{amsmath}
\usepackage[T1]{fontenc}
\usepackage[latin1]{inputenc}
\usepackage{color}
\usepackage{graphicx}
\usepackage{nicefrac}
\usepackage[total={7.0in,9.5in},
top=1.in, left=0.75in, includefoot]{geometry}

\newcommand{\qns}{\ensuremath{q_{{\rm ns}}}}
\newcommand{\qs}{\ensuremath{q_{{\rm s}}}}
\newcommand{\cqns}[1]{\ensuremath{C_{#1,q}^{{\rm ns}}}}
\newcommand{\cqps}[1]{\ensuremath{C_{#1,q}^{{\rm ps}}}}
\newcommand{\cqs}[1]{\ensuremath{C_{#1,q}^{{\rm s}}}}
\newcommand{\cg}[1]{\ensuremath{C_{#1,g}}}

\def\msbar{\ensuremath{{\rm{\overline{MS}}}}}

\newcommand{\Ord}{\ensuremath{{\cal O}}}

\newcommand{\aplus}{\ensuremath{a^+(n_f)}}

\providecommand{\tabularnewline}{\\}

\newcommand{\ZMVFNS}{\ensuremath{\text{ZM-VFNS}}}

\makeatletter

\@ifundefined{textcolor}{}
{%
 \definecolor{BLACK}{gray}{0}
 \definecolor{WHITE}{gray}{1}
 \definecolor{RED}{rgb}{1,0,0}
 \definecolor{GREEN}{rgb}{0,1,0}
 \definecolor{BLUE}{rgb}{0,0,1}
 \definecolor{CYAN}{cmyk}{1,0,0,0}
 \definecolor{MAGENTA}{cmyk}{0,1,0,0}
 \definecolor{YELLOW}{cmyk}{0,0,1,0}
 }

\@ifundefined{showcaptionsetup}{}{%
 \PassOptionsToPackage{caption=false}{subfig}}

\makeatother

\usepackage[justification=centerlast]{caption,subfig}

\usepackage{babel}

\begin{document}
\def\msbar{\overline{MS}}
\def\gsim{\mathrel{\rlap{\lower4pt\hbox{\hskip1pt$\sim$}} \raise1pt\hbox{$>$}}} 
\def\lsim{\mathrel{\rlap{\lower4pt\hbox{\hskip1pt$\sim$}} \raise1pt\hbox{$<$}}}


\preprint{
\vbox{
\null \vspace{0.3in}
\hbox{LPSC 12-048}
\hbox{SMU-HEP-12-05}
\hbox{KA-TP-09-2012}
}
}

\title{\null \vspace{0.3in}
Heavy Quark Production in the ACOT Scheme at NNLO and N$^{3}$LO}

\author{T.~Stavreva}
\thanks{stavreva@lpsc.in2p3.fr}
\affiliation{Laboratoire de Physique Subatomique et de Cosmologie, Universit\'e
Joseph Fourier/CNRS-IN2P3/INPG, \\
 53 Avenue des Martyrs, 38026 Grenoble, France}

\author{F.~I.~Olness}
\thanks{olness@smu.edu}
\affiliation{Southern Methodist University, Dallas, TX 75275, USA}

\author{I.~Schienbein}
\thanks{schien@lpsc.in2p3.fr}
\affiliation{Laboratoire de Physique Subatomique et de Cosmologie, Universit\'e
Joseph Fourier/CNRS-IN2P3/INPG, \\
 53 Avenue des Martyrs, 38026 Grenoble, France}

\author{T.~Je\v{z}o}
\thanks{jezo@lpsc.in2p3.fr}
\affiliation{Laboratoire de Physique Subatomique et de Cosmologie, Universit\'e
Joseph Fourier/CNRS-IN2P3/INPG, \\
 53 Avenue des Martyrs, 38026 Grenoble, France}

\author{A.~Kusina}
\thanks{akusina@smu.edu}
\affiliation{Southern Methodist University, Dallas, TX 75275, USA}

\author{K.~Kova\v{r}\'{\i}k}
\thanks{kovarik@particle.uni-karlsruhe.de}
\affiliation{Institute for Theoretical Physics, Karlsruhe Institute of Technology, 
Karlsruhe, D-76128, Germany}

\author{J.~Y.~Yu}
\thanks{yu@physics.smu.edu}
\affiliation{Southern Methodist University, Dallas, TX 75275, USA}
\affiliation{Laboratoire de Physique Subatomique et de Cosmologie, Universit\'e
Joseph Fourier/CNRS-IN2P3/INPG, \\
 53 Avenue des Martyrs, 38026 Grenoble, France}



\begin{abstract}
We analyze the properties of the ACOT scheme for heavy quark production
and make use of the $\msbar$ massless results at NNLO and N$^{3}$LO
for the structure functions $F_{2}$ and $F_{L}$ in neutral current 
deep-inelastic scattering to estimate the higher order corrections. 
For this purpose we decouple the heavy quark mass entering the phase space
from the one entering the dynamics of the short distance cross section.
We show numerically that the phase space mass is generally more important.
Therefore, the dominant heavy quark mass effects at higher orders 
can be taken into account using the massless Wilson coefficients together with
an appropriate slow-rescaling prescription implementing the phase space constraints.
Combining the exact ACOT scheme at NLO with these expressions should provide
a  good approximation to the missing full calculation in the ACOT scheme 
at NNLO and N$^{3}$LO. 
\end{abstract}
\maketitle
\tableofcontents{}

\clearpage
\section{Introduction\label{sec:intro}}

\subsection{Motivation}

The production of heavy quarks in high energy processes has become
an increasingly important subject of study both theoretically and
experimentally. The theory of heavy quark production in perturbative
Quantum Chromodynamics (pQCD) is more challenging than that of light
parton (jet) production because of the new physics issues brought
about by the additional heavy quark mass scale. The correct theory
must properly take into account the changing role of the heavy quark
over the full kinematic range of the relevant process from the threshold
region (where the quark behaves like a typical {}``heavy particle'')
to the asymptotic region (where the same quark behaves effectively
like a parton, similar to the well known light quarks $\{u,d,s\}$).

With the ever-increasing precision of experimental data and
the progression of theoretical calculations and parton distribution function (PDF) 
evolution to next-to-next-to-leading order (NNLO) of QCD there is a clear
need to formulate and also implement the heavy quark schemes at
this order and beyond.
The most important case is arguably the heavy quark treatment
in inclusive deep-inelastic scattering (DIS) since the very precise
HERA data for DIS structure functions and cross sections form the backbone
of any modern global analysis of PDFs. Here, the heavy quarks
contribute up to 30\% or 40\% to the structure functions at small momentum
fractions $x$.
Extending the heavy quark schemes to higher orders is therefore 
necessary for extracting precise PDFs and hence for precise predictions
of observables at the LHC.
However, we would like to also stress the theoretical importance of having
a general pQCD framework including heavy quarks which is valid to all orders in perturbation theory
over a wide range of hard energy scales and which is also applicable to other observables 
than inclusive DIS in a straightforward manner.

\begin{figure}
\includegraphics[clip,width=0.45\textwidth]{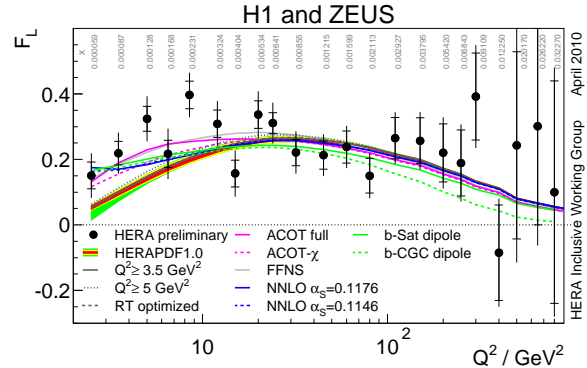}
\caption{$F_{L}$ vs. $Q$ from combined HERA-I inclusive deep inelastic
cross sections measured by the H1 and ZEUS collaborations.
Figure taken from Ref.~\cite{hera:FL}.
\label{fig:slac-4-4}}
\end{figure}

An example, where higher order corrections are particularly important 
is the structure function $F_L$ in DIS.
The leading order (${\cal O}(\alpha_S^0)$) contribution to this structure function
vanishes for massless quarks due to helicity conservation (Callan-Gross relation).
This has several consequences:
\begin{itemize}
\item $F_L$ is useful for constraining the gluon PDF via the dominant subprocess $\gamma^* g \to q \bar q$.

\item The heavy quark mass effects of order ${\cal O}(\tfrac{m^2}{Q^2})$ are relatively more 
pronounced.\footnote{%
Similar considerations also hold for target mass corrections (TMC) and higher twist terms.
We focus here mainly on the kinematic region $x<0.1$ where TMC are small \cite{Schienbein:2007gr}.
An inclusion of higher twist terms is beyond the scope of this study.}

\item Since the first non-vanishing contribution to $F_L$ is  
next-to-leading order (up to mass effects), the 
NNLO and N$^3$LO corrections are more important than for $F_2$.
\end{itemize}
In Fig.~\ref{fig:slac-4-4} we show a comparison of different theoretical
calculations of $F_L$ with preliminary HERA data~\cite{hera:FL}.
As can be seen, in particular at small $Q^2$ (i.e.\ small $x$),
there are considerable differences between the predictions.\footnote{%
An updated analysis
of the H1 measurements extending down to even lower $Q^2$ values
has been published in Ref.~\cite{Collaboration:2010ry},
and a combined analysis with ZEUS is in progress.
}

%

The purpose of this paper is to calculate the leading twist neutral current DIS structure functions
$F_2$ and $F_L$ in the ACOT factorization scheme up to order ${\cal O}(\alpha_S^3)$ (N$^3$LO)
and to estimate the error due to approximating the heavy quark mass terms 
${\cal O}(\alpha_S^2 \times \tfrac{m^2}{Q^2})$ 
and ${\cal O}(\alpha_S^3 \times \tfrac{m^2}{Q^2})$ 
in the higher order corrections.
The results of this study form the basis for using the ACOT scheme in NNLO global analyses
and for future comparisons with precision data for DIS structure functions.

\subsection{Outline of Paper}

The rest of this paper is organized as follows.
In Sec.\ \ref{sec:schemes} we review theoretical approaches to include heavy flavors
in QCD calculations. Particular emphasis is put on the ACOT scheme which is the
minimal extension of the $\msbar$ scheme in the sense that the observables in the
ACOT scheme reduce to the ones in the $\msbar$ scheme in the limit $m \to 0$
without any finite renormalizations. 
In this discussion we explicitly distinguish between the heavy quark/heavy meson
mass entering the final state phase space which we will call ``phase space mass''
and the heavy quark mass entering the dynamics of the short distance cross section
denoted ``dynamic mass.''  We show numerically using the exact ACOT scheme 
at ${\cal O}(\alpha_S)$ (NLO) that the effects of the phase space mass are more important than 
the ones due to the dynamic mass.
We use this observation to construct in Sec.\ \ref{sec:ho} the NC DIS structure functions
in the ACOT scheme up to ${\cal O}(\alpha_S^3)$. 
The corresponding numerical results are presented in Sec.\ \ref{sec:numerics}.
Finally, in Sec.\ \ref{sec:conclusion} we summarize the main results.
%

\section{Review of Theoretical Methods}
\label{sec:schemes}

We review theoretical methods which have been advanced to improve
existing QCD calculations of heavy quark production, and the impact
on recent experimental results.

\subsection{ACOT Scheme \label{subsec:acot}}

%
\begin{figure}[t]
\includegraphics[width=0.27\columnwidth]{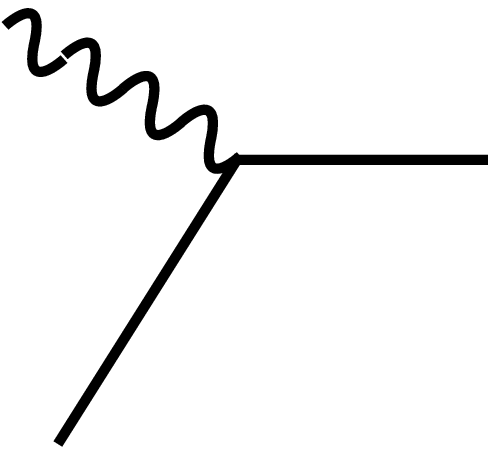}\quad{}
\includegraphics[width=0.25\columnwidth]{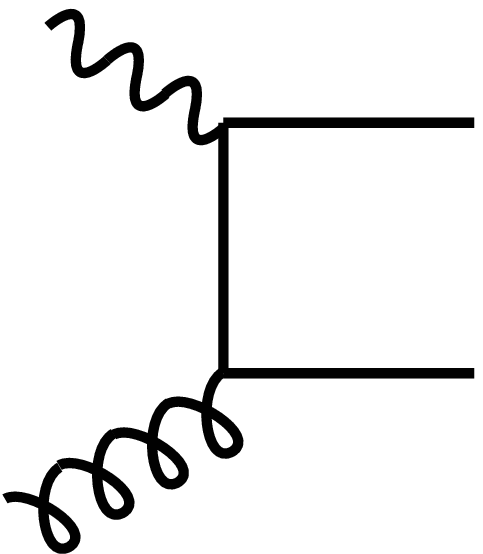}\quad{}
\includegraphics[width=0.25\columnwidth]{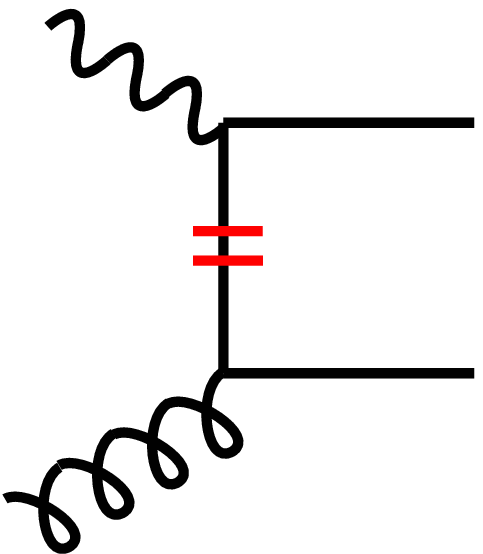}
\caption{Characteristic Feynman graphs which contribute to DIS heavy quark
production in the ACOT scheme: a)~the LO ${\cal O}(\alpha_{S}^{0})$
quark-boson scattering $QV\to Q$, b)~the NLO ${\cal O}(\alpha_{S}^{1})$
gluon-boson scattering $gV\to Q\bar{Q}$, and c)~the corresponding
subtraction term (SUB) $(g\to Q\bar{Q})\otimes(Q\to gQ)$.
\label{fig:diagrams}}
\end{figure}

The ACOT renormalization scheme \cite{Aivazis:1993pi} provides a mechanism
to incorporate the heavy quark mass into the theoretical calculation
of heavy quark production both kinematically and dynamically. In 1998
Collins \cite{Collins:1998rz} extended the factorization theorem to
address the case of heavy quarks; this work provided the theoretical
foundation that allows us to reliably compute heavy quark processes
throughout the full kinematic realm.

Figure~\ref{fig:diagrams} displays characteristic Feynman graphs
for the first two orders of DIS heavy quark production. If we consider
the DIS production of heavy quarks at ${\cal O}(\alpha_{S}^{1})$
this involves the LO $QV\to Q$ process and the NLO $gV\to Q\bar{Q}$
process.%
\footnote{At NLO, there are corresponding quark-initiated terms; for simplicity
we do not display them here, but they are fully contained in our calculations
\protect\cite{Kretzer:1998ju}. %
}

The key ingredient provided by the ACOT scheme is the subtraction
term (SUB) which removes the {}``double counting'' arising from
the regions of phase space where the LO and NLO contributions overlap.
Specifically, at NLO order, we can express the total result as a sum
of
\begin{equation}
\sigma_{TOT}=\sigma_{LO}+\left\{ \sigma_{NLO}-\sigma_{SUB}\right\} 
\label{eq:acot}
\end{equation}
where the subtraction term for the gluon-initiated processes is
\begin{equation}
\sigma_{SUB}=f_{g}\otimes\tilde{P}_{g\to Q}\otimes\sigma_{QV\to Q}\, .
\label{eq:sub}
\end{equation}
$\sigma_{SUB}$ represents a gluon emitted from a proton ($f_{g}$)
which undergoes a collinear splitting to a heavy quark $(\tilde{P}_{g\to Q})$
convoluted with the LO quark-boson scattering $\sigma_{QV\to Q}$.
Here, $\tilde{P}_{g\to Q}(x,\mu)=\frac{\alpha_{s}}{2\pi}\,\ln(\mu^{2}/m^{2})\, P_{g\to Q}(x)$
where $P_{g\to Q}(x)$ is the usual $\overline{MS}$ splitting kernel,
$m$ is the quark mass and $\mu$ is the renormalization scale\footnote{%
In this subsection we will  distinguish $\mu$ and $Q$; 
in the following, we will set  $\mu=Q$ and display the results 
as a function of $Q$.
}
which we typically choose to be $\mu=Q$.

An important feature of the ACOT scheme is that it reduces to the
appropriate limit both as $m\to0$ and $m\to\infty$ as we illustrate
below.

\subsubsection{Fixed-Flavor-Number-Scheme (FFNS) Limit}
Specifically, in the limit where the quark $Q$ is relatively heavy compared
to the characteristic energy scale $(\mu\lsim m)$, we find $\sigma_{LO}\sim\sigma_{SUB}$
such that $\sigma_{TOT}\sim\sigma_{NLO}$. In this limit, the ACOT
result naturally reduces to the Fixed-Flavor-Number-Scheme (FFNS)
result. In the FFNS, the heavy quark is treated as being extrinsic
to the hadron, and there is no corresponding heavy quark PDF ($f_{Q}\sim0$);
thus $\sigma_{LO}\sim0$. We also have $\sigma_{SUB}\sim0$ because
this is proportional to $\ln(\mu^{2}/m^{2})$. Thus, when the
quark $Q$ is heavy relative to the characteristic energy scale $\mu$,
the ACOT result reduces to $\sigma_{TOT}\sim\sigma_{NLO}$.

\subsubsection{Zero-Mass Variable-Flavor-Number-Scheme (ZM-VFNS) Limit}
Conversely, in the limit where the quark $Q$ is relatively light compared
to the characteristic energy scale $(\mu\gsim m)$, we find that
$\sigma_{LO}$ yields the dominant part of the result, and the ``formal''
NLO ${\cal O}(\alpha_{S})$ contribution $\left\{ \sigma_{NLO}-\sigma_{SUB}\right\} $
is an ${\cal {\cal O}}(\alpha_{S})$ correction.

In the limit $m/\mu\to0$, the ACOT result will reduce to the
$\msbar$ Zero-Mass Variable-Flavor-Number-Scheme (ZM-VFNS) limit
exactly without any finite renormalizations. In this limit, the quark
mass $m$ no longer plays any dynamical role and purely serves
as a regulator. The $\sigma_{NLO}$ term diverges due to the internal
exchange of the quark $Q$, and this singularity will be canceled by
$\sigma_{SUB}$.

\subsubsection{ACOT as a minimal extension of $\msbar$}
%
\begin{figure}
\includegraphics[clip,width=0.35\textwidth]{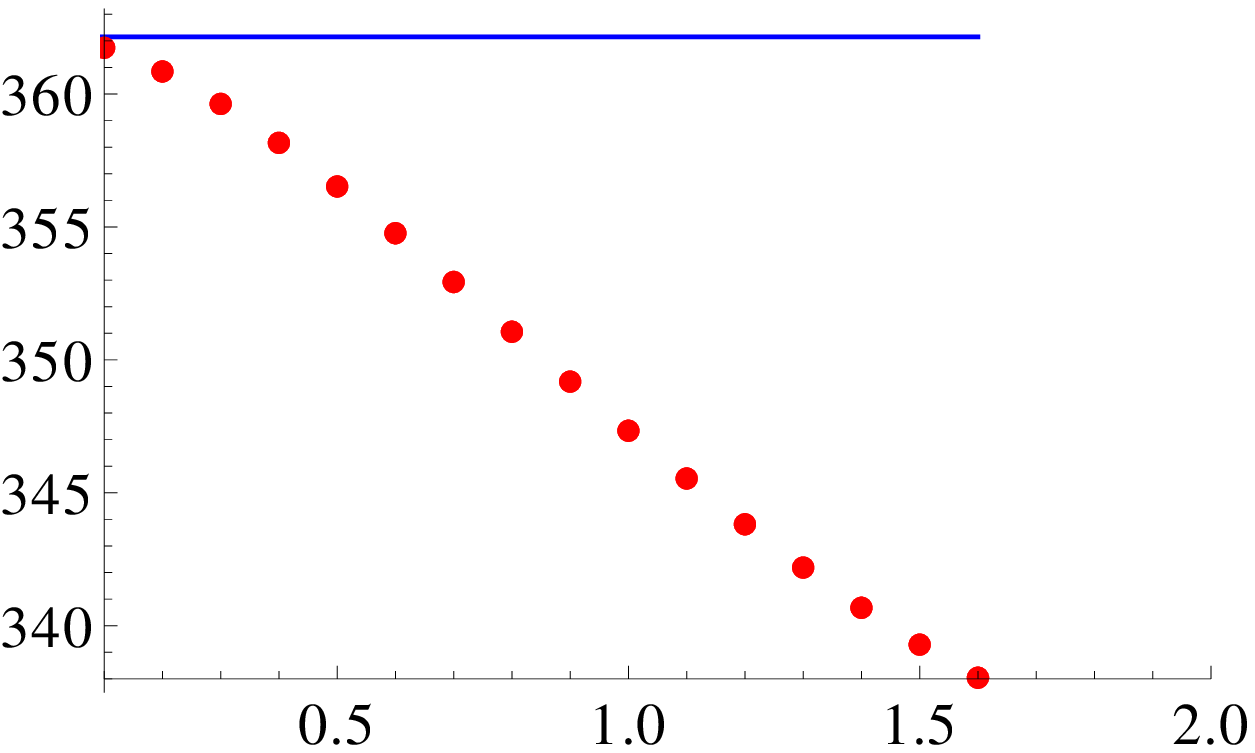}
\includegraphics[clip,width=0.35\textwidth]{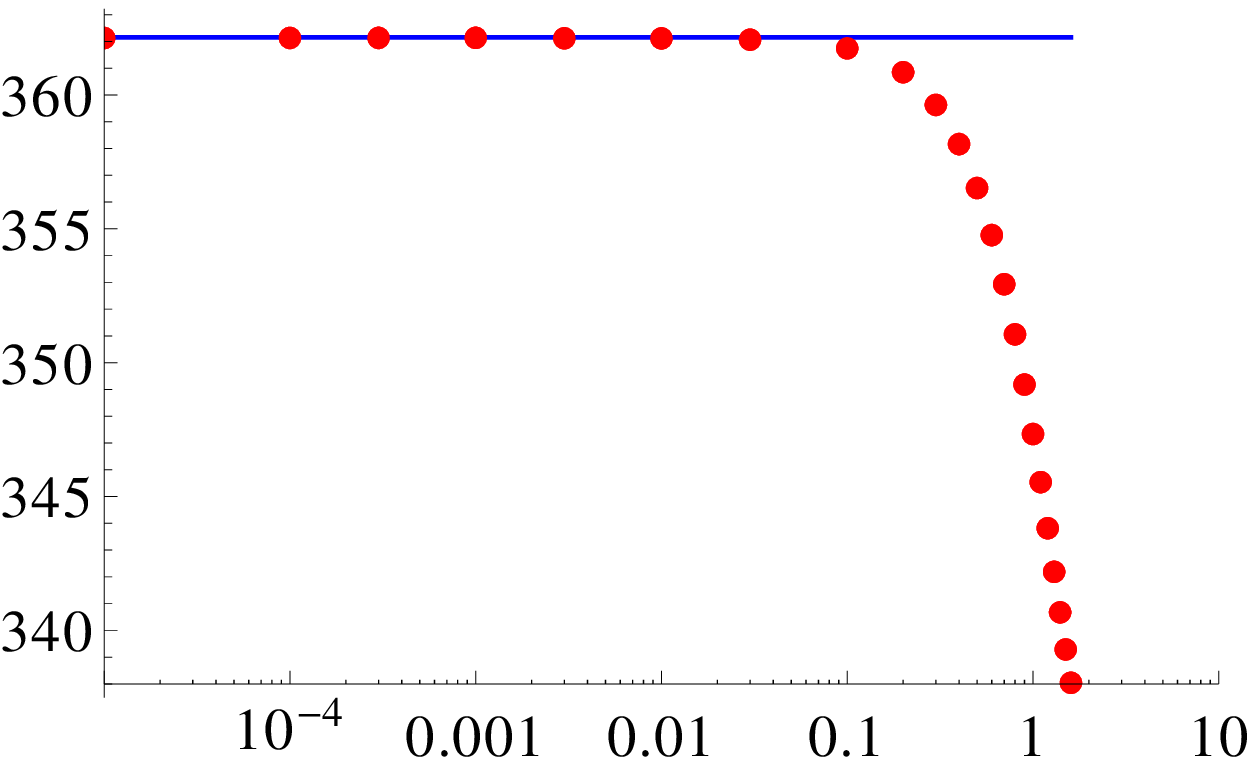}
\caption{Comparison of $F_{2}^{c}(x,Q)$ (scaled by $10^4$) vs. the quark mass $m$ in GeV
for fixed $x=0.1$ and $Q=10$~GeV. The red dots are the full ACOT
result, and the blue line is the massless $\msbar$ result.
The logarithmic plot demonstrates this result holds precisely in
the $m\to 0$ limit.
\label{fig:acotLimit}}
\end{figure}

We illustrate the versatile role of the quark mass 
in Fig.~\ref{fig:acotLimit}-a where we
display the $\msbar$ ZM-VFNS and the ACOT result as a function of
the quark mass $m$.

We observe that when $m$ is within a decade or two of $\mu$
that the quark mass plays a dynamic role; however, for $m\ll\mu$,
the quark mass purely serves as a regulator and the specific value
is not important. Operationally, it means we can obtain the $\msbar$
ZM-VFNS result either by i) computing the terms using dimensional
regularization and setting the regulator to zero, or ii) by computing
the terms using the quark mass as the regulator and then setting this
to zero.%
\footnote{If we were to compute this process in the $\msbar$ scheme,
the $\ln\left(m^{2}/Q^{2}\right)$
in the SUB term would simply be replaced by a $1/\varepsilon$
pole which would cancel the corresponding singularity in the NLO
contribution. %
}
To demonstrate this point explicitly, in Fig.~\ref{fig:acotLimit}-b 
we again display the $\msbar$ ZM-VFNS and the ACOT results but this
time with a logarithmic scale to highlight the small $m$ region.
We clearly see that  ACOT  reduces the  $\msbar$ ZM-VFNS exactly  
in this limit without any additional  finite renormalization 
contributions.\footnote{It is possible to define other  massive schemes 
that could include additional matching parameters or 
extra observable--dependent contributions.
For example, the calculation of $F_2^c$ in the
original RT scheme~\cite{Thorne:1997uu}
included extra higher-order contributions
that do not vanish as $Q/m\to \infty$.
}

The ACOT scheme is minimal in the sense that the construction of the
massive short distance cross sections does not need any
observable--dependent extra contributions or any regulators to smooth
the transition between the high and low scale regions.  The ACOT
prescription is to just calculate the massive partonic cross sections
and perform the factorization using the quark mass as regulator.

It is in this sense that we claim the ACOT scheme is the minimal
massive extension of the $\msbar$ ZM-VFNS. In the limit $m/\mu\to0$
it reduces exactly to the $\msbar$ ZM-VFNS, in the limit $m/\mu\gsim1$
the heavy quark decouples from the PDFs and we obtain exactly the FFNS
for $m/\mu\gg1$ and no finite renormalizations
or additional parameters are needed.

\subsubsection{When do we need Heavy Quark PDFs}
%
\begin{figure}
\includegraphics[clip,width=0.3\textwidth]{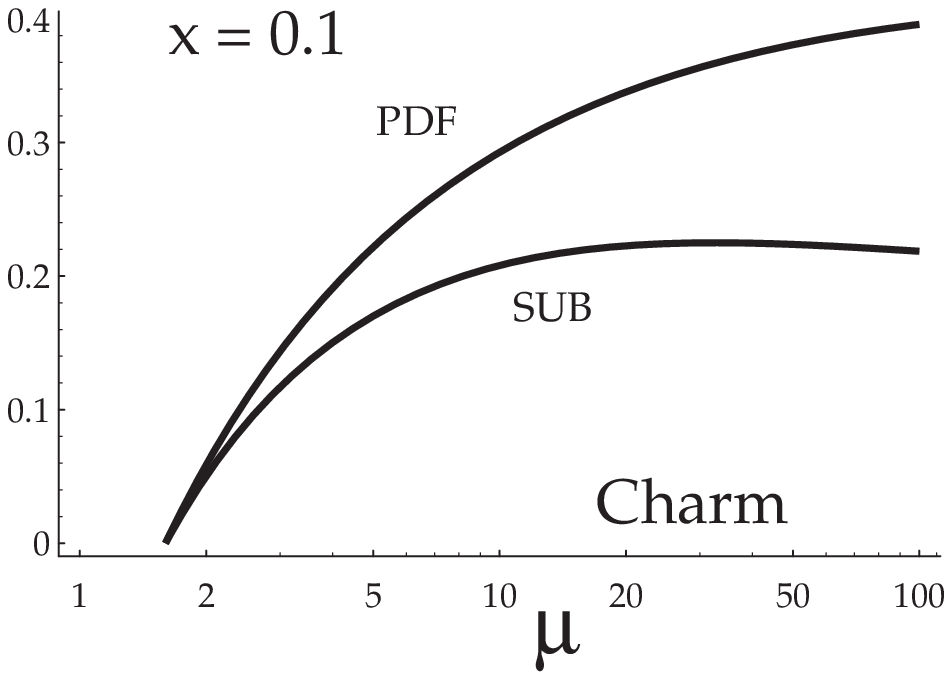}
\includegraphics[clip,width=0.3\textwidth]{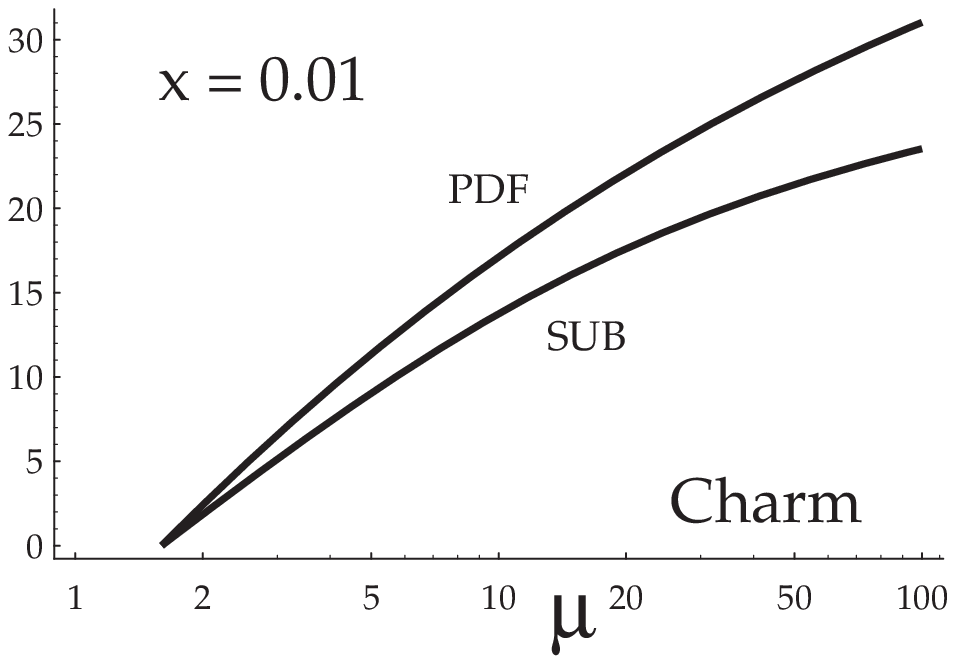}
\caption{Comparison of the DGLAP evolved charm PDF $f_{c}(x,\mu)$
with the perturbatively computed single splitting (SUB) 
$\widetilde{f}_{c}(x,\mu)=f_{g}(x,\mu)\otimes\widetilde{P}_{g\to c}$
vs. $\mu$ in GeV for two representative values of $x$.
\label{fig:charmPDF} }
\end{figure}

The novel ingredient in the above calculation is the inclusion of
the heavy quark PDF contribution which resums logs of
$\alpha_{S}\ln(\mu^{2}/m^{2})$.
An obvious question is when do we need to consider such terms, and
how large are their contributions? The answer is illustrated in Fig.~\ref{fig:charmPDF}
where we compare the DGLAP evolved PDF $f_{Q}(x,\mu)$ with the single
splitting perturbative result $\tilde{f}_{Q}(x,\mu)$.

The DGLAP PDF evolution sums a non-perturbative infinite tower of
logs which are contained in $\sigma_{LO}$ while the $\sigma_{SUB}$
contribution removes the perturbative single splitting component which
is already included in the $\sigma_{NLO}$ contribution. Hence, at
the PDF level the difference between the heavy quark DGLAP evolved
PDF $f_{Q}$ and the single-splitting perturbative $\tilde{f}_{Q}$
will indicate the contribution of the higher order logs which are
resummed into the heavy quark PDF.
Here, $\tilde{f}_{Q}=f_{g}\otimes\tilde{P}_{g\to Q}$ represents
the PDF of a heavy quark $Q$ generated from a single perturbative
splitting.

For $\mu\sim m$ we see that $f_{Q}$ and $\tilde{f}_{Q}$ match
quite closely, whereas they  differ significantly
for $\mu$ values a few times $m$. While the details will depend
on the specific process, in general we find that for $\mu$-scales
a few times $m$ the terms resummed by the heavy quark PDF can
be significant.
Additionally, the difference between $f_{Q}$ and $\tilde{f}_{Q}$
will be reduced at higher orders as more perturbative splittings
are included in $\tilde{f}_{Q}$.

Note that these scales are much lower than one might estimate using
the naive criterion $\frac{\alpha_{S}}{2\pi}\ln(\mu^{2}/m^{2})\sim1$; in particular,
the ACOT calculation often yields reduced $\mu$-dependence as the
quark dominated $\sigma_{LO}$ contributions typically have behavior which
is complementary to the gluon-initiated $\sigma_{NLO}$ terms.

\subsection{S-ACOT}

%
\begin{figure}[t]
\includegraphics[width=0.40\textwidth]{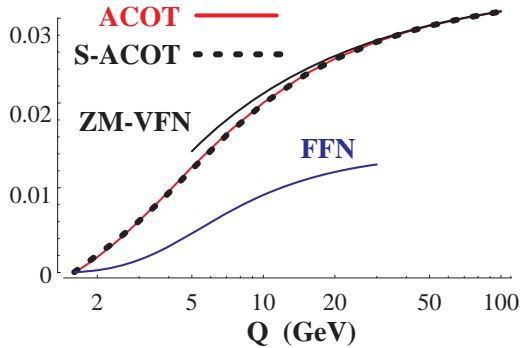}
\caption{$F_2^c$ for $x=0.1$ for NLO DIS heavy quark production
as a function of $Q$. We display calculations using the ACOT,
S-ACOT, Fixed-Flavor Number Scheme (FFNS),
and Zero-Mass Variable Flavor Number Scheme (ZM-VFNS).
The ACOT and S-ACOT results are virtually identical.
\label{fig:sacot}}
\end{figure}

In a corresponding application, it was observed that the heavy quark
mass could be set to zero in certain pieces of the hard scattering
terms without any loss of accuracy. This modification of the ACOT
scheme goes by the name Simplified-ACOT (S-ACOT) and can be summarized
as follows~\cite{Kramer:2000hn}.

\begin{quote}
\textbf{S-ACOT}: For hard-scattering processes with incoming
heavy quarks or with internal on-shell cuts on a heavy quark line,
the heavy quark mass can be set to zero ($m=0$) for these pieces.
\end{quote}

If we consider the case of NLO DIS heavy quark production, this means
we can set $m=0$ for the LO terms ($QV\to Q$) as this involves
an incoming heavy quark, and we can set $m=0$ for the SUB terms
as this has an on-shell cut on an internal heavy quark line. Hence,
the only contribution which requires calculation with $m$ retained
is the NLO $gV\to Q\bar{Q}$ process. Figure~\ref{fig:sacot} displays
a comparison of a calculation using the ACOT scheme with all masses
retained vs. the S-ACOT scheme; as expected, these two results match
throughout the full kinematic region.

It is important to note that the S-ACOT scheme is not an approximation;
this is an exact renormalization scheme, extensible to all orders.

\subsection{ACOT and $\chi$-Rescaling}

As we have illustrated in Sec.~\ref{subsec:acot} above, in the limit $Q^2\gg m^{2}$ the mass
simply plays the role of a regulator.
In contrast, for $Q^2 \sim m^{2}$ the value of the mass is of consequence for the physics.
The mass can enter dynamically in the hard-scattering matrix element,
and can enter kinematically in the phase space of the process.

We will demonstrate that for the processes of interest the primary
role of the mass is kinematic and not dynamic. It was this idea which
was behind the original slow-rescaling prescription of \cite{Barnett:1976ak}
which considered DIS charm production (e.g., $\gamma c\to c)$ introducing
the shift
\begin{equation}
x\to\chi=x\left[1+\left(\frac{m_{c}}{Q}\right)^{2}\right]\, 
\label{eq:barnett}.
\end{equation}
This prescription accounted for the charm quark mass by effectively
reducing the phase space for the final state by an amount proportional
to $(m_{c}/Q)^{2}$.

This idea was extended in the $\chi$-scheme by realizing that 
in addition to the observed final-state charm quark, 
if the beam has a  charm-flavor quantum number of zero (such as a proton beam)
then
there
is also an anti-charm quark in the beam fragments 
because
all the charm
quarks are ultimately produced by gluon splitting ($g\to c\overline{c}$)
into a charm pair.%
\footnote{%
If the beam has non-zero charm-flavor quantum number, such as 
a $D$-meson, this argument would be incorrect. 
Technically,  $\chi$-scaling violates factorization as we are
presuming the mass of the beam fragments; if we perform a thought
experiment with a beam of $D$-mesons, charm quark need not be associated with an
anti-charm quark.
}
For this case the scaling variable becomes
\begin{equation}
\chi=x\left[1+\left(\frac{2m_{c}}{Q}\right)^{2}\right]\, .
\label{eq:chi}
\end{equation}
This rescaling is implemented in the ACOT$_{\chi}$ scheme, for
example~\cite{Tung:2001mv,Kretzer:2003it,Guzzi:2011ew}.
The factor $(1+(2m_{c})^{2}/Q^{2})$ represents a kinematic suppression
factor which will suppress the charm process relative to the lighter
quarks.
Additionally, the $\chi$-scaling ensures the threshold kinematics ($W^2>4m_c^2+M^2$) are satisfied;
while it is important 
to satisfy this condition for large $x$, this may prove too restrictive  at small $x$ where 
the HERA data are especially precise.\footnote{We sketch the relevant kinematics in Appendix~\ref{subsec:w}.}

To encompass all the above results, we can define a general scaling
variable $\chi(n)$ as
\begin{equation}
\chi(n)=x\left[1+\left(\frac{n\: m_{c}}{Q}\right)^{2}\right]
\label{eq:chin}
\end{equation}
where $n=\{0,1,2\}$. 
Here, $n=0$ corresponds to the massless result without rescaling, 
$n=1$ corresponds to the original Barnett slow-rescaling,
and $n=2$ corresponds to the $\chi$-rescaling.

\subsection{Phase Space (Kinematic) \& Dynamic Mass}
\label{sec:kinMass}

%
\begin{figure*}
\subfloat[
Comparison of $F_{2}^{c}(x,Q)$ vs. $Q$ for the NLO ACOT calculation for $x=\{10^{-1},10^{-3},10^{-5}\}$ (left to right) using zero dynamic mass {[}$\widehat{\sigma}(m=0)${]} to show the effect of $n$ scaling;
from top to bottom $n=\{0,1,2\}$ (pink, black, purple).]
{
\includegraphics[width=0.3\textwidth]{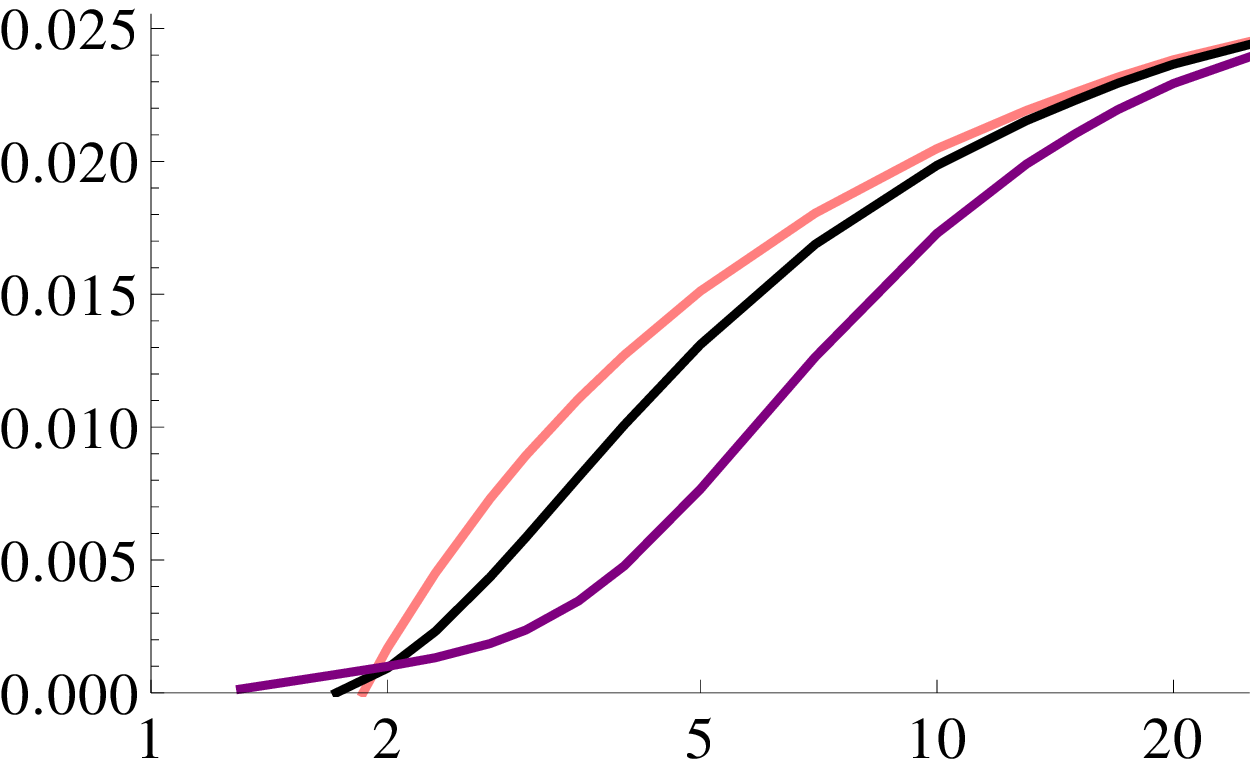}
\hfil
\includegraphics[width=0.3\textwidth]{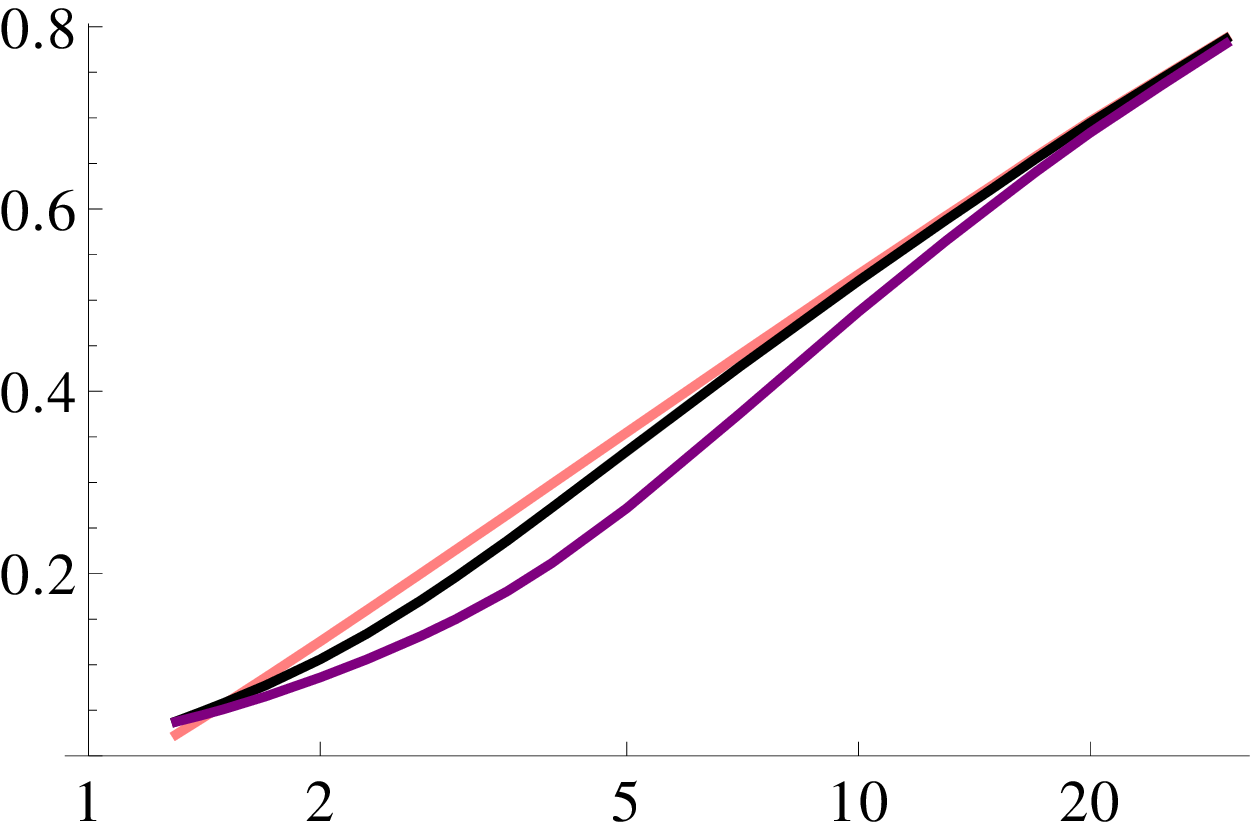}
\hfil
\includegraphics[width=0.3\textwidth]{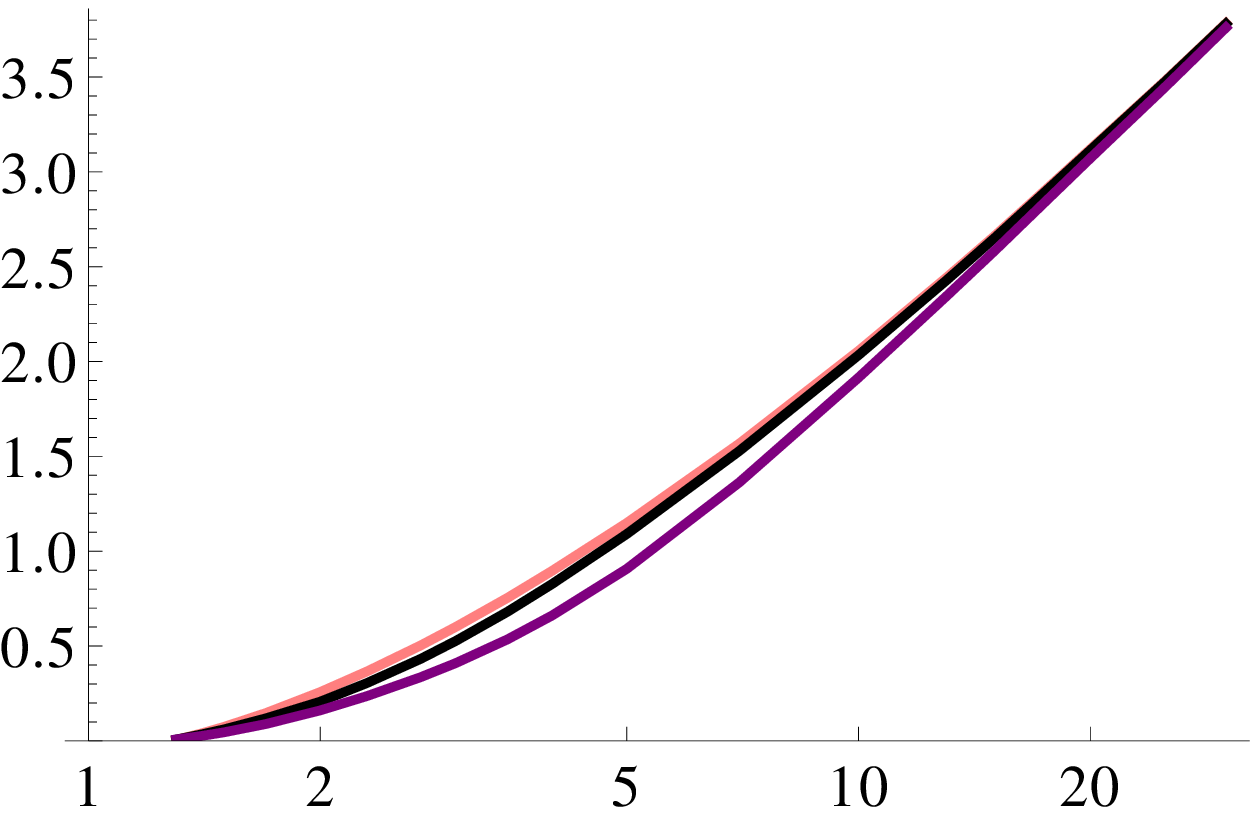}
\label{fig:nscale}
}

\subfloat[
Comparison of $F_{2}^{c}(x,Q)$ vs. $Q$ for the NLO ACOT calculation for
$x=\{10^{-1},10^{-3},10^{-5}\}$ (left to right).
Here we keep the scaling fixed $n=2$ and compare the effect of varying
the dynamic mass in the Wilson coefficient.
The upper (cyan) curve uses a non-zero dynamic mass {[}$\widehat{\sigma}(m=1.3)${]}
and the lower (purple) curve uses a zero dynamic mass {[}$\widehat{\sigma}(m=0)${]}.
]
{
\includegraphics[width=0.3\textwidth]{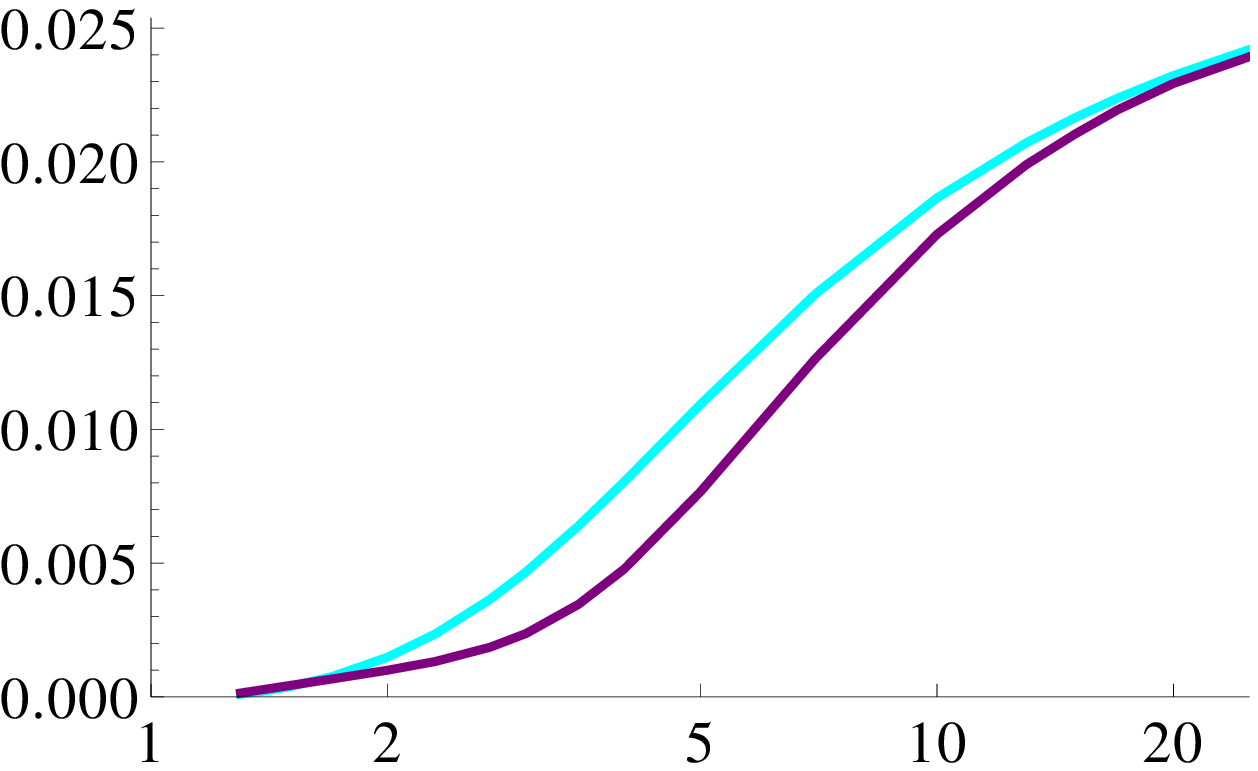}
\hfil
\includegraphics[width=0.3\textwidth]{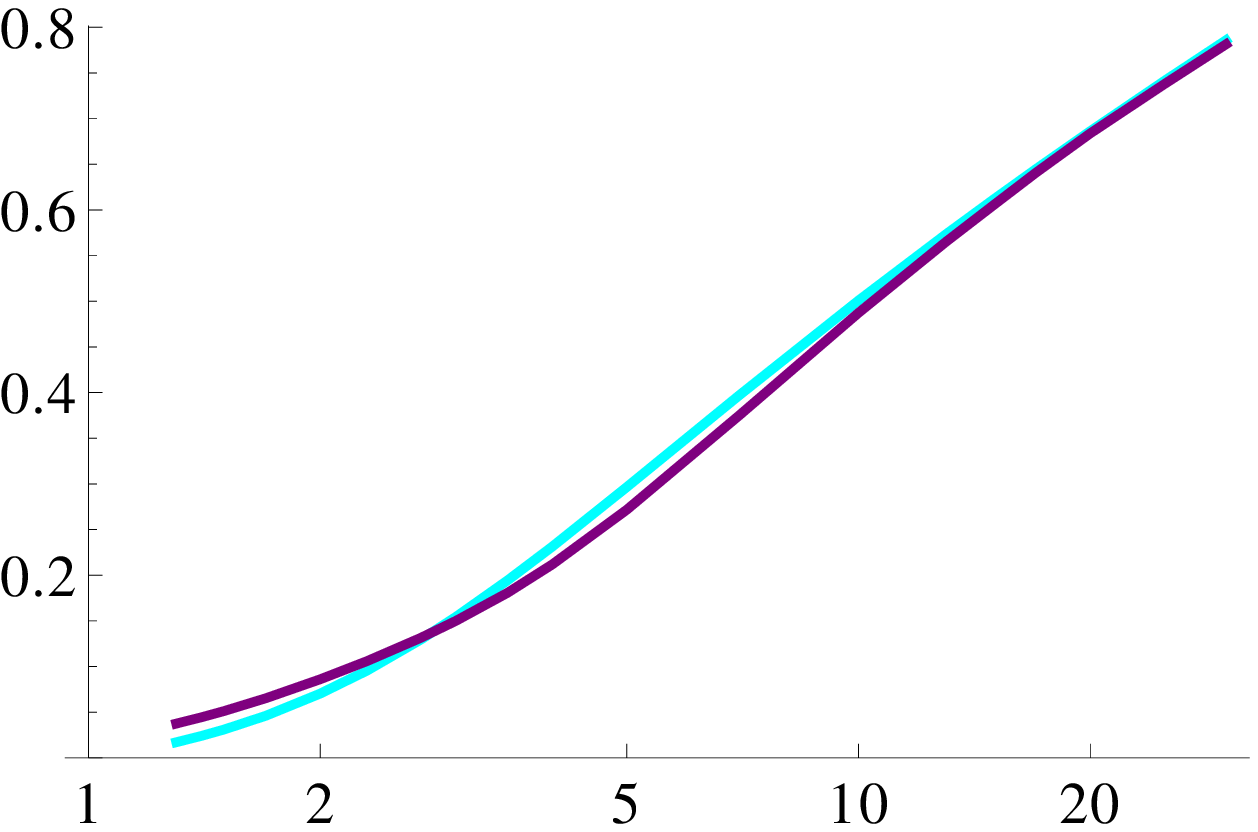}
\hfil
\includegraphics[width=0.3\textwidth]{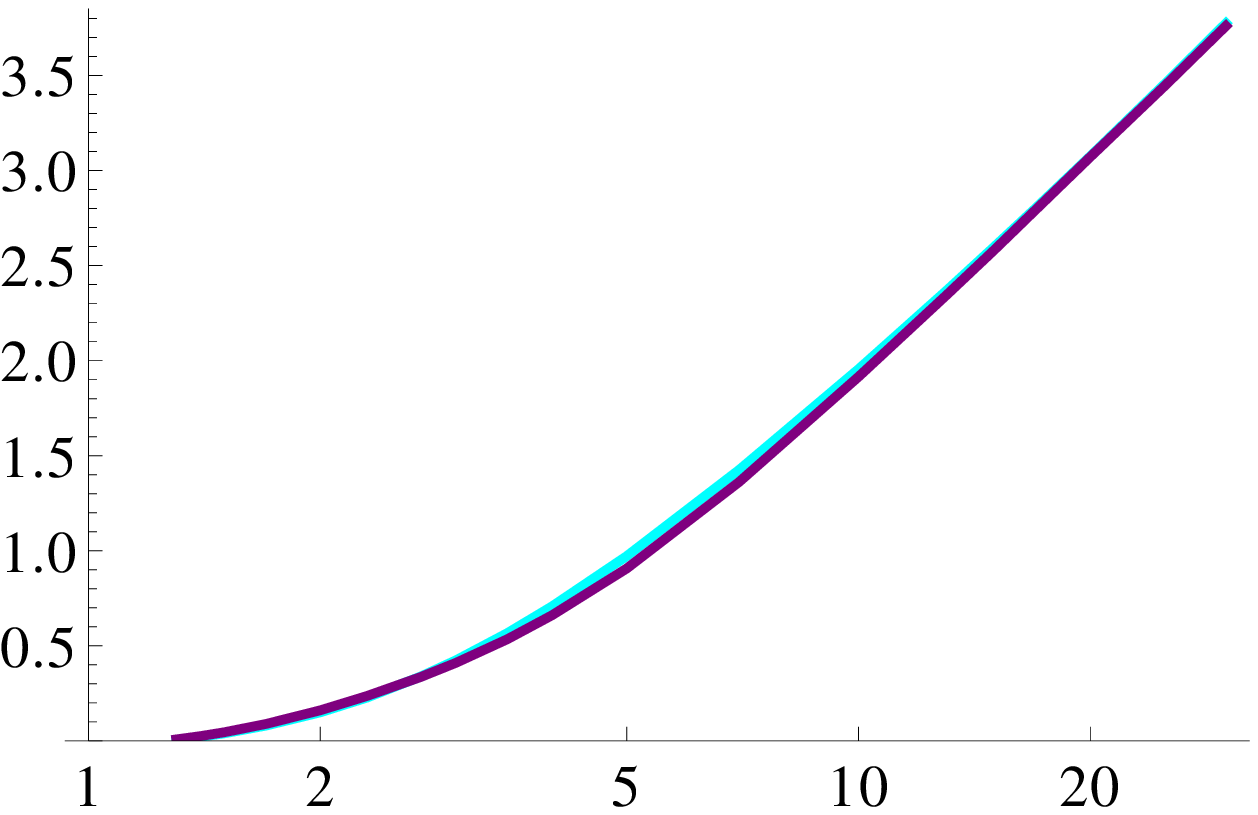}
\label{fig:dynamic}
}
\caption{Comparison of phase space (kinematic) \& dynamic mass effects}
\label{fig:massEffects}
\end{figure*}

\begin{figure*}
\includegraphics[width=0.3\textwidth]{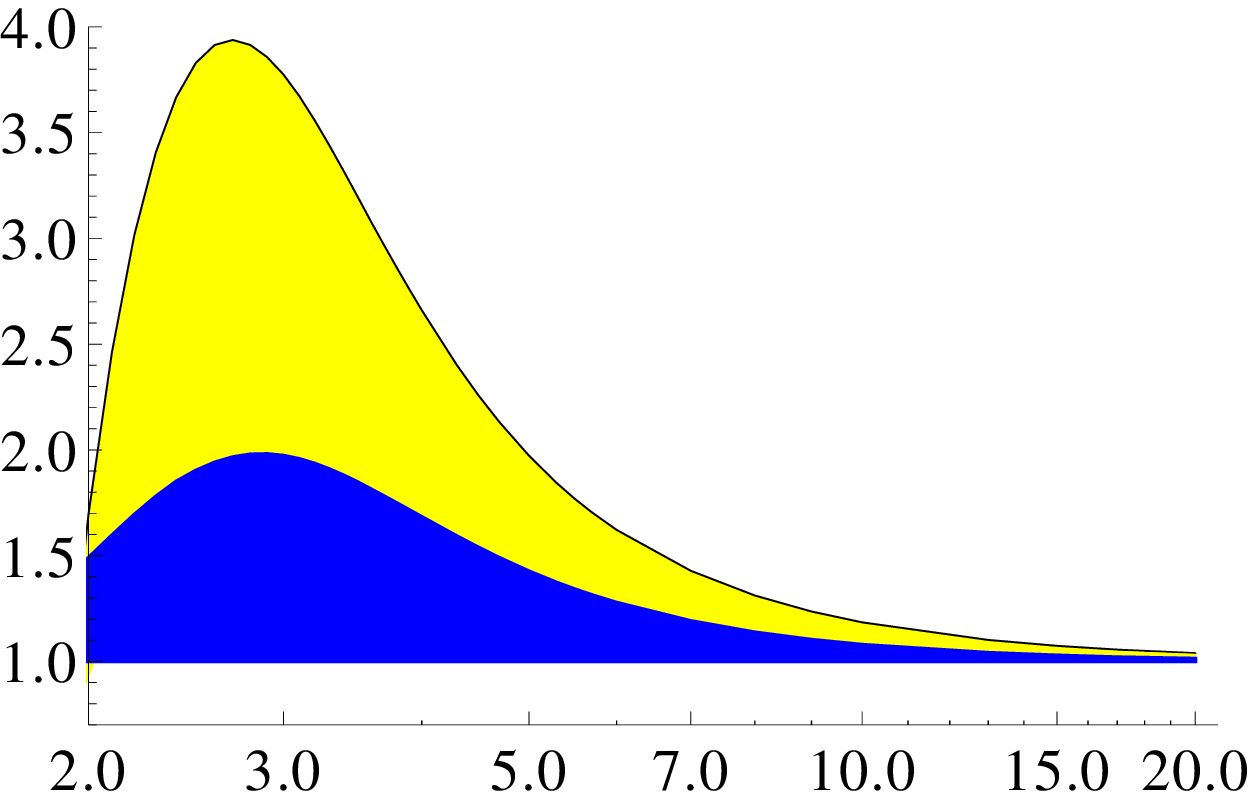}
\hfil
\includegraphics[width=0.3\textwidth]{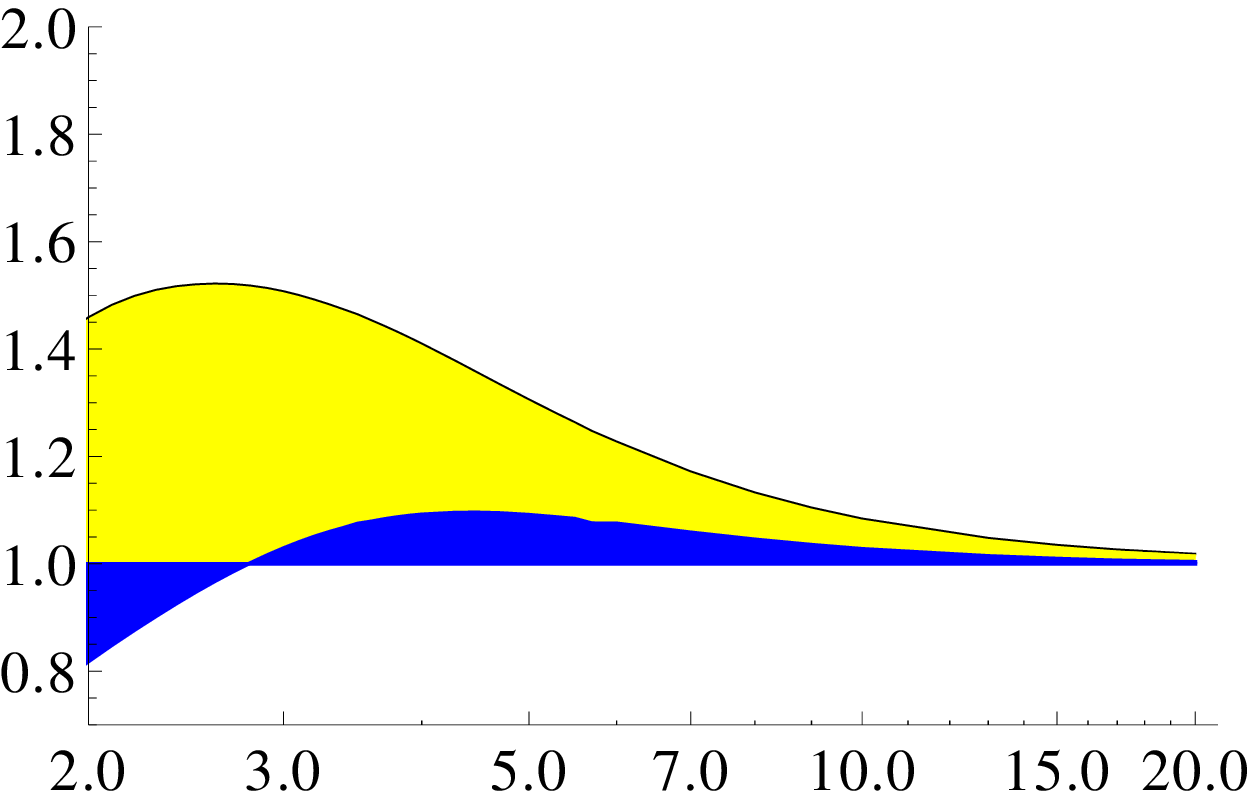}
\hfil
\includegraphics[width=0.3\textwidth]{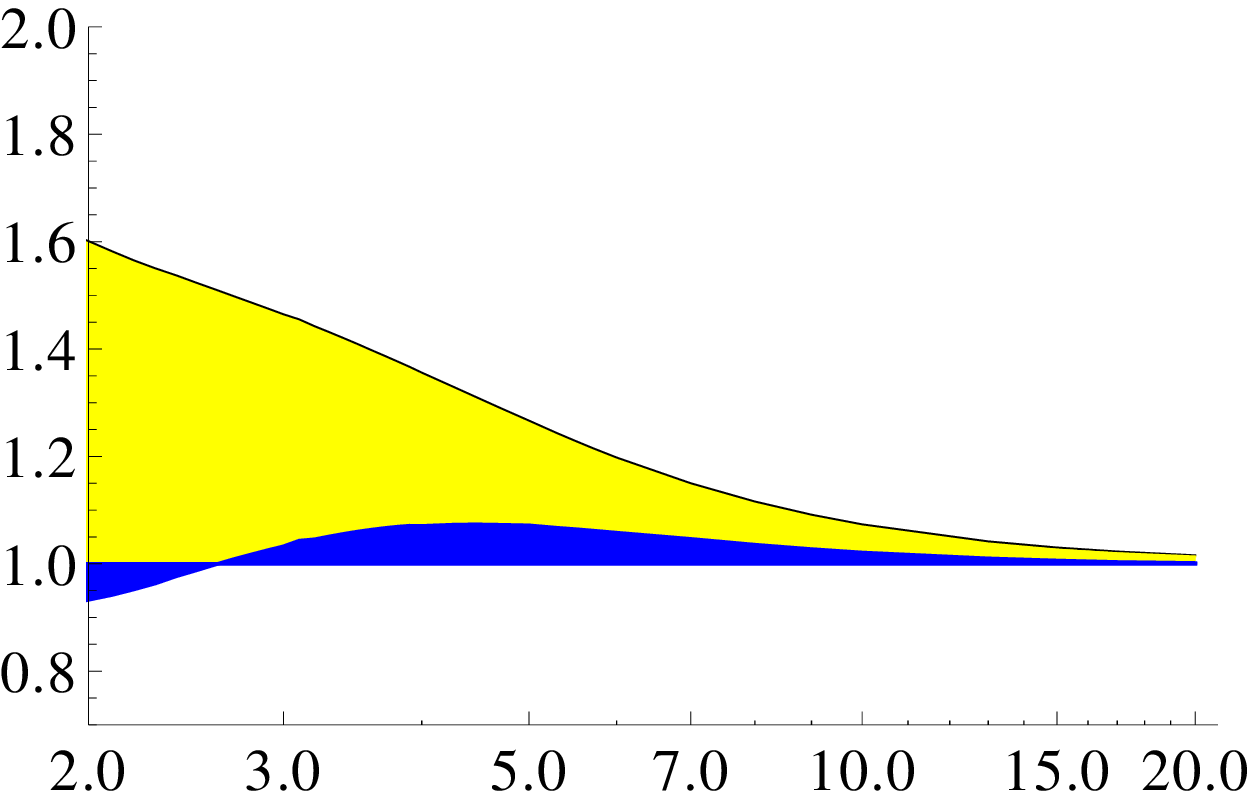}
\caption{Comparison of kinematic \& dynamic mass effects for
$F_{2}^{c}(x,Q)$ vs. $Q$ for the NLO ACOT calculation for
$x=\{10^{-1},10^{-3},10^{-5}\}$ (left to right).
The curves are scaled by the massless $n=2$ result. 
The wider (yellow) band represents the variation of the  kinematic mass 
of Fig.~\ref{fig:nscale};  
note this band extends down to a ratio of 1.0.
The narrower (blue) band 
is overlaid on the plot and 
represents the variation of the  dynamic mass 
of Fig.~\ref{fig:dynamic}.
}
\label{fig:massEffects_ratio}
\end{figure*}


We now investigate the effects of separately varying the mass entering
the $\chi(n)$ variable taking into account the phase space constraints and
the mass value entering the hard scattering cross section $\widehat{\sigma}(m)$.
We call the former mass parameter ``phase space (kinematic) mass'' and the
latter ``dynamic mass''\footnote{Note that the finite mass terms $(m^2/Q^2)^n$
in $\widehat{\sigma}(m)$ receive contributions from both, masses in the heavy quark propagators
and masses in the phase space. Still we refer to them as dynamic mass terms and show
that they are numerically less important than the mass terms in the slow rescaling variable
$\chi(n)$ which are of purely kinematic origin.}.

In Fig.~\ref{fig:nscale} we display $F_{2}^{c}(x,Q)$ vs. $Q$. 
The family of 3 curves shows the   NLO ACOT calculation
with $\chi(n)$ scaling using a zero dynamic mass for the hard scattering. 
We compare this with 
 Fig.~\ref{fig:dynamic} which shows
$F_2^c(x,Q)$ in the NLO ACOT scheme using a fixed  $n=2$ scaling, but 
varying the mass used in the  hard-scattering cross section.
The upper (cyan) curves use a non-zero dynamic mass {[}$\widehat{\sigma}(m_c=1.3)${]}
and the lower (purple) curves have been obtained with a vanishing dynamic mass {[}$\widehat{\sigma}(m_c=0)${]}.
We observe that the effect of the `dynamic mass' in $\widehat{\sigma}(m_c)$ is only
of consequence in the limited region $Q\gsim m,$ and even in this
region the effect is minimal.
In contrast, the influence of the phase space
(kinematic) mass 
shown in  Fig.~\ref{fig:nscale}
is larger than the dynamic mass
shown in  Fig.~\ref{fig:dynamic}.
To highlight these differences,
we scale the curves in  Fig.~\ref{fig:massEffects_ratio}
by  the massless  $n=2$ scaling result and plot
bands that represent the variation of the dynamic and kinematic masses.

In conclusion, we have shown that (up to ${\cal O}(\alpha_S)$)
the phase space mass dependence is generally the dominant
contribution to the DIS structure functions.
Assuming that this observation remains true at higher orders, 
it is possible to obtain a good approximation of the structure functions
in the  ACOT scheme at NNLO and N$^3$LO using the massless Wilson 
coefficients together with a non-zero phase space mass entering via the
$\chi(n)$-prescription.

\section{ACOT scheme beyond NLO}
\label{sec:ho}

%
\begin{figure}[t]
\includegraphics[clip,width=0.12\textwidth]{eps/feyn_LO_1}
\includegraphics[clip,width=0.10\textwidth]{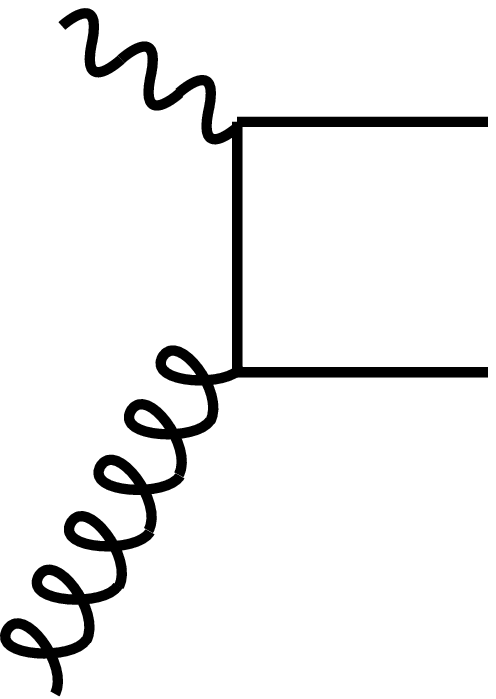}
\includegraphics[clip,width=0.10\textwidth]{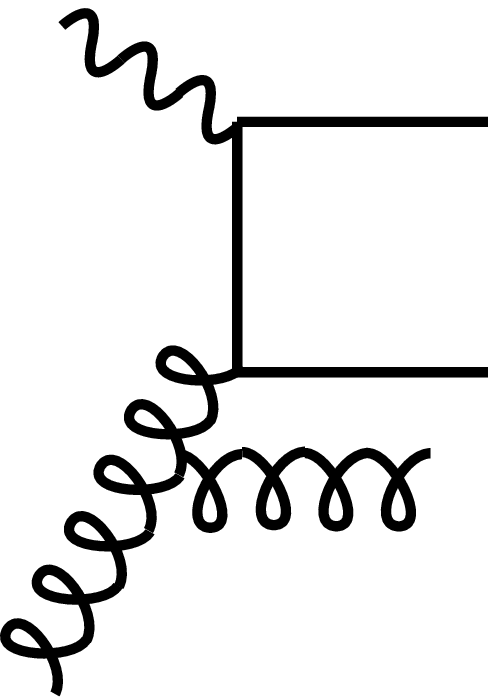}
\includegraphics[clip,width=0.10\textwidth]{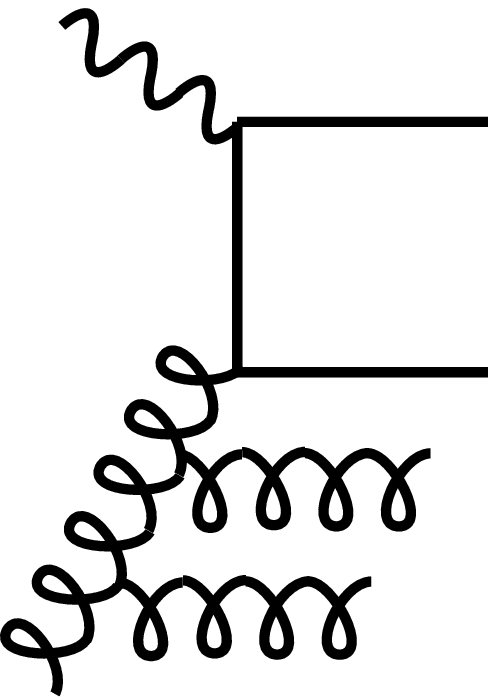}
\caption{Sample Feynman diagrams contributing to DIS heavy quark production
(from left): 
LO ${\cal O}(\alpha_{S}^{0})$ quark-boson scattering $QV\to Q$, 
NLO ${\cal O}(\alpha_{S}^{1})$ gluon-boson scattering $gV\to Q\bar{Q}$, 
NNLO ${\cal O}(\alpha_{S}^{2})$ boson-gluon scattering $gV\to gQ\bar{Q}$, 
and 
N$^{3}$LO ${\cal O}(\alpha_{S}^{3})$ boson-gluon scattering $gV\to ggQ\bar{Q}$.
\label{fig:slac-4-3-1}}
\end{figure}

We have shown using the  NLO  full ACOT scheme that the dominant mass effects 
are those coming from the phase space which can be  taken into account
via a generalized slow-rescaling $\chi(n)$-prescription.
Assuming that a similar relation remains true at higher orders, one can construct
the following approximation to the ACOT result  up to N$^3$LO ($\Ord(\alpha_S^3)$):
\begin{eqnarray}
&&{\rm ACOT} [\Ord(\alpha_S^{0+1+2+3})] 
\ \simeq \ 
\nonumber \\
&& {\rm ACOT} [\Ord(\alpha_S^{0+1})] + \ZMVFNS_{\chi(n)} [\Ord(\alpha_S^{2+3})]
\label{eq:approx}
\end{eqnarray}
In this equation, 
``ACOT'' generically represents any variant of the 
ACOT scheme (ACOT, \hbox{S-ACOT}, \hbox{S-ACOT${}_\chi$});
for the results presented in Sec.~\ref{sec:numerics}, we will use the fully massive 
ACOT scheme with all masses retained out to NLO. 
The
$\ZMVFNS_{\chi(n)}$ term uses the massless Wilson coefficients at ${\cal O}(\alpha\,\alpha_{S}^{2})$
and ${\cal O}(\alpha\,\alpha_{S}^{3})$ with the specified $\chi(n)$-scaling.\footnote{%
In Sec.II.A.2 we demonstrated that the ACOT calculation reduces to the $\ZMVFNS$ result
in the massless limit. We will address the choice of the $\chi(n)$-rescaling in the Sec.~III.A.}
Sample processes which contribute at this order are displayed in Fig.~\ref{fig:slac-4-3-1}.

We use the $\ZMVFNS_{\chi(n)}$ result 
in Eq.~(\ref{eq:approx})
to approximate the higher-order terms because 
not all the necessary massive Wilson coefficients at ${\cal O}(\alpha\,\alpha_{S}^{2})$
and ${\cal O}(\alpha\,\alpha_{S}^{3})$ have been computed.
There has been a calculation of neutral current electroproduction 
(equal quark masses, vector coupling) 
of 
heavy quarks at this order 
by Smith \& VanNeerven \protect\cite{Laenen:1992zk} 
in the FFNS which could be used to obtain the massive Wilson coefficients in the 
S-ACOT scheme by applying appropriate collinear subtraction terms.
However, for the 
original ACOT scheme it would then still be necessary to compute the
massive Wilson coefficients for the heavy quark initiated subprocess at 
${\cal O}(\alpha\,\alpha_{S}^{2})$.
See Refs.~\cite{Guzzi:2011ew,Guzzi:2011dk} for details.

Using the result of Ref.~\cite{Laenen:1992zk}, Thorne and Roberts
developed an NLO VFNS~\cite{Thorne:1997uu,Thorne:1997ga}, and an
improved NNLO formulation was presented in Ref.~\cite{Thorne:2006qt}.
The FONLL formalism was outlined in Ref.~\cite{Cacciari:1998it} and
this was used to construct matched expressions for structure functions
to NNLO~\cite{Forte:2010ta}; implications of these results in the
context of the NNPDF analysis were presented in Ref.~\cite{Ball:2011mu}.
An overview and comparison of these analyses was presented in the 2009
Les Houches report~\cite{Binoth:2010ra}. 
More recently, an NNLO S-ACOT-$\chi$ calculation was developed in 
Refs.~\cite{Guzzi:2011ew,Guzzi:2011dk}.  
For charge current case massive calculations are available
at order ${\cal O}(\alpha\,\alpha_{S})$~\cite{PhysRevD.23.56,Gluck:1996ve,Blumlein:2011zu}
and partial results at order
${\cal O}(\alpha\,\alpha_{S}^2)$~\cite{Buza:1997mg}.
Comparative analyses of these schemes are under investigation; 
however, this is beyond the scope of this paper.%

Here, we argue  that the massless Wilson coefficients at
${\cal O}(\alpha\,\alpha_{S}^{2})$ together with a $\chi(n)$-prescription
provide a very good approximation of the exact result.
At worst, the maximum error would be of order $\Ord(\alpha\, \alpha_{S}^{2}\times[m^{2}/Q^{2}])$.
However, based on the arguments of Sec.~\ref{sec:kinMass} we expect the inclusion
of the phase space mass effects to contain the dominant higher order contributions
so that the actual error should be substantially smaller.

The massless higher order coefficient functions for the DIS structure function
$F_2$ via photon exchange can be found in
Refs.~\cite{Furmanski:1981cw,Bardeen:1978yd,Altarelli:1978id}
for ${\cal O}(\alpha_S^1)$,
Refs.~\cite{vanNeerven:1991nn,Zijlstra:1991qc,Zijlstra:1992qd}
for ${\cal O}(\alpha_S^2)$,
and Ref.~\cite{Vermaseren:2005qc} for ${\cal O}(\alpha_S^3)$.
For our numerical code we have used the $x$-space parameterization provided
in Refs.~\cite{vanNeerven:1999ca,vanNeerven:2000uj} for ${\cal O}(\alpha_S^2)$,
and Refs.~\cite{Moch:2002sn,Vermaseren:2005qc} for ${\cal O}(\alpha_S^3)$.

The expressions for the structure function $F_L$ have been calculated
in Refs.~\cite{SanchezGuillen:1990iq,Zijlstra:1991qc}
for ${\cal O}(\alpha_S^2)$, and
Ref.~\cite{Vermaseren:2005qc} for ${\cal O}(\alpha_S^3)$.
In our FORTRAN code we have used the $x$-space parameterization provided in
Refs.~\cite{vanNeerven:1999ca,Moch:2004xu} for ${\cal O}(\alpha_S^2)$ and
Ref.~\cite{Moch:2004xu} for ${\cal O}(\alpha_S^3)$.

In order to calculate the inclusive structure
functions $F_2$ and $F_L$ in the $\ZMVFNS_\chi$ 
using these Wilson coefficients, plus- and delta-distributions have to be evaluated
which is in principle straightforward. However, for the implementation of the
slow-rescaling prescription it is necessary to decompose the Wilson coefficients
into the contributions from different parton flavors. This step is non-trivial
at $\Ord(\alpha_S^2)$ and beyond, and we therefore provide some details of
our calculation in the Appendix~\ref{app:decomposition}.

\subsection{Choice of $\chi(n)$-Rescaling}

%
\begin{table*}
    \begin{tabular}{|c|c|c|c|c|}
\hline
$\xi$  & General  & $m_{1}=0$  & $m_{1}=m_{2}=m$  & $\chi$-scheme:
\tabularnewline
\hline
\hline
\raisebox{-0.1cm}{\includegraphics[scale=0.2]{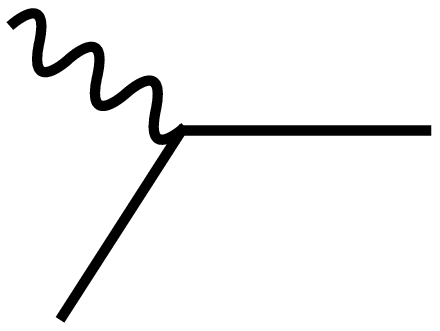}} &
\raisebox{0.15cm}{$\eta\,\left[\frac{Q^{2}-m_{1}^{2}+m_{2}^{2}+\Delta[-Q^{2},m_{1}^{2},m_{2}^{2}]}{2Q^{2}}\right]$}
& \raisebox{0.15cm}{$\eta\,\left[1+\frac{m_{2}^{2}}{Q^{2}}\right]$}  &
\raisebox{0.15cm}{$\eta\,\left[1+\frac{m^{2}}{Q^{2}}\right]$}  &
\raisebox{0.15cm}{$\eta\,\left[1+\frac{\left(2m\right)^{2}}{Q^{2}}\right]$}\tabularnewline
\hline
\raisebox{-0.1cm}{\includegraphics[scale=0.15]{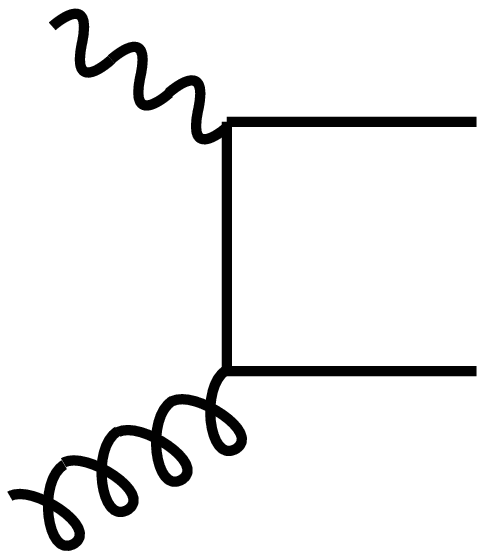}}  &
\raisebox{0.23cm}{$\eta\,\left[1+\left(\frac{m_{1}+m_{2}}{Q}\right)^{2}\right]$}
& \raisebox{0.23cm}{$\eta\,\left[1+\frac{m_{2}^{2}}{Q^{2}}\right]$}  &
\raisebox{0.23cm}{$\eta\,\left[1+\frac{(2m)^{2}}{Q^{2}}\right]$}  &
\raisebox{0.23cm}{$\eta\,\left[1+\frac{\left(2m\right)^{2}}{Q^{2}}\right]$}\tabularnewline
\hline
\end{tabular}\caption{The massive rescaling factor for the LO
quark-initiated process ($Vq_{1}\to q_{2}$),
and
the NLO gluon-initiated process ($Vg\to q_{1}\overline{q}_{2}$).
The quarks $q_{1,2}$ have mass $m_{1,2}$, respectively, and $V$
represents the vector boson; $\gamma/Z$ for neutral current processes
($m_{1}=m_{2}$), and $W^{\pm}$ for charged current processes
($m_{1}\not=m_{2}$).
$\eta$ is the scaling factor which depends on the hadronic mass $M$; see
Appendix~\ref{subsec:tmc} for details.
The triangle-function is defined as:
$\Delta[a,b,c]=\sqrt{a^{2}+b^{2}+c^{2}-2(ab+bc+ca)}$.}
\label{tab:scaling}
\end{table*}

We now consider our choice for the appropriate  generalized $\chi(n)$-rescaling variable.

In Table~\ref{tab:scaling} we display the various rescalings of $\xi$
for the LO $\gamma Q\to Q$ process and the NLO $\gamma g\to Q\overline{Q}$
process. The ``general'' result is obtained by working out the
detailed kinematics for the corresponding process~\cite{Aivazis:1993kh}.

The factor $\eta$ is the rescaling due to the hadronic mass $M$;
notice that this factors out from the partonic mass dependence as
it should~\cite{Schienbein:2007gr}. 
For details see Appendix~\ref{subsec:tmc}.

The LO case with full massive kinematics has been computed in Ref.~\cite{Aivazis:1993kh}.
In the limit where the initial mass  is small ($m_{1}\to0$), we recover the Barnett~\cite{Barnett:1976ak}
slow-rescaling result. Additionally, we obtain the curious result
that for a neutral current equal mass case ($m_{1}=m_{2}$) the rescaling
is this same factor.

For the NLO gluon-induced process, the interpretation of the rescaling
is straightforward; the phase space is simply suppressed by the total
invariant mass of the final state $(m_{1}+m_{2})$ compared to the
scale $Q$. For the charged current case where we neglect $m_{1}$,
we again obtain the standard rescaling factor. However, for the neutral
current case ($m_{1}=m_{2}$) we obtain a rescaling factor which is
analogous to the $\chi$-scaling factor.

For the purposes of this study, we will vary the phase space mass 
using the $\chi(n)$ rescaling with $n=\{0,1,2\}$.
While $n=0$ corresponds to the massless case (no rescaling), 
it is not obvious whether $n=1$ or $n=2$ is the preferred rescaling choice
for higher orders. Thus, we will use the range between  $n=1$ and  $n=2$
as a measure of our theoretical uncertainty arising from this ambiguity.

\section{Results}
\label{sec:numerics}


We now present the results of our calculation extending the ACOT scheme to 
NNLO and N${}^3$LO. 
As outlined in Eq.~(\ref{eq:approx}), we will use the fully massive ACOT scheme 
for the LO and  NLO contributions, 
and combine this with the $\ZMVFNS$  supplemented with the $\chi$-rescaling prescription
to approximate the higher order terms. 
We will use the QCDNUM program \cite{Botje:2010ay}
with the VFNS evolved with the DGLAP kernels at NNLO to generate our PDFs 
from an initial distribution based on the Les Houches benchmark set~\cite{Giele:2002hx};
this ensures that our heavy quark PDFs are consistently evolved so that the 
heavy quark initiated LO terms properly match the corresponding SUB contribution. 
At NNLO the proper matching conditions across flavor thresholds
introduces discontinuities in the PDFs which are incorporated in the
QCDNUM program; we discuss this in detail in Appendix~C.
We choose $m_c=1.3$~GeV,  $m_b=4.5$~GeV,  $\alpha_S(M_Z)=0.118$.
We note that the \hbox{QCDNUM} ZM-STFN package has the massless Wilson coefficients computed
up to  N${}^3$LO; we cross checked our implementation of ACOT  in the massless limit with QCDNUM, 
and they agree precisely.

\subsection{Effect of $\chi(n)$-Scaling}

%
\begin{figure*}
\subfloat[$F_{2}$ vs. $Q$.
\label{fig:f2x135}]{
\includegraphics[width=0.3\textwidth]{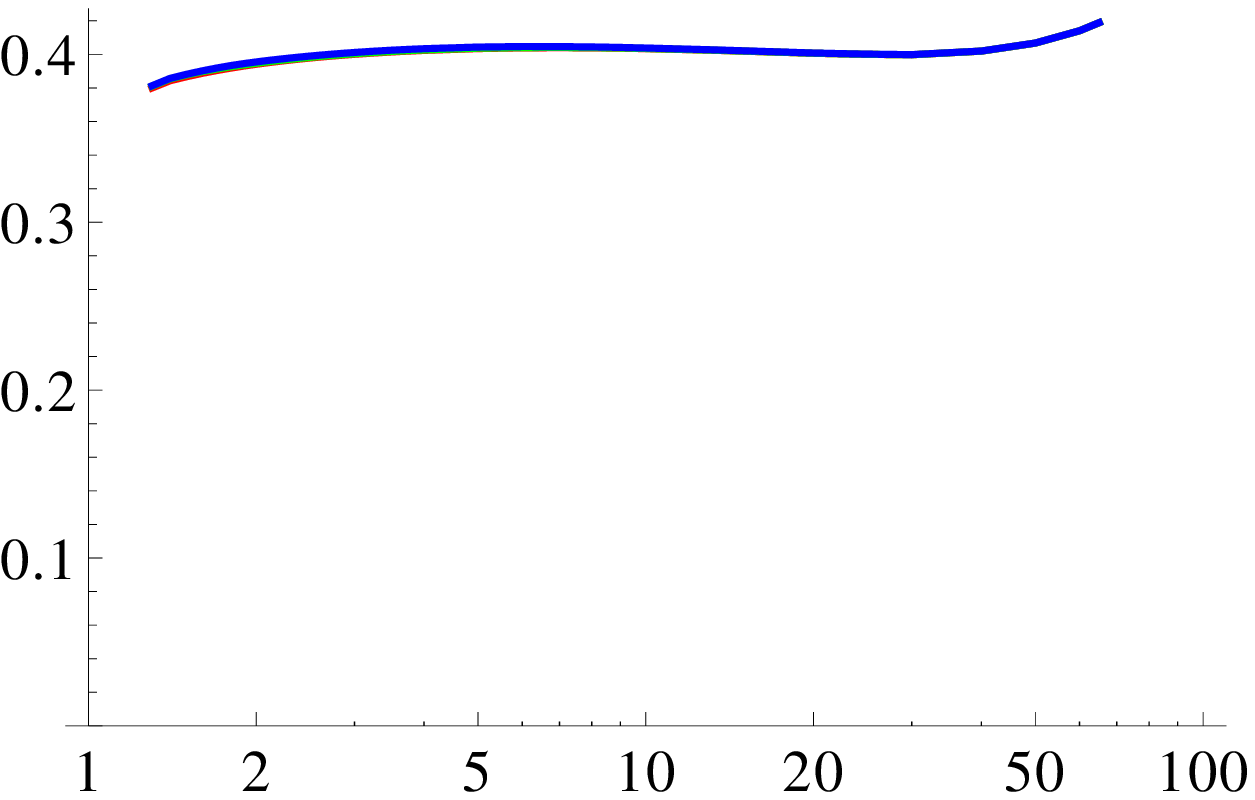}
\hfil
\includegraphics[width=0.3\textwidth]{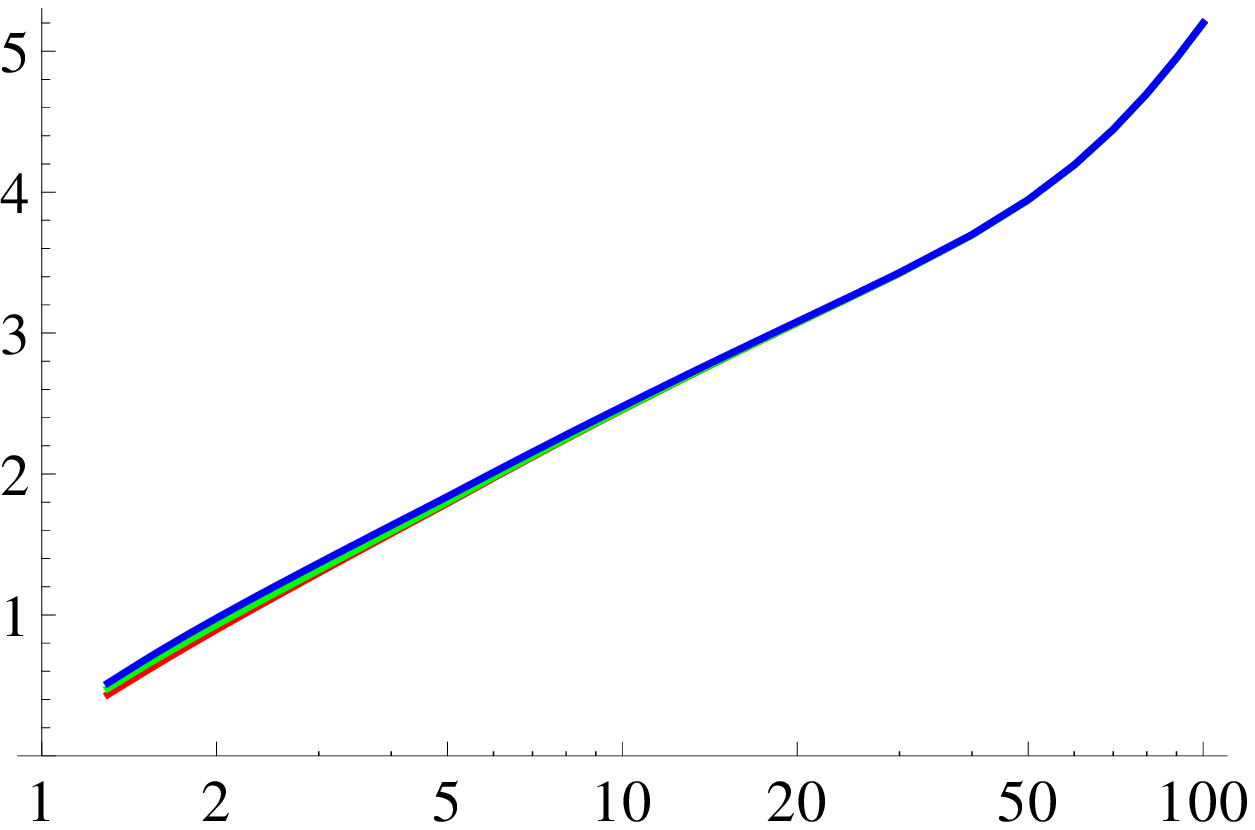}
\hfil
\includegraphics[width=0.3\textwidth]{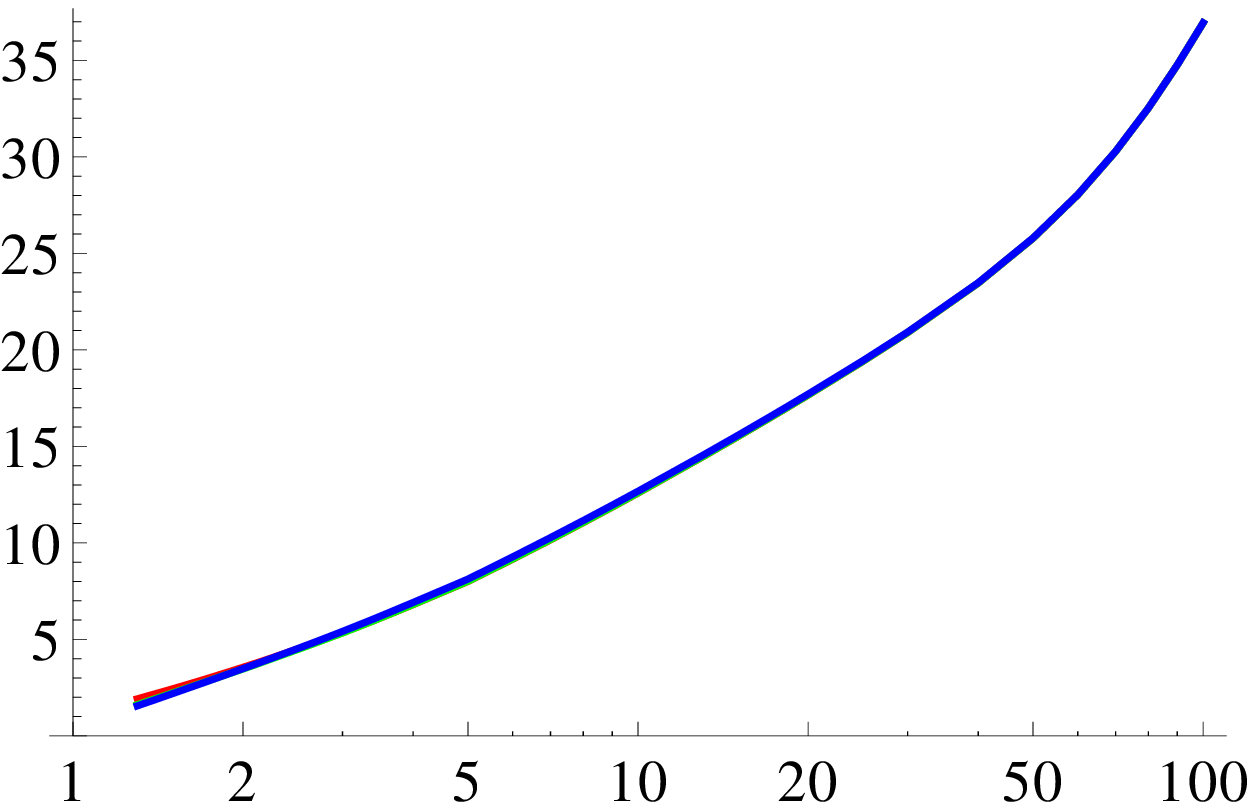}
}

\subfloat[$F_{L}$ vs. $Q$.
\label{fig:fLx135}]{
\includegraphics[width=0.3\textwidth]{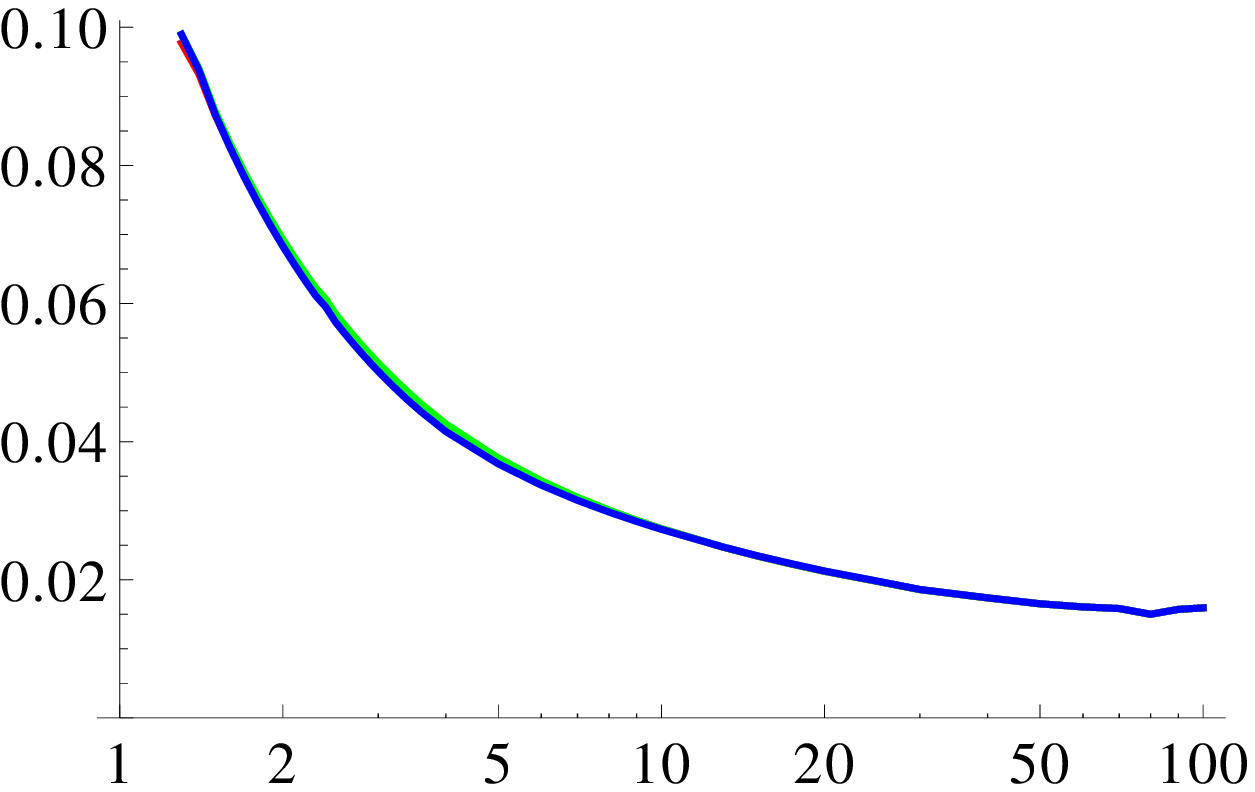}
\hfil
\includegraphics[width=0.3\textwidth]{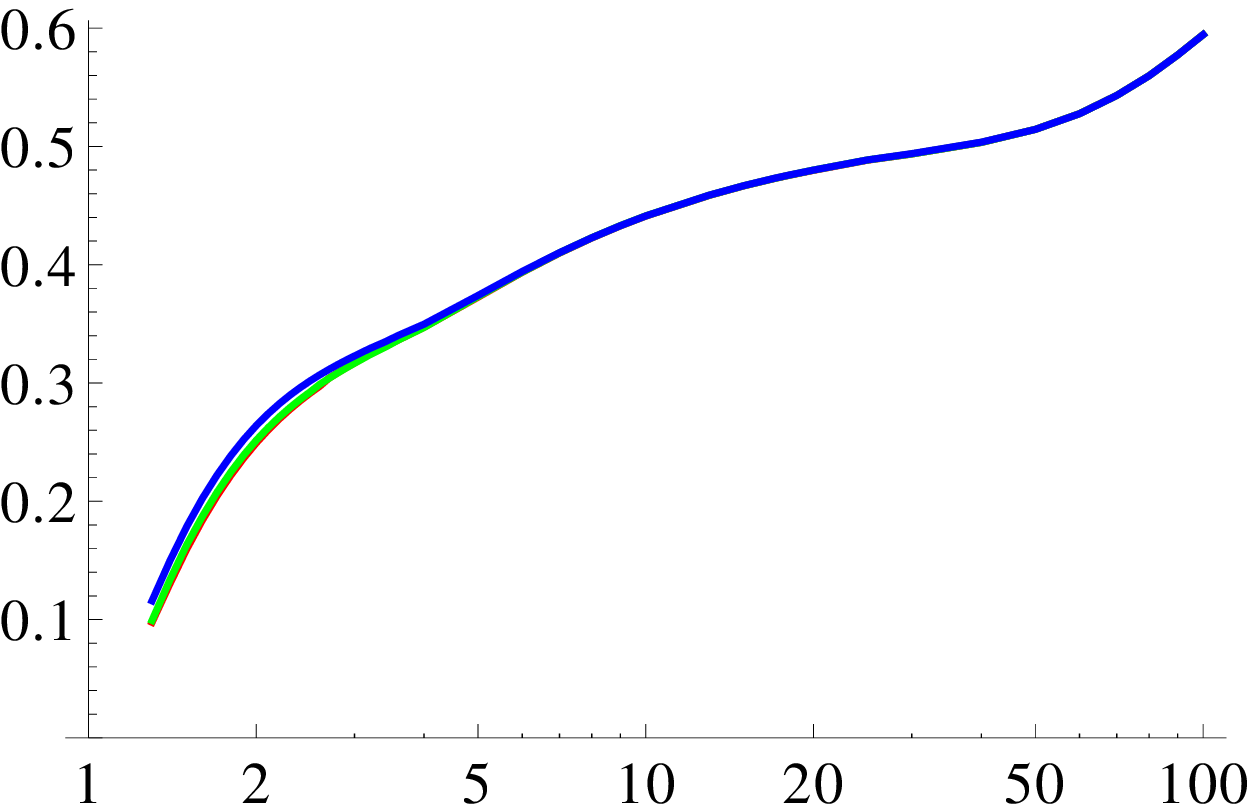}
\hfil
\includegraphics[width=0.3\textwidth]{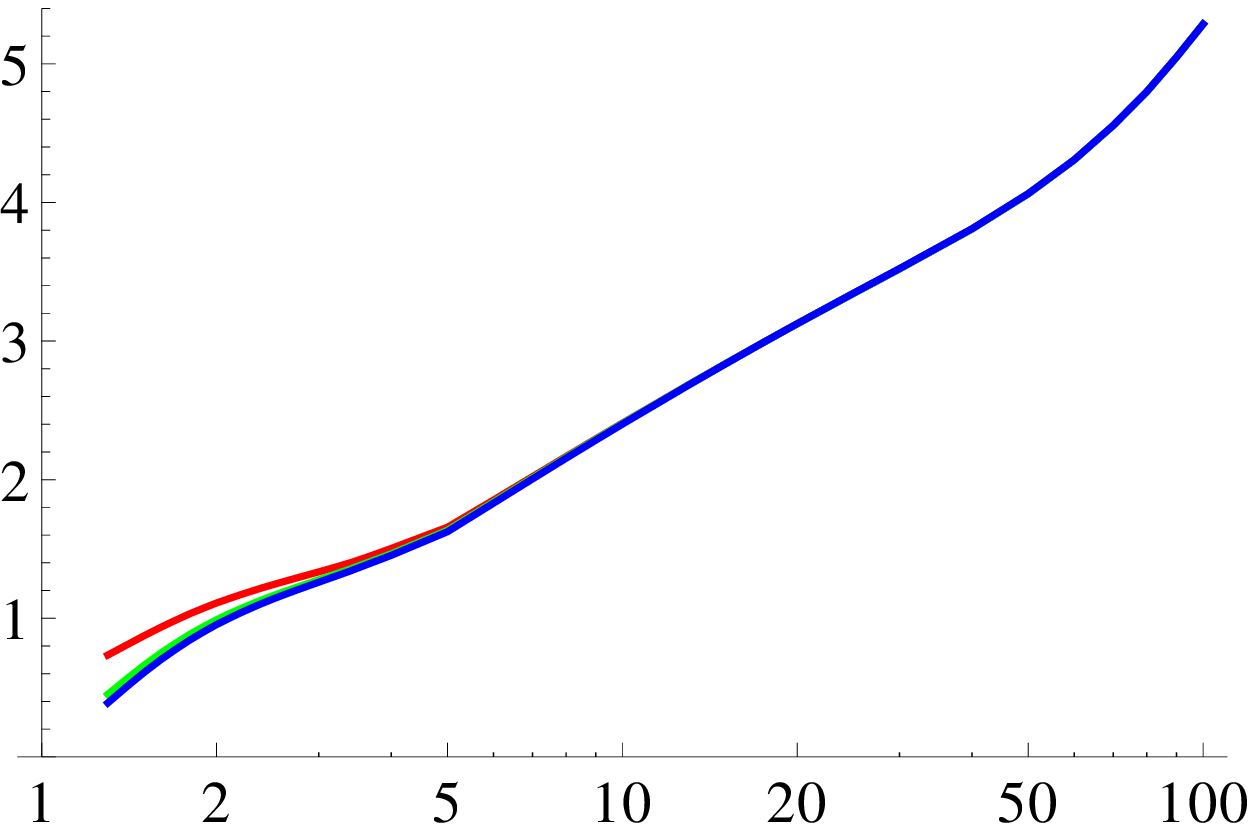}
}
\caption{$F_{2,L}$ vs. $Q$ at N$^3$LO for fixed $x=\{10^{-1},10^{-3},10^{-5}\}$
(left to right). The three lines show the scaling variable: $n=\{0,1,2\}$
(red, green, blue). We observe the effect of the $n$-scaling is negligible
except for very small $Q$ values.}
\end{figure*}

%
\begin{figure}
\includegraphics[width=0.3\textwidth]{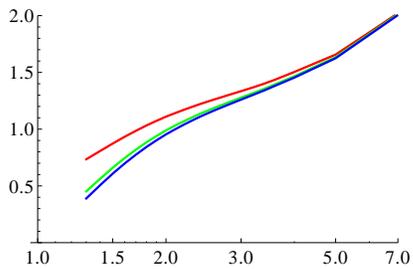}
\caption{Enlargement of Fig.~\ref{fig:fLx135} for $x=10^{-5}$ showing
the small $Q$ region. Here we can distinguish plots for different
scalings; from top to bottom we have $n=\{0,1,2\}$ (red, green, blue).}
\label{fig:clip}
\end{figure}

In Figures~\ref{fig:f2x135} and~\ref{fig:fLx135} we display the
structure functions $F_{2}$ and $F_{L}$, respectively, for selected
$x$ values as a function of $Q$. Each plot has three curves which
are computed using $n$-scalings of $\{0,1,2\}$. We observe that
the effect of the $n$-scaling is negligible except for very small $Q$
values. This result is in part because the heavy quarks are only a fraction
of the total structure function, and the effects of the $n$-scaling
are reduced at larger $Q$ values.

In Fig.~\ref{fig:clip} we magnify the small $Q$ region of $F_L$ of
Fig.~\ref{fig:fLx135} for $x=10^{-5}$, where the effects of using different
scalings are largest. We can see that for  inclusive observables, the  $n=1$ and $n=2$
scalings give nearly identical results, but they differ from the massless
case ($n=0$).
This result,  together with the observation that at NLO kinematic mass
effects are dominant, suggests that the error we have in our approach
is relatively small and approximated  by the band between $n=1$ and $n=2$ results.

\subsection{Flavor Decomposition of $\chi(n)$ Scaling}

%
\begin{figure*}
\subfloat[$F_2^j/F_2$ vs. $Q$.
\label{fig:f2RatN123}]{
\includegraphics[width=0.3\textwidth]{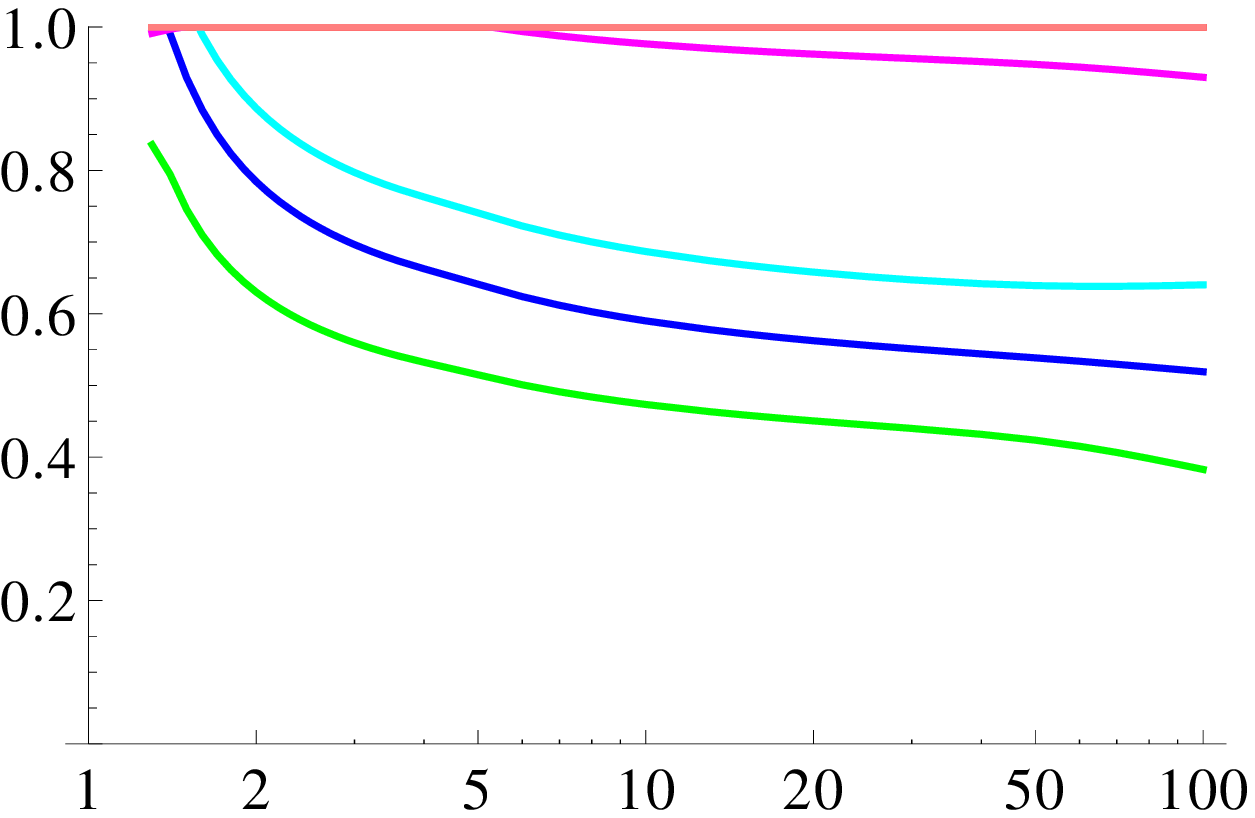}
\hfil
\includegraphics[width=0.3\textwidth]{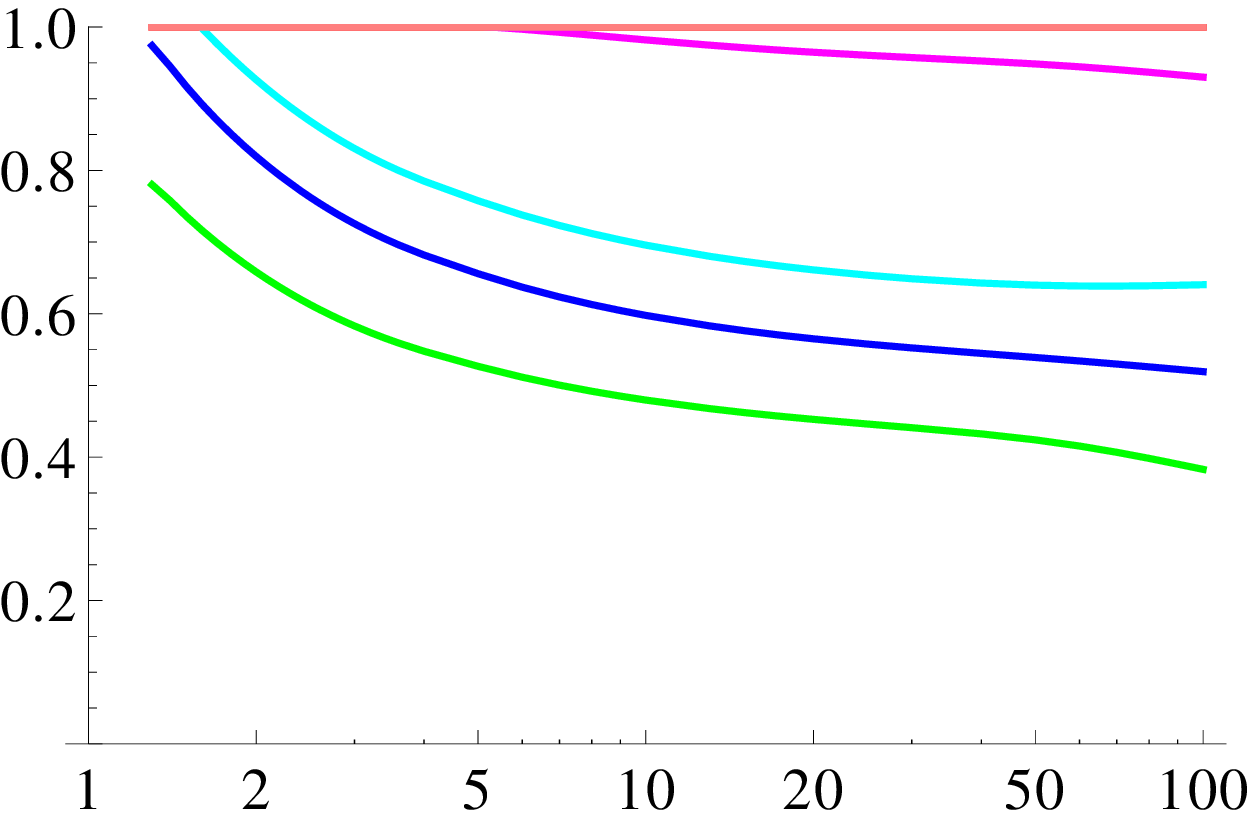}
\hfil
\includegraphics[width=0.3\textwidth]{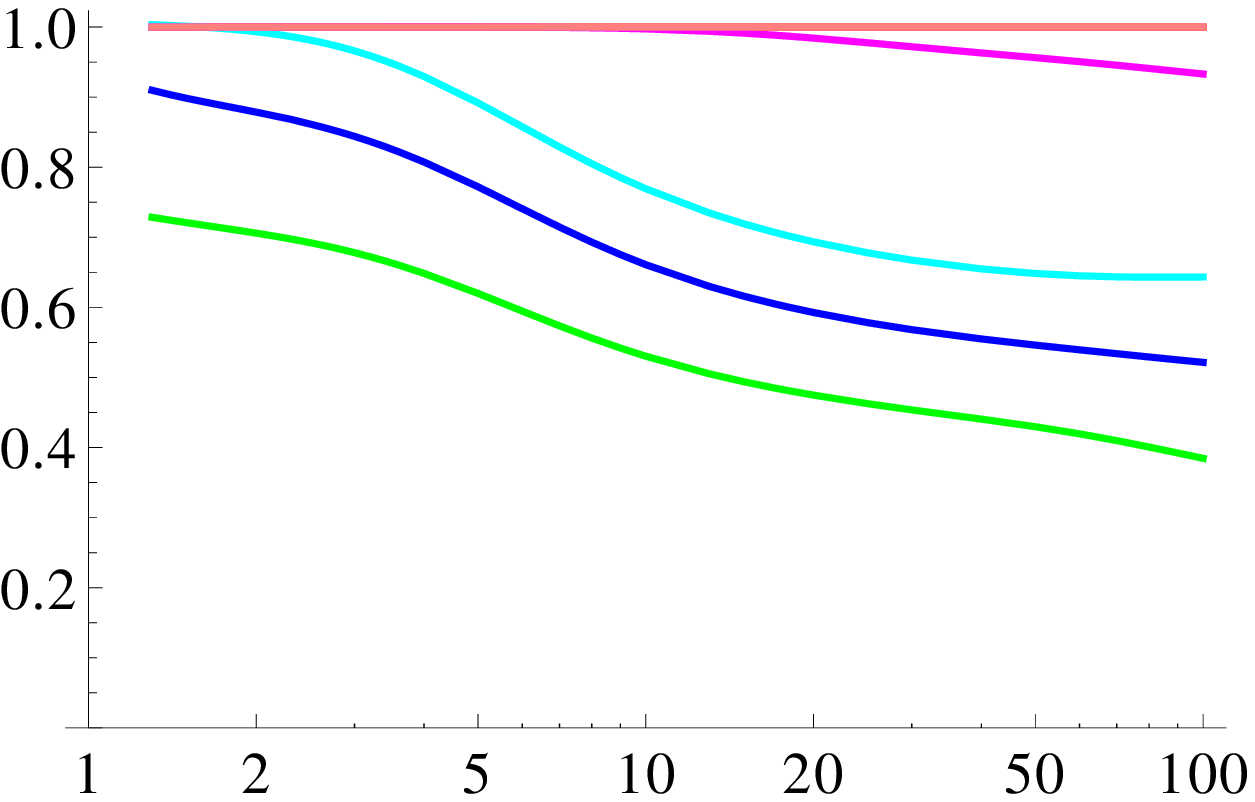}
}

\subfloat[$F_L^j/F_L$ vs. $Q$.
\label{fig:fLRatN123}]{
\includegraphics[width=0.3\textwidth]{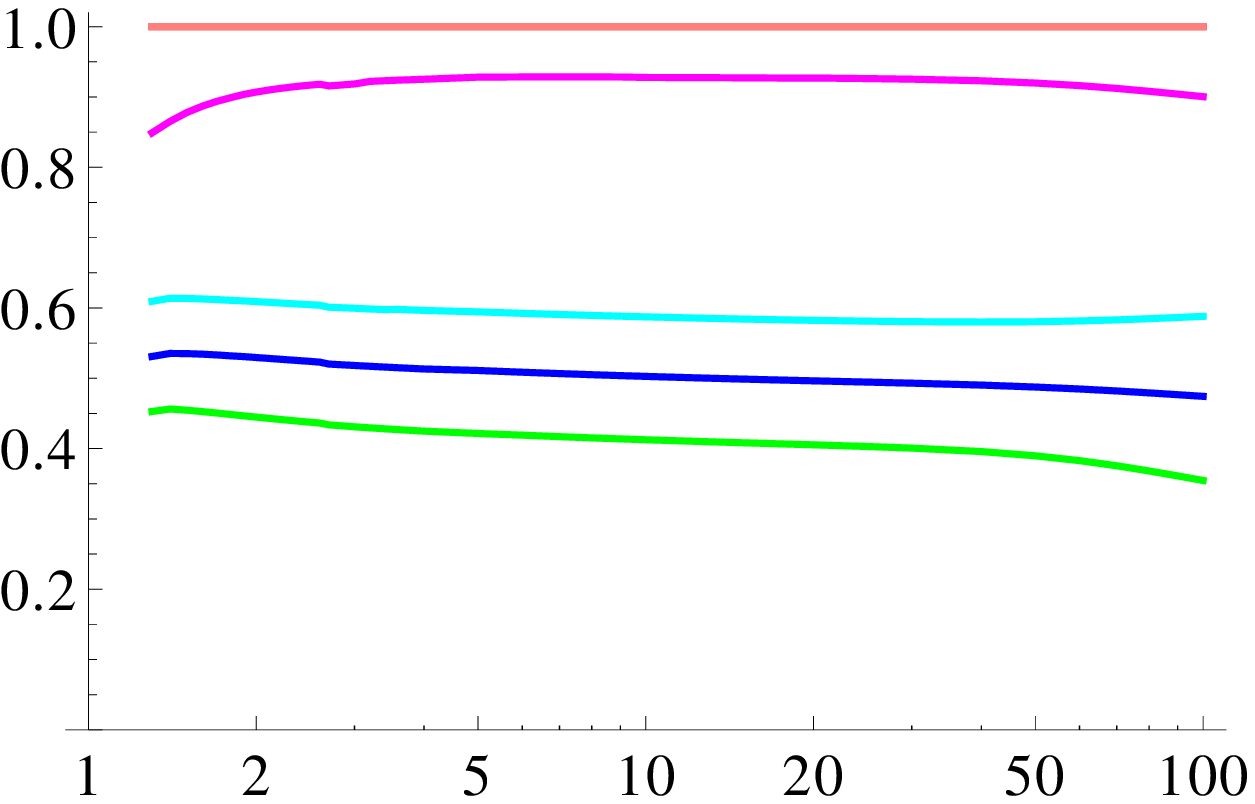}
\hfil
\includegraphics[width=0.3\textwidth]{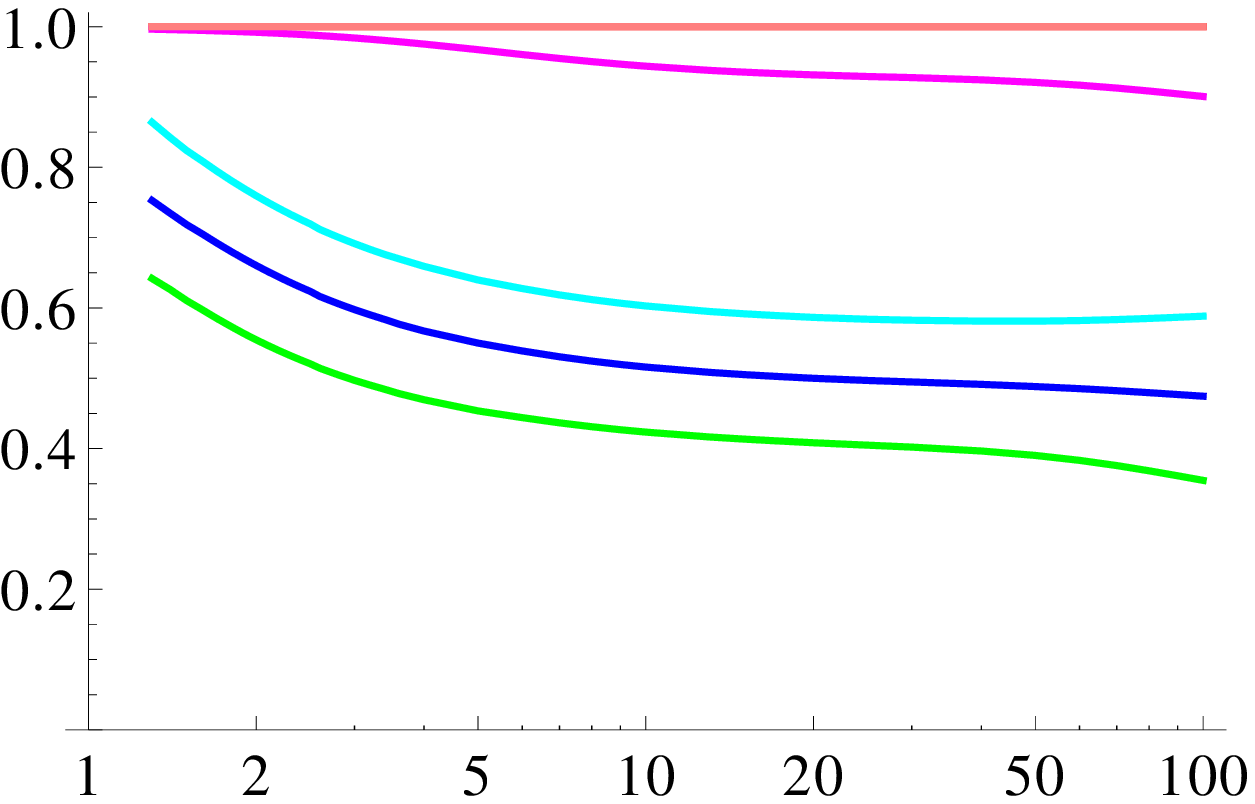}
\hfil
\includegraphics[width=0.3\textwidth]{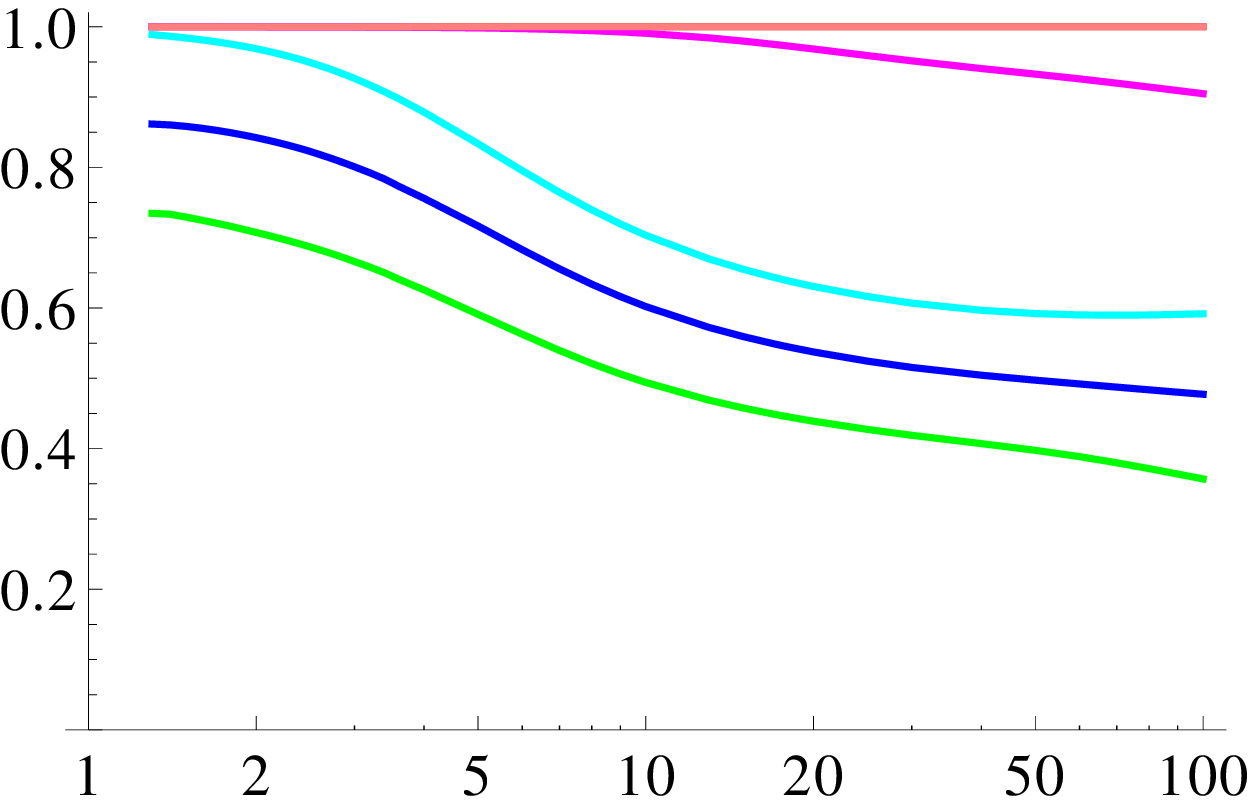}
}
\caption{
Effect of $\chi(n)$-scaling for $n=\{0,1,2\}$ (left to right)
at N$^3$LO for fixed $x=\{10^{-3}\}$.
Reading from the bottom we have fractional contribution for each
quark flavor to $F_{2,L}^j/F_{2,L}$ vs. $Q$
from $\{u,d,s,c,b\}$ (green, blue, cyan, magenta, pink).}
\end{figure*}

We can investigate the effects of the $\chi(n)$-scaling in more details by 
examining the flavor decomposition of the structure functions. 

In Figures~\ref{fig:f2RatN123} and~\ref{fig:fLRatN123} we display
the fractional contributions of quark flavors
to the structure functions $F_{2,L}$ for selected $n$-scaling values as a function
of $Q$.
Flavor decomposition of inclusive structure functions is defined
in appendix~\ref{app:decomposition} in
Eqs.~\eqref{F:decomp} and~\eqref{F:decompFc}.
We observe the $n$-scaling reduces the relative contributions
of charm and bottom at low $Q$ scales. For example, without any $n$-scaling
($n=0$) we find the charm and bottom quarks contribute an unusually
large fraction  at very low scales $(Q\sim m_{c})$
as they are (incorrectly) treated as massless partons in this region.
The result of the different $n$-scalings ($n=1,2$) is to introduce
a kinematic penalty which properly suppresses the contribution of
these heavy quarks in the low $Q$ region. 
In the following, we will generally use the $n=2$ scaling for our comparisons.

\subsection{$F_{2,L}$ Initial-State Flavor Decomposition}

\begin{figure*} 
\subfloat[
$F_2^i/F_2$ vs. $Q$.
\label{fig:F2trans}
]
{
\includegraphics[width=0.3\textwidth]{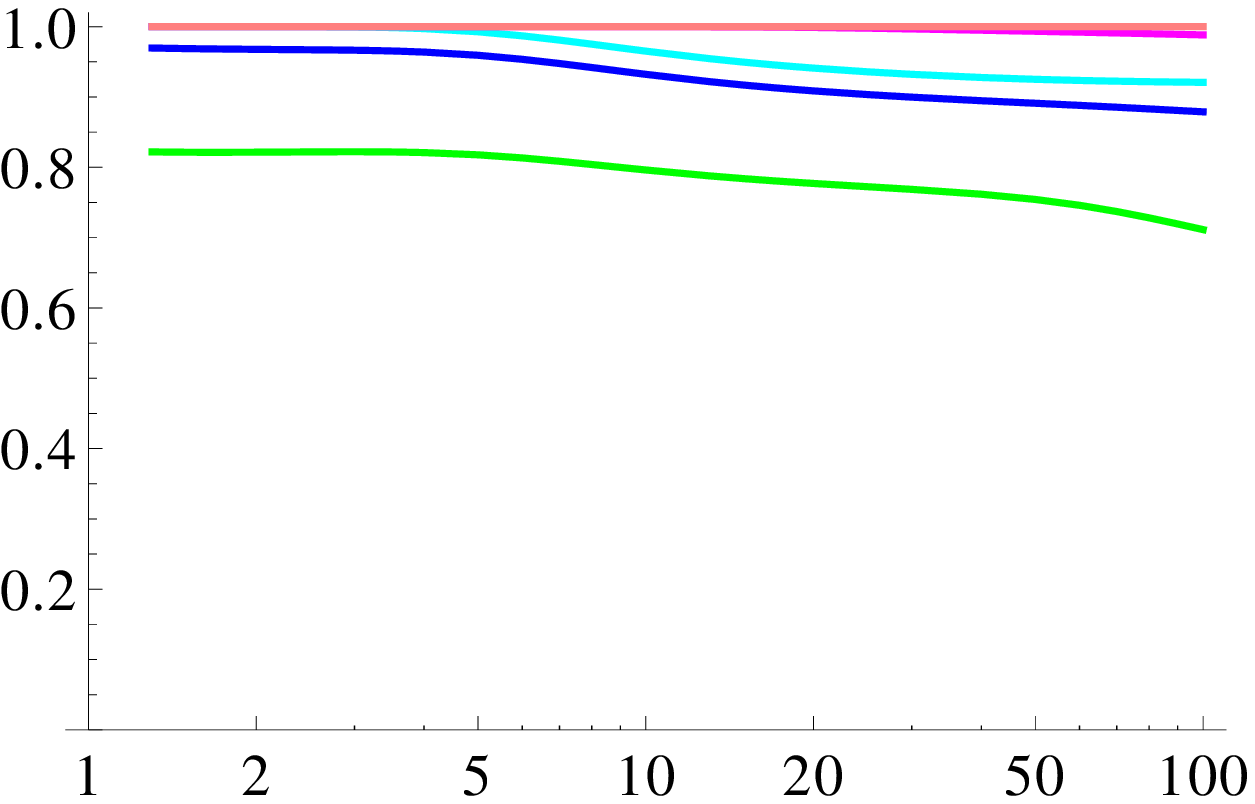}
\hfil
\includegraphics[width=0.3\textwidth]{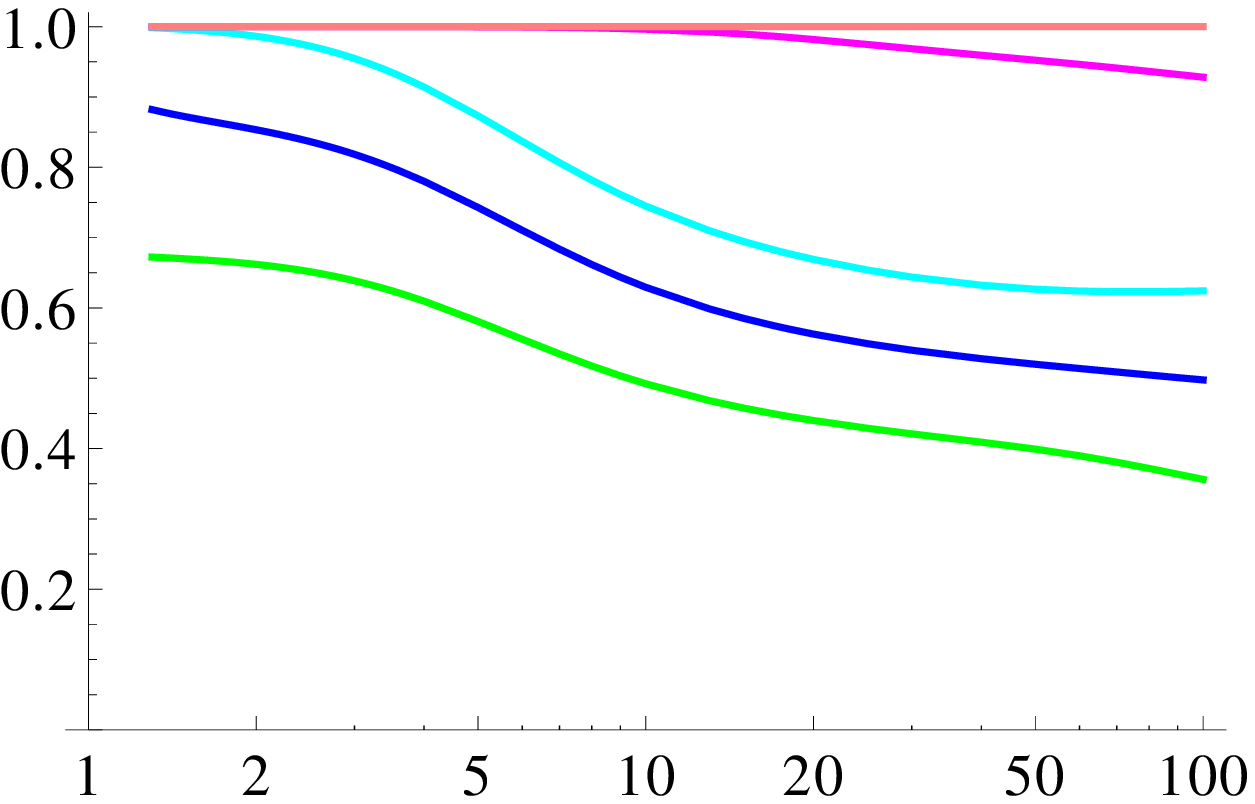}
\hfil
\includegraphics[width=0.3\textwidth]{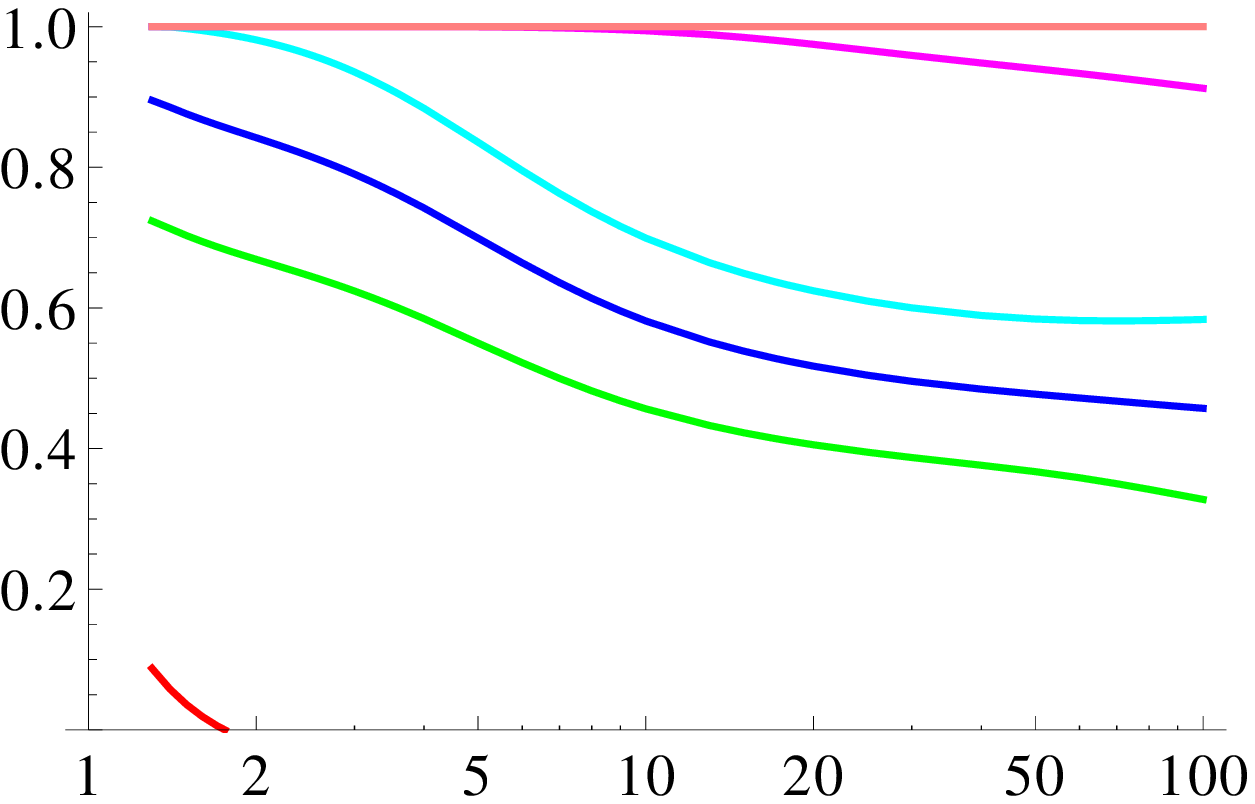}
}

\subfloat[
$F_L^i/F_L$ vs. $Q$.
\label{fig:FLtrans}
]
{
\includegraphics[width=0.3\textwidth]{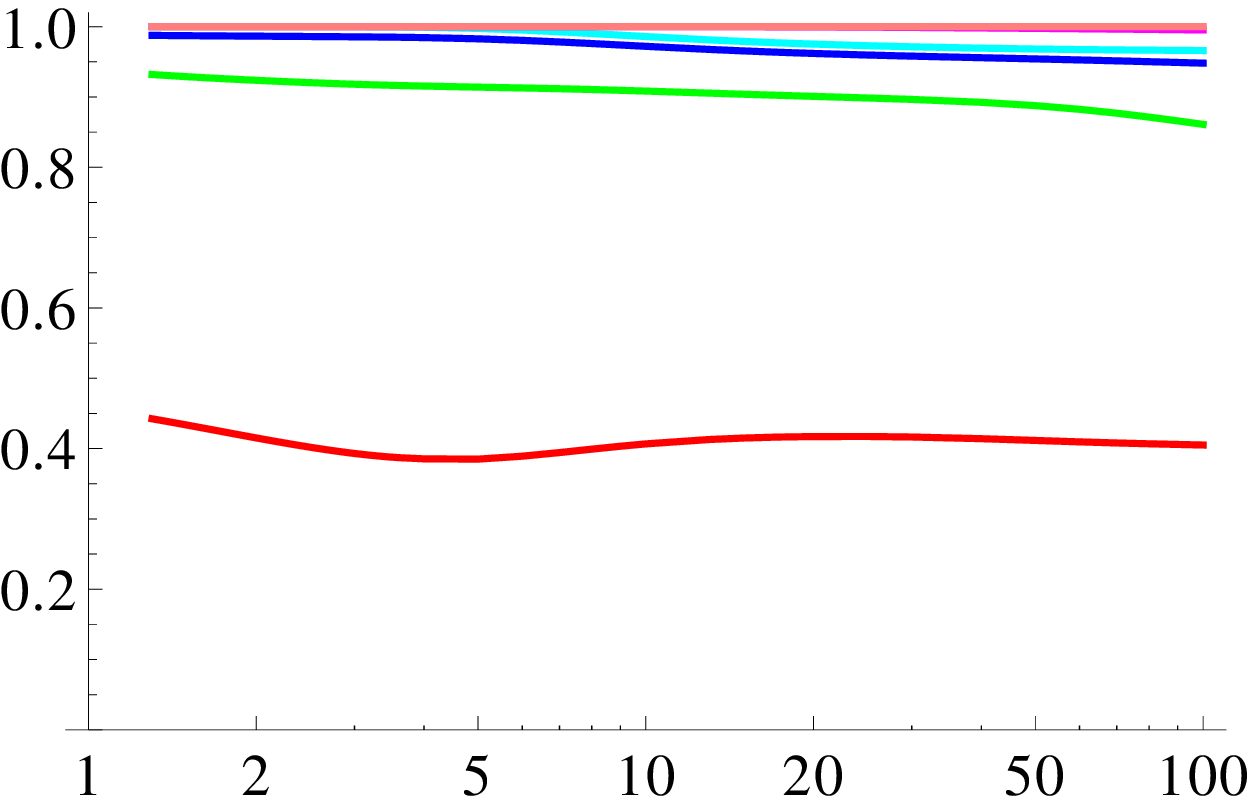}
\hfil
\includegraphics[width=0.3\textwidth]{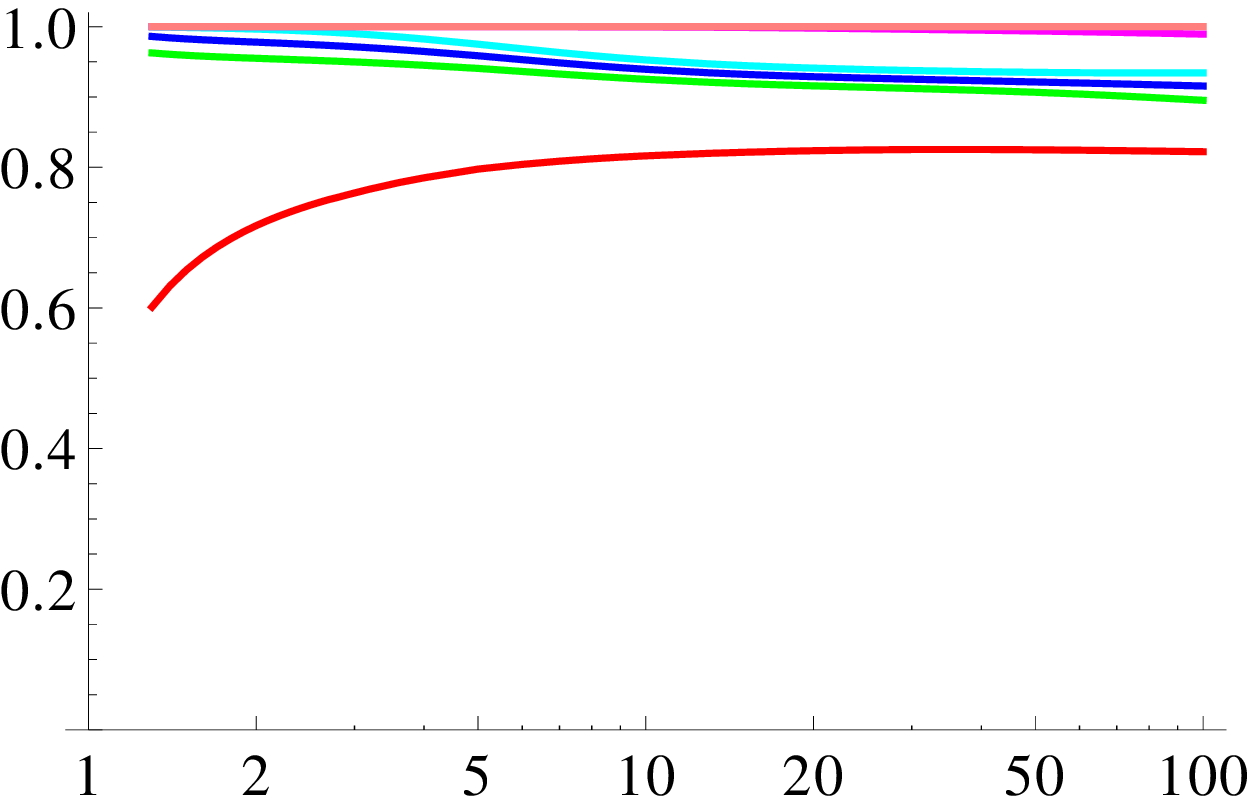}
\hfil
\includegraphics[width=0.3\textwidth]{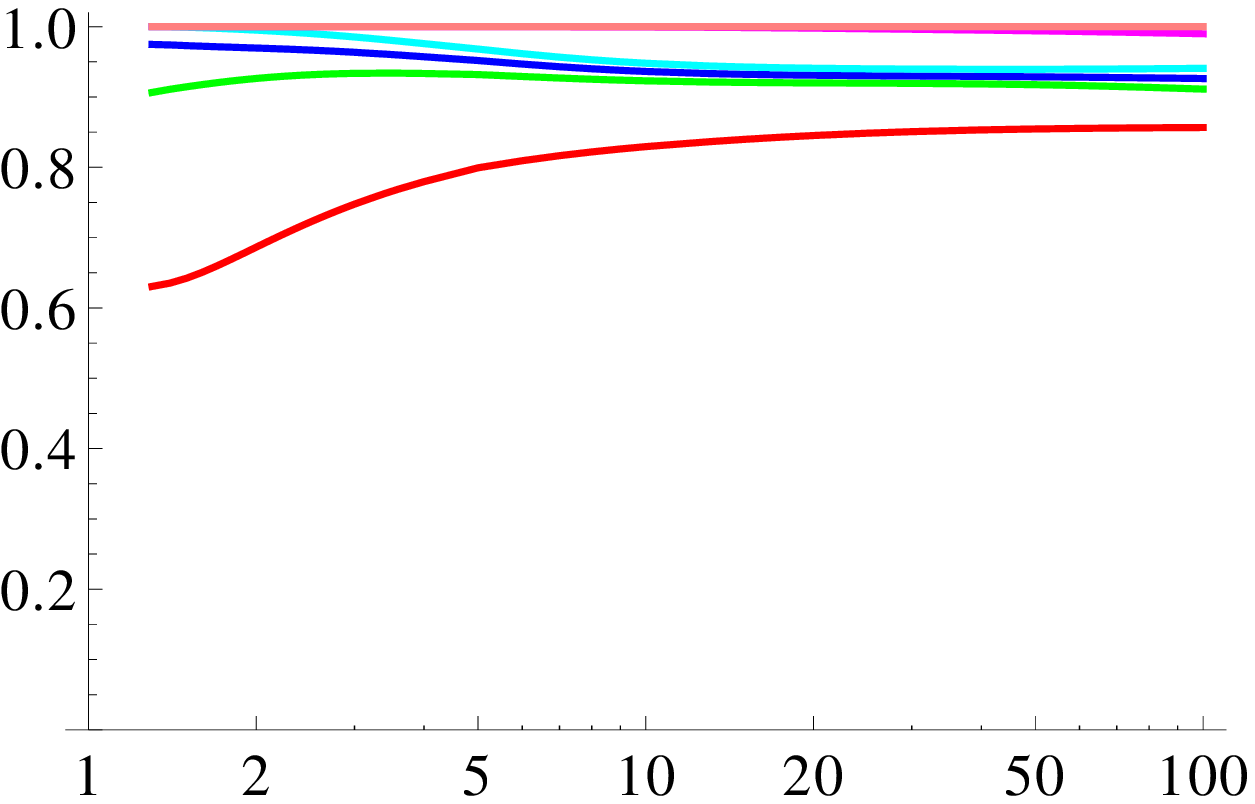}
}
\caption{Fractional flavor decomposition of ``initial-state'' $F_{2,L}^i/F_{2,L}$
vs. $Q$ at N$^3$LO for $x=\{10^{-1},10^{-3},10^{-5}\}$ (left to right)
for $n=2$  scaling.
Reading from the bottom, we plot the cumulative contributions to $F_{2,L}$
from $\{g,u,d,s,c,b\}$, (red, green, blue, cyan, magenta, pink).}
\label{fig:tmp3}
\end{figure*} 

In Figures~\ref{fig:F2trans}  and~\ref{fig:FLtrans} we display
the fractional contributions for the initial-state quarks ($i$)
to the structure functions $F_{2}$ and $F_{L}$,%
\footnote{Fractional decomposition of ``initial-state''
structure functions is understood as
$F_{2,L}^i = \sum_{j=1}^6 F_{2,L}^{ij}$.
}
respectively, for selected $x$ values as a function of
$Q$; here we have used $n=2$ scaling. Reading from the bottom, we
have the cumulative contributions from the $\{g,u,d,s,c,b\}$. 
Although this decomposition is not physically observable,
it is instructive to see which PDFs are dominantly influencing the result.
We observe that for
large $x$ and low $Q$ the heavy flavor contributions are minimal. For
example, for $x=10^{-1}$ we see the contribution of the $u$-quark
comprises $\sim80\%$ of the $F_2$ structure function at low $Q$. In contrast, at $x=10^{-5}$
and large $Q$ we see the $F_2$ contributions of the $u$-quark and $c$-quark
are comparable (as they both couple with a factor 4/9), and the $d$-quark
and $s$-quark are comparable (as they both couple with a factor 1/9).

It is notable that the gluon contribution to $F_{L}$ is significant. 
For  $x=10^{-1}$ this is roughly 40\% throughout the $Q$ range, and can 
be even larger for smaller $x$ values.

\subsection{$F_{2,L}$ Final-State Flavor Decomposition}

%
\begin{figure*}
\subfloat[$F_2^j/F_2$ vs. $Q$.
\label{fig:f2RatX135-1}]{
\includegraphics[width=0.3\textwidth]{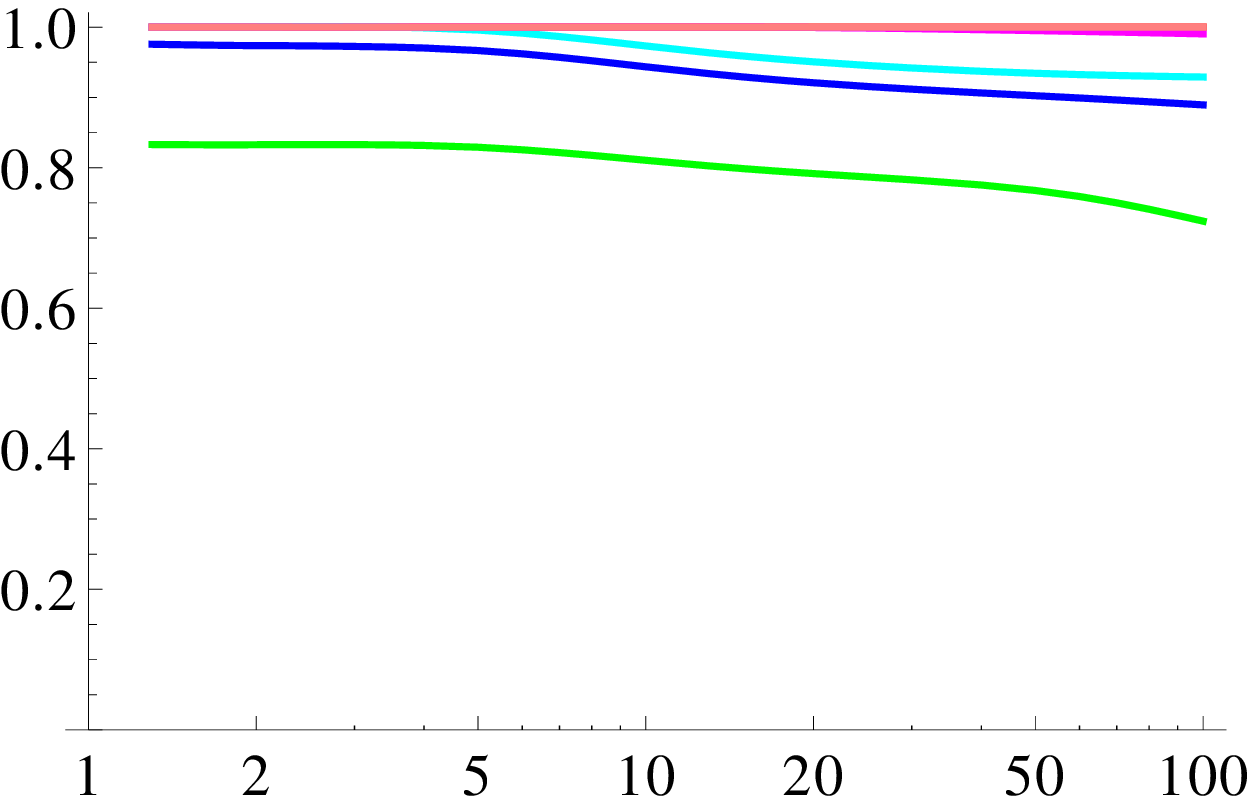}
\hfil
\includegraphics[width=0.3\textwidth]{eps/figN3LO_F2eq11Rat_x3_n2}
\hfil
\includegraphics[width=0.3\textwidth]{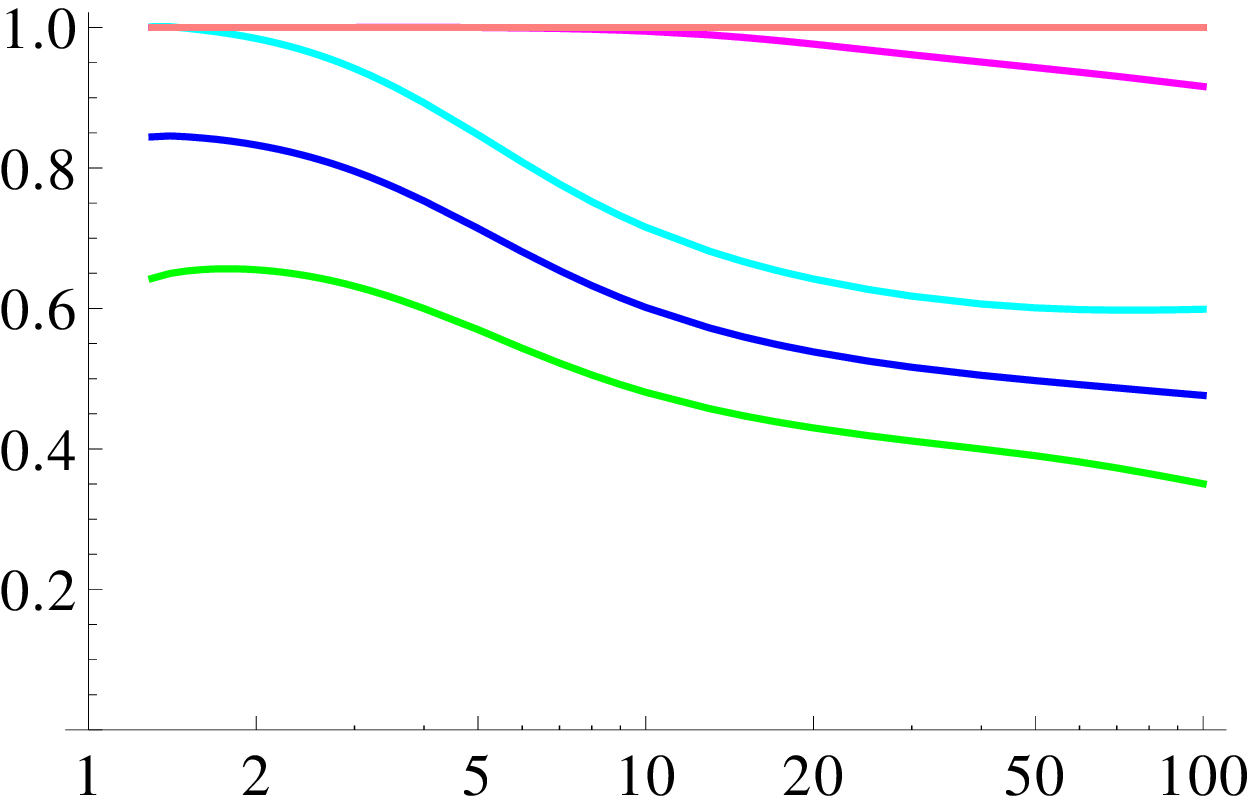}
}

\subfloat[ $F_L^j/F_L$ vs. $Q$.
\label{fig:fLRatX135-1}]{
\includegraphics[width=0.3\textwidth]{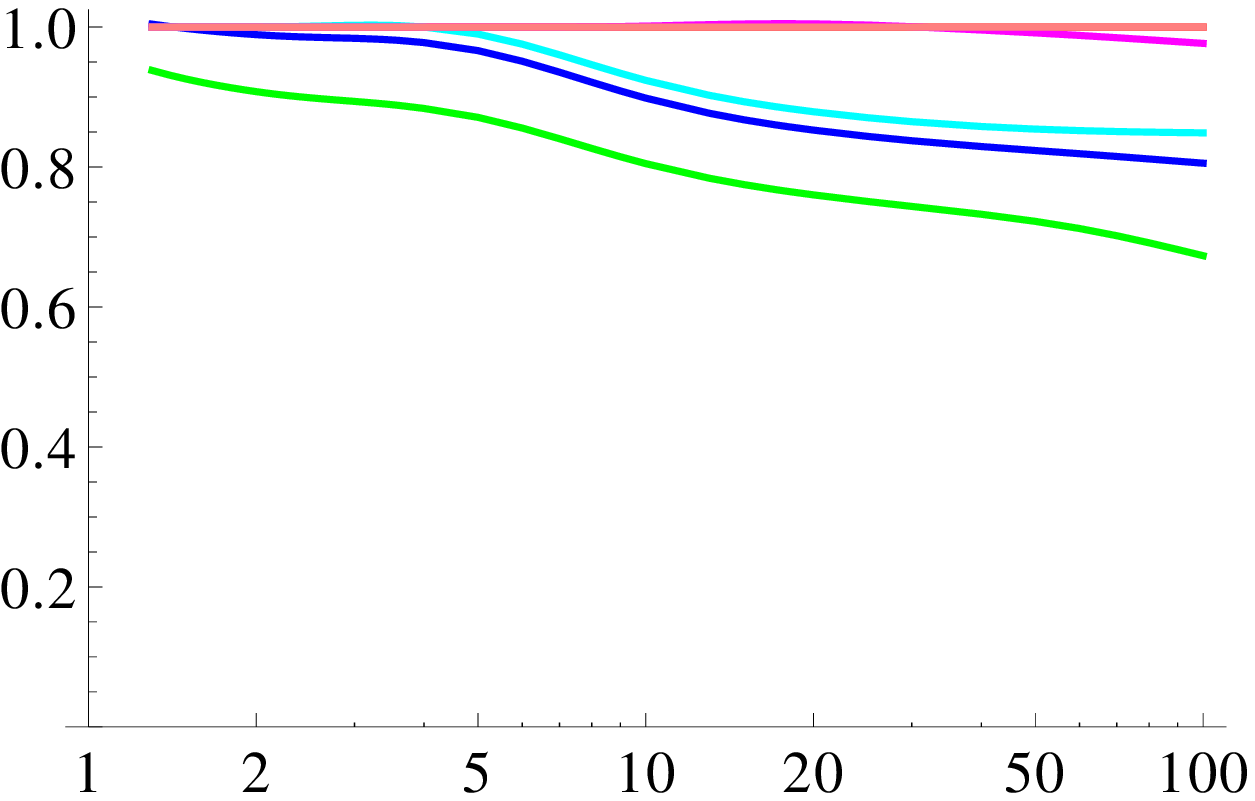}
\hfil
\includegraphics[width=0.3\textwidth]{eps/figN3LO_fLeq11Rat_x3_n2}
\hfil
\includegraphics[width=0.3\textwidth]{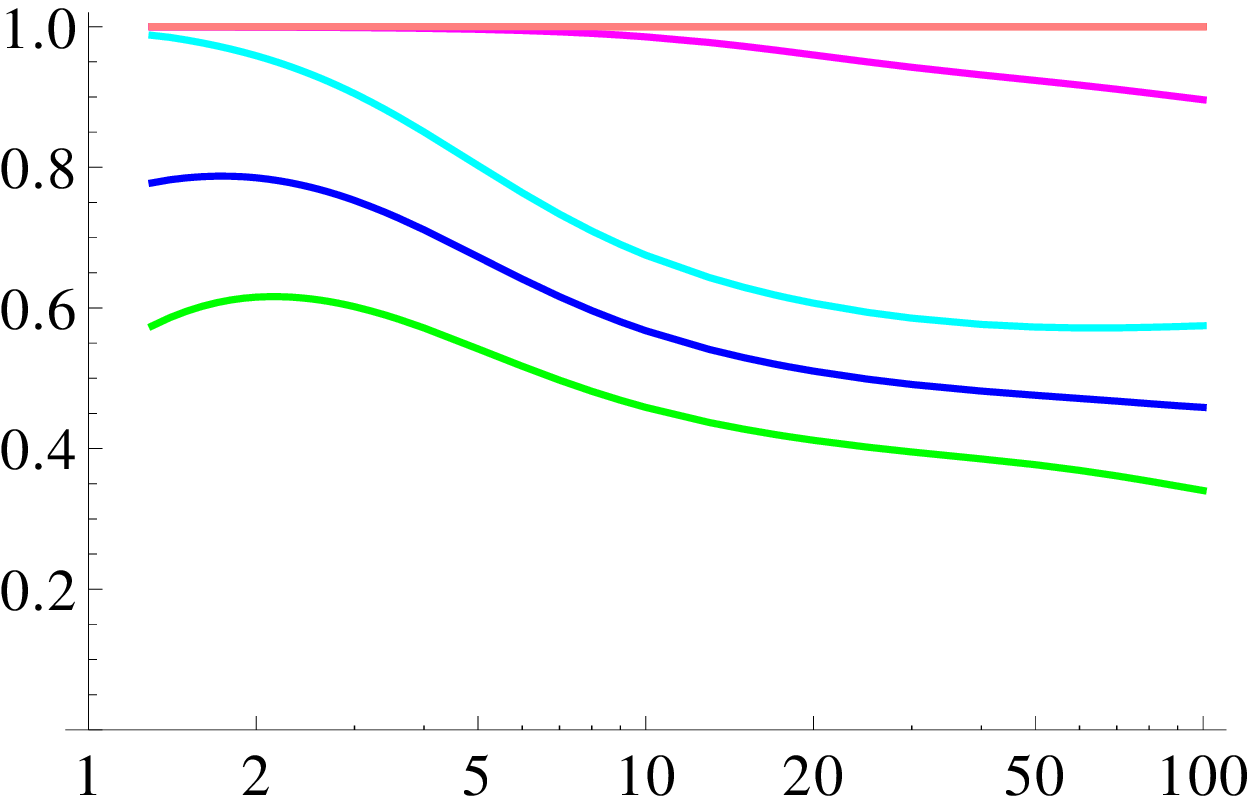}
}
\caption{Fractional contribution for each quark flavor to
$F_{2,L}^j/F_{2,L}$ vs. $Q$ at N$^3$LO for fixed $x=\{10^{-1},10^{-3},10^{-5}\}$
(left to right). Results are displayed for $n=2$ scaling.
Reading from the bottom, we have the cumulative contributions from
the $\{u,d,s,c,b\}$ (green, blue, cyan, magenta, pink).}
\end{figure*}

In Figures \ref{fig:f2RatX135-1} and \ref{fig:fLRatX135-1} we display
the fractional contributions for the final-state quarks ($j$)
 to the structure functions $F_{2}$ and
$F_{L}$, respectively, for selected $x$ values as a function of
$Q$; here we have used $n=2$ scaling. Reading from the bottom, we
have the cumulative contributions from the $\{u,d,s,c,b\}$. 
Again, we observe that for
large $x$ and low $Q$ the heavy flavor contributions are minimal, 
but these can grow quickly as we move to smaller $x$ and larger $Q$.

\subsection{Comparison of LO, NLO, NNLO, N$^3$LO}

%
\begin{figure*}
\subfloat[$F_{2}$ vs. $Q$.
\label{fig:f2orders}]{
\includegraphics[width=0.3\textwidth]{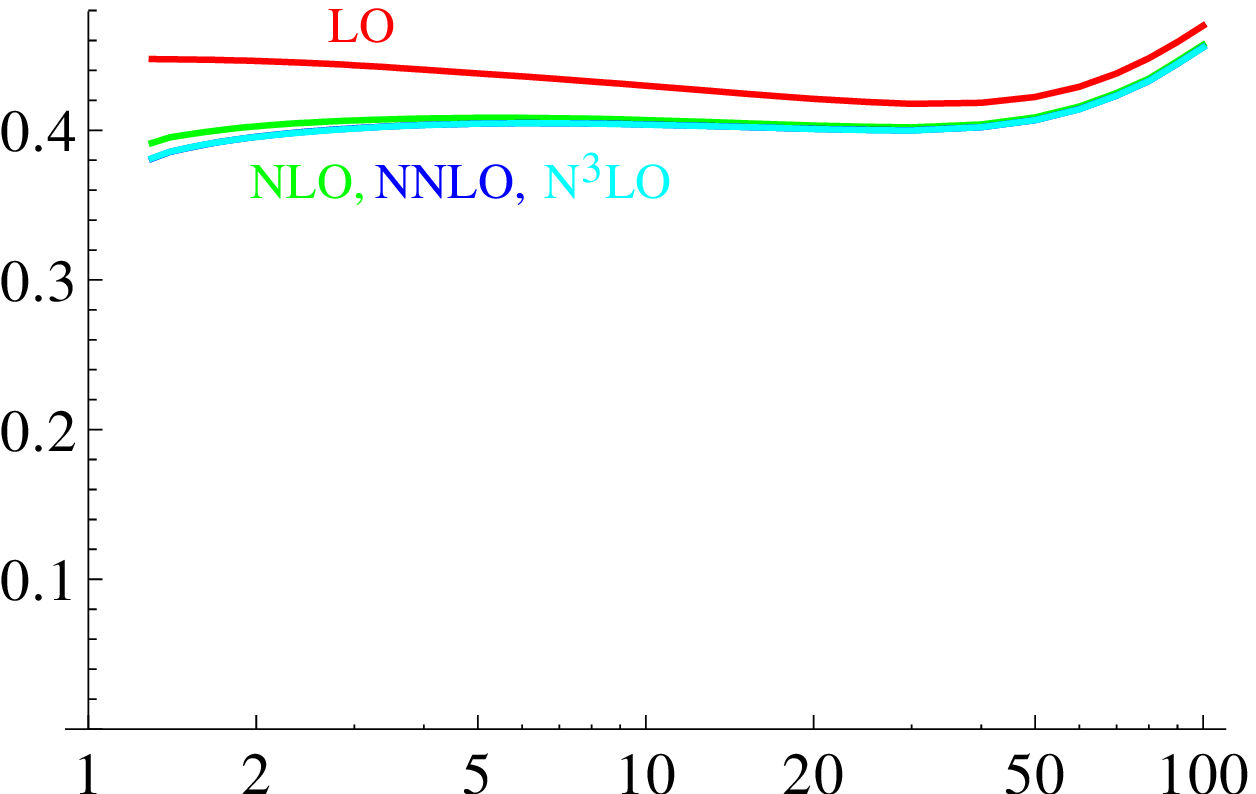}
\hfil
\includegraphics[width=0.3\textwidth]{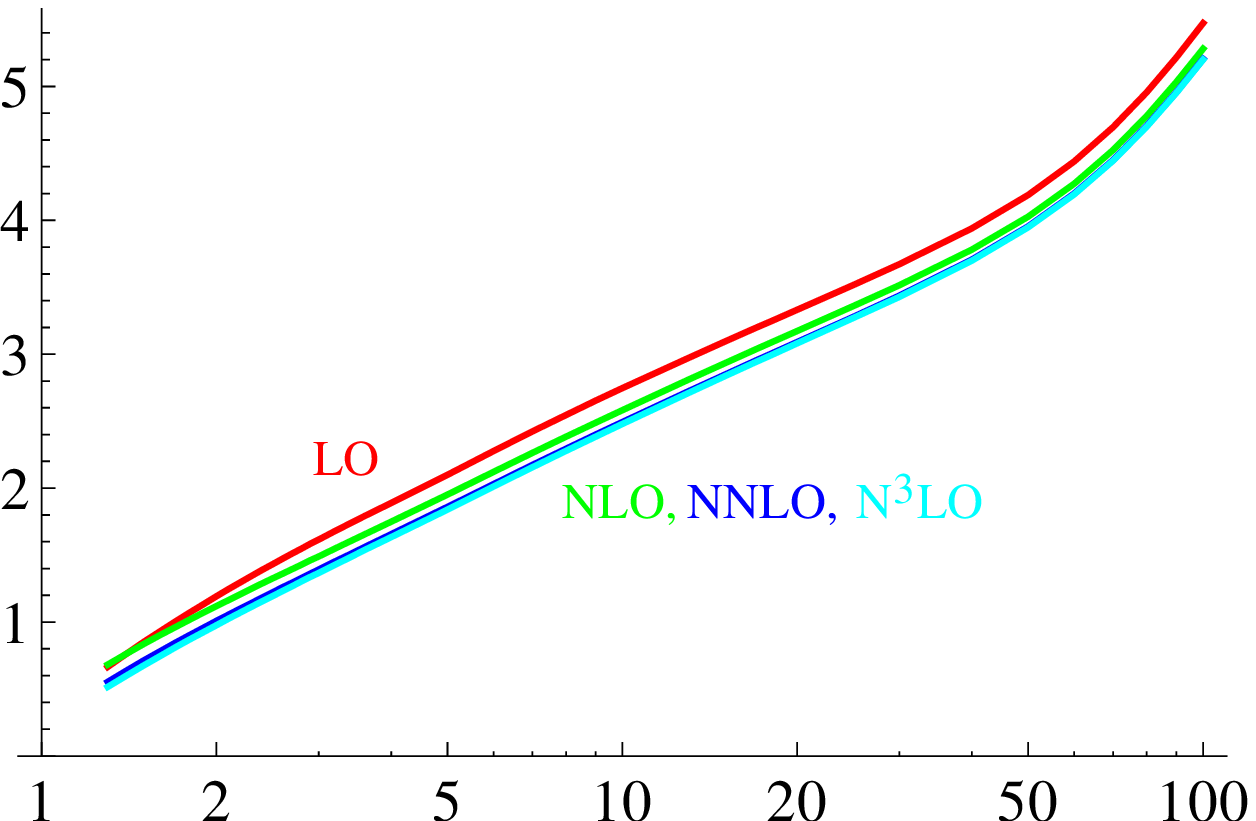}
\hfil
\includegraphics[width=0.3\textwidth]{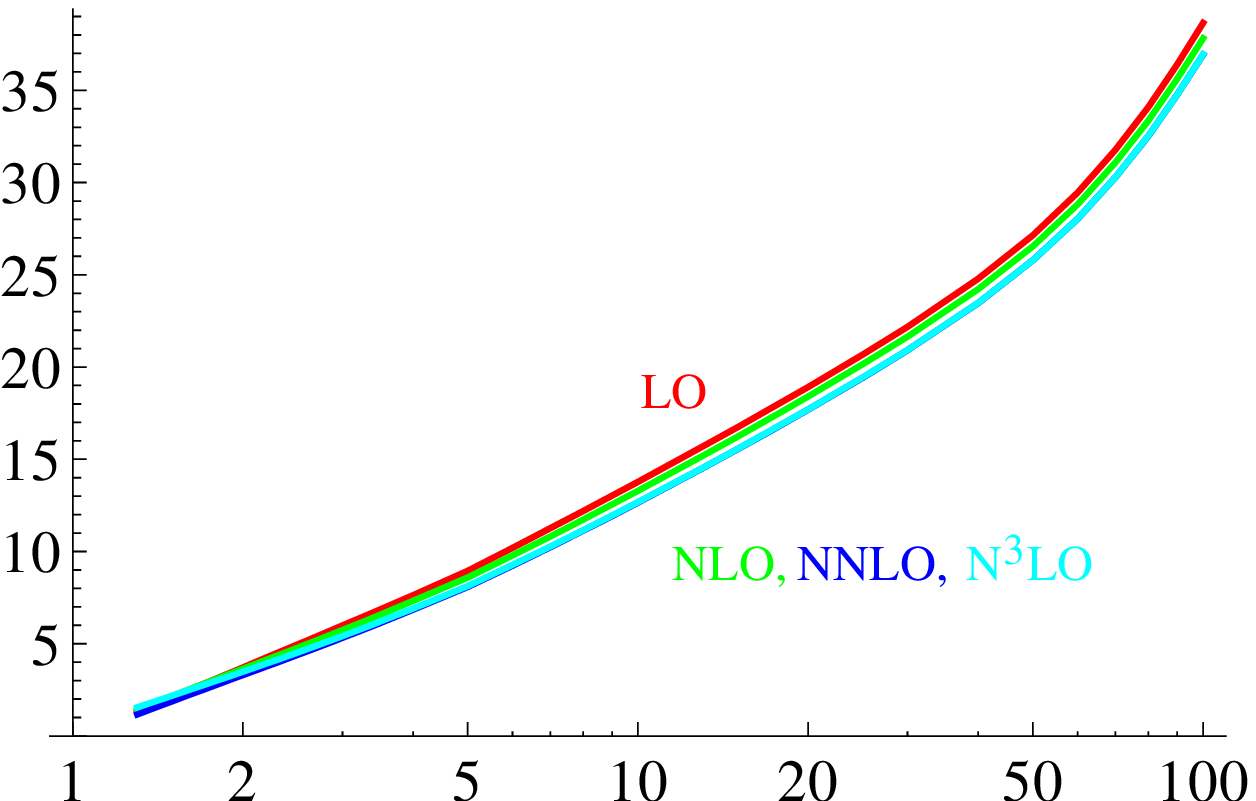}
}

\subfloat[$F_{L}$ vs. $Q$.
\label{fig:fLorders}]{
\includegraphics[width=0.3\textwidth]{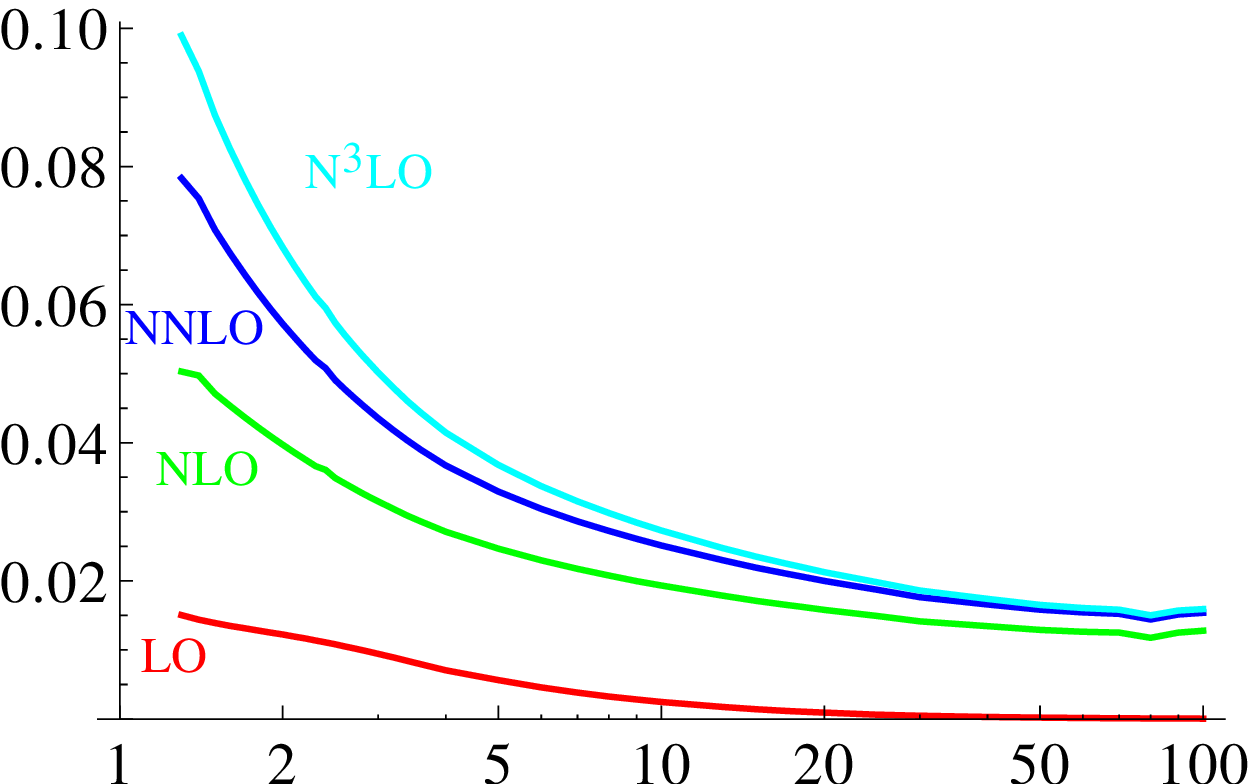}
\hfil
\includegraphics[width=0.3\textwidth]{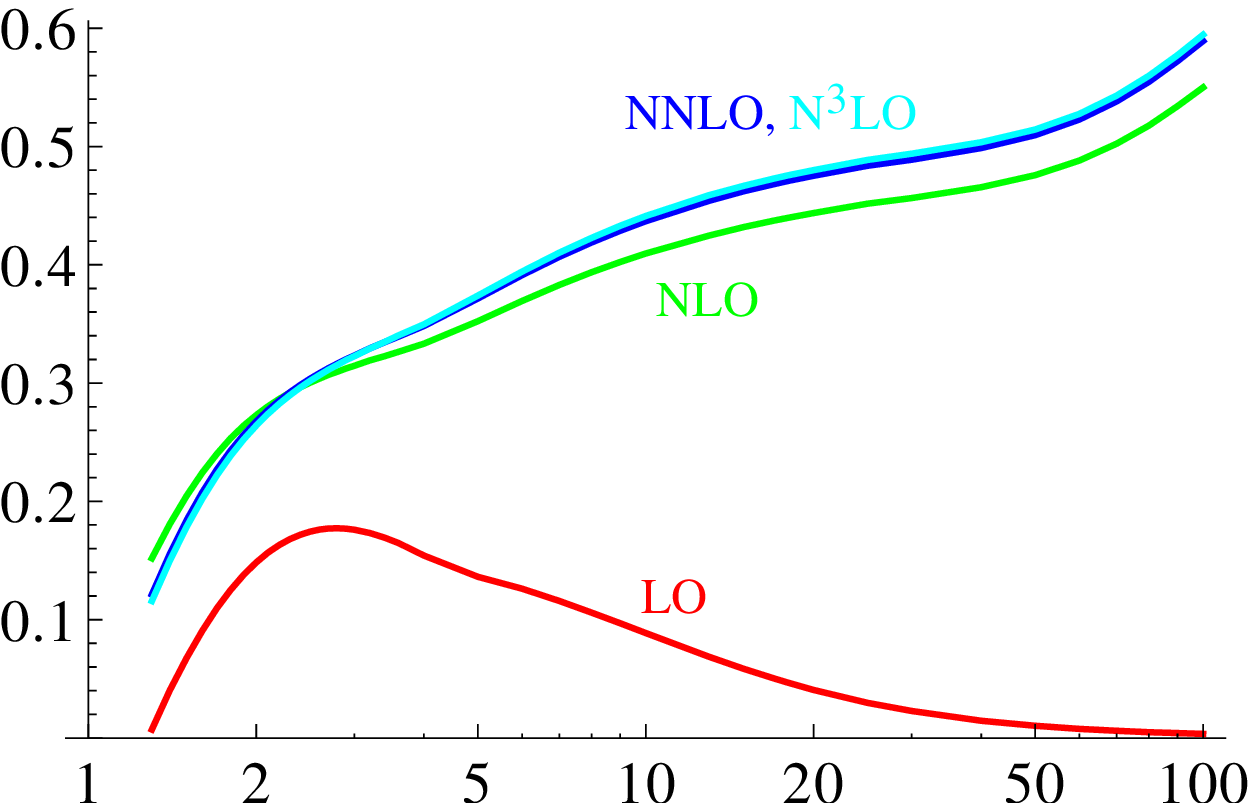}
\hfil
\includegraphics[width=0.3\textwidth]{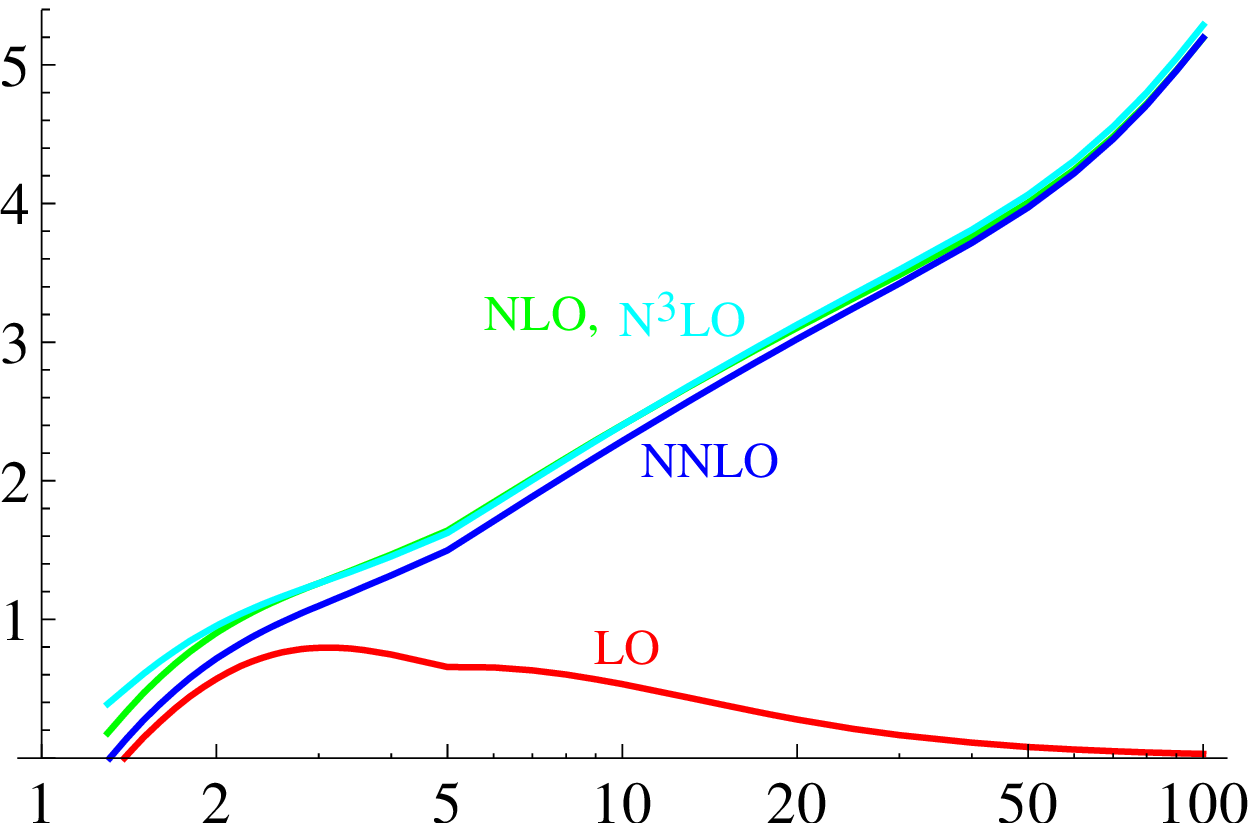}
}

\caption{$F_{2,L}$ vs. $Q$ at \{LO, NLO, NNLO, N$^3$LO\}
(red, green, blue, cyan) for fixed $x=\{10^{-1},10^{-3},10^{-5}\}$
(left to right) for $n=2$ scaling.}
\end{figure*}

In Figure~\ref{fig:f2orders} we display the results for $F_{2}$
vs. $Q$ computed at various orders. For large $x$ (c.f. $x=0.1$) we
find the perturbative calculation is particularly stable; we see
that the LO result is within 20\% of the others at small $Q$, and
within 5\% at large $Q$. The NLO is within 2\% at small $Q$, and
indistinguishable from the NNLO and N$^3$LO for $Q$ values above $\sim10$~GeV.
The NNLO and N$^3$LO results are essentially identical throughout the
kinematic range. For smaller $x$ values ($10^{-3}$, $10^{-5}$) the
contribution of the higher order terms increases. Here, the
NNLO and N$^3$LO coincide for $Q$ values above $\sim5$~GeV, but the NLO
result can differ by $\sim5\%$.

In Figure~\ref{fig:fLorders} we display the results for $F_{L}$
vs. $Q$ computed at various orders. In contrast to $F_{2}$,
we find the NLO corrections are large for $F_L$; this is because the LO
$F_{L}$ contribution (which violates the Callan-Gross relation) is suppressed
by $(m^{2}/Q^{2})$ compared to the dominant gluon contributions which
enter at NLO. Consequently, we observe (as expected) that the LO result
for $F_{L}$ receives large contributions from the higher order 
terms.\footnote{%
Because we use the  fully massive ACOT scheme to LO and NLO, 
 the LO result in  Fig.~\ref{fig:fLorders}  
contains the  $(m^{2}/Q^{2})$ helicity-violating contributions $\sim{\cal O}(\alpha_s^0)$;
hence, it is non-zero.
In the S-ACOT scheme, the LO result for $F_L$ vanishes, but
the NLO result is comparable to the NLO  ACOT result. 
}
Essentially, the NLO is the first non-trivial order for $F_{L}$, and
the subsequent contributions then converge. For example, at large
$x$ (c.f. $x=0.1$) for $Q\sim10$~GeV we find the NLO result yields
$\sim60$ to $80\%$ of the total, the NNLO is a $\sim20\%$ correction,
and the N$^3$LO is a $\sim10\%$ correction. For lower $x$ values ($10^{-3}$,
$10^{-5}$) the convergence of the perturbative series improves, and
the NLO results is within $\sim10\%$ of the N$^3$LO result. Curiously,
for $x=10^{-5}$ the NNLO and N$^3$LO roughly compensate each other so
that the NLO and the N$^3$LO match quite closely for $Q\geq2$~GeV.

While the calculation of $F_L$ is certainly more challenging,
examining Fig.~\ref{fig:slac-4-4} we see that for most of the relevant
kinematic range probed by HERA the theoretical calculation is quite
stable.
For example, in the high $Q^2$ region where HERA is probing
intermediate $x$ values ($x\sim 10^{-3}$) the spread of the $\chi(n)$
scalings is small. The challenge arises in the low $Q$ region ($Q\sim
2$~GeV) where the $x$ values are $\sim 10^{-4}$; in this region, there
is some spread between the various curves at the lowest $x$ value
$(\sim 10^{-5})$, but for $x\sim 10^{-3}$ this is greatly reduced.

\section{Conclusions\label{sec:conclusion} }

We extended the ACOT calculation for DIS structure functions 
to  N$^3$LO by combining  the exact ACOT
scheme at NLO with a $\chi(n)$-rescaling; this allows us to
include the leading mass dependence at NNLO and N$^3$LO.
Using the full ACOT calculation at NLO, we demonstrated that the heavy
quarks mass dependence for the DIS structure functions is dominated by
the kinematic mass contributions, and this can be implemented via a
generalized $\chi(n)$-rescaling prescription.

We studied the $F_2$ and $F_L$ structure functions as a function of
$x$ and $Q$.  We examined the flavor decomposition of these structure
functions, and verified that the heavy quarks were appropriately
suppressed in the low $Q$ region.
We found the results for $F_2$ were very stable across the full kinematic range
for $\{x,Q\}$, and the contributions from the NNLO and N${}^3$LO terms
were small.
For $F_L$, the higher order terms gave a proportionally larger
contribution (due to the suppression of the LO term from the
Callan-Gross relation); nevertheless, the contributions from the NNLO
and N${}^3$LO terms were generally small in the region probed by HERA.

The result of this calculation was to obtain precise predictions for
the inclusive $F_2$ and $F_L$ structure functions which can be used
to analyze the HERA data.

\appendix
\section{Kinematic Relations \label{subsec:tmc}}

\subsection{Target Mass Contributions}

In the DIS process, 
the effect of the target mass ($M$) on the scaling variable is a multiplicative correction 
factor
\begin{eqnarray}
\eta &=&
\frac{2x}{1+\sqrt{1+\frac{4x^{2}M^{2}}{Q^{2}}}}
\nonumber \\
&\displaystyle\mathop{\longrightarrow}_{M\to0}&
 x\left[1-\left(\frac{xM}{Q}\right)^{2}\right]+... \quad .
\label{eq:eta}
\end{eqnarray}
This is used in Table~\ref{tab:scaling} to modify the scaling variable~\cite{Aivazis:1993kh,Schienbein:2007gr}.

\subsection{Barnett Scaling}

If we consider the charged-current DIS process for charm production, 
this takes place via the subprocess
$W^+(q)\, s(\xi P) \to c(k)$.
If we impose  4-momentum conservation, we have 
$\left(q+\xi P\right)^{2}=k^2=m_{c}^{2}$. 
Defining $q^2=-Q^2$ and  $x=Q^{2}/(2p\cdot q)$, 
we obtain the traditional ``slow rescaling'' relation~\cite{Barnett:1976ak}
\[
\xi=x\left(1+\frac{m_{c}^{2}}{Q^{2}}\right)
\]
which was used in Eq.~\eqref{eq:barnett}.

\subsection{$\widehat{W}$ constraints \label{subsec:w}}

If we compute the invariant mass $\widehat{W}$ of a boson of momentum $q$ scattering from 
a light parton $a$ of momentum $p_a=\xi P$, we find~\cite{Guzzi:2011dk}
\begin{equation}
\widehat{W} = (p_a + q)^2 = Q^2 (\xi/x - 1) \quad .
\end{equation}
If the partonic final state has a minimum invariant mass  $\widehat{W}_{min}=4 m^2$, then 
$\xi$ is constrained by
\begin{equation}
1 \geq \xi \geq \chi \geq x
\end{equation}
where 
$\chi=x(1+4m^2/Q^2)$. 
This is the relation used in Eq.~\eqref{eq:chi}.
This choice will ensure  $\widehat{W}\geq \widehat{W}_{min}$ is satisfied. 
While this constraint is important in the large $x$ region, this may be too restrictive in the 
small $x$ region---especially as this is the region where the HERA data is very precise.

\section{Decomposition of the Wilson coefficients}
\label{app:decomposition} 

\begin{figure}[t]
 \begin{center} 
  \includegraphics[scale=0.23]{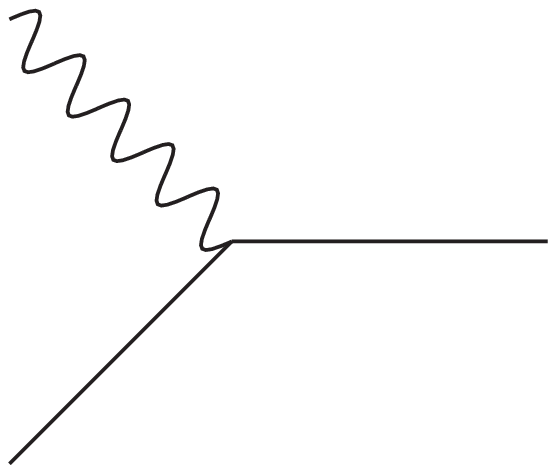} 
  \caption{${\cal O} (\alpha_S^0)$ - $\gamma^*q_i\rightarrow q_i$.
Contributes to $\cqns{a}$ (and hence to $\cqs{a}$) but not to $\cqps{a}$.}
  \label{fig:q2q_0}
 \end{center}
\end{figure}
\begin{figure}[t]
 \begin{center} 
  \includegraphics[scale=0.23]{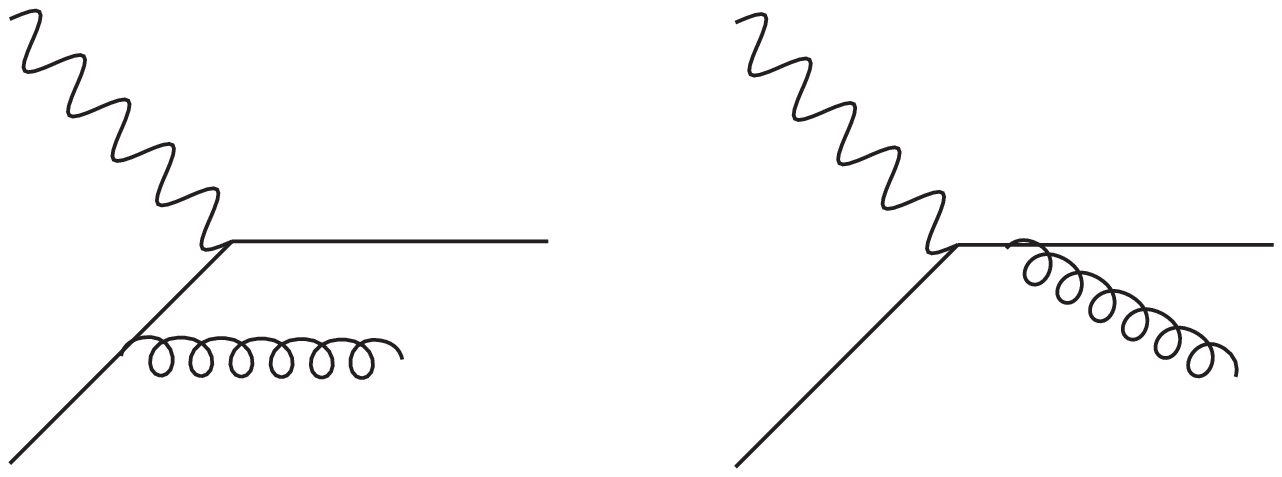} 
  \caption{${\cal O} (\alpha_S^1)$ - $\gamma^*q_i\rightarrow q_i g$.
Contributes to $\cqns{a}$ (and hence to $\cqs{a}$) but not to $\cqps{a}$.
This contribution does not depend on $n_f$.
}
  \label{fig:q2qg_1}
 \end{center}
\end{figure}
\begin{figure}[t]
 \begin{center} 
  \includegraphics[scale=0.23]{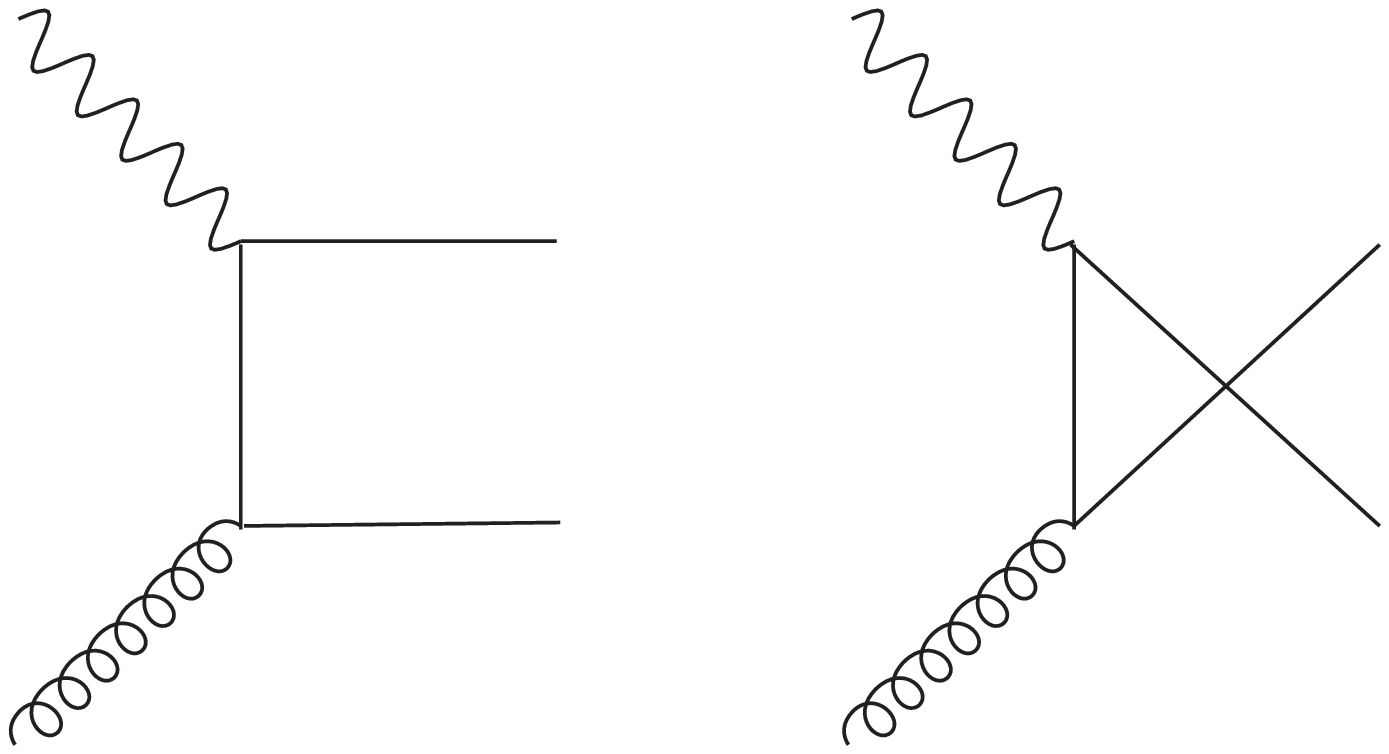}  
 \caption{${\cal O} (\alpha_S^1)$ - $\gamma^*g\rightarrow q_j\bar q_j$. }
  \label{fig:g2qqb_1}
 \end{center}
\end{figure}
\begin{figure}[t]
 \begin{center} 
  \includegraphics[scale=0.23]{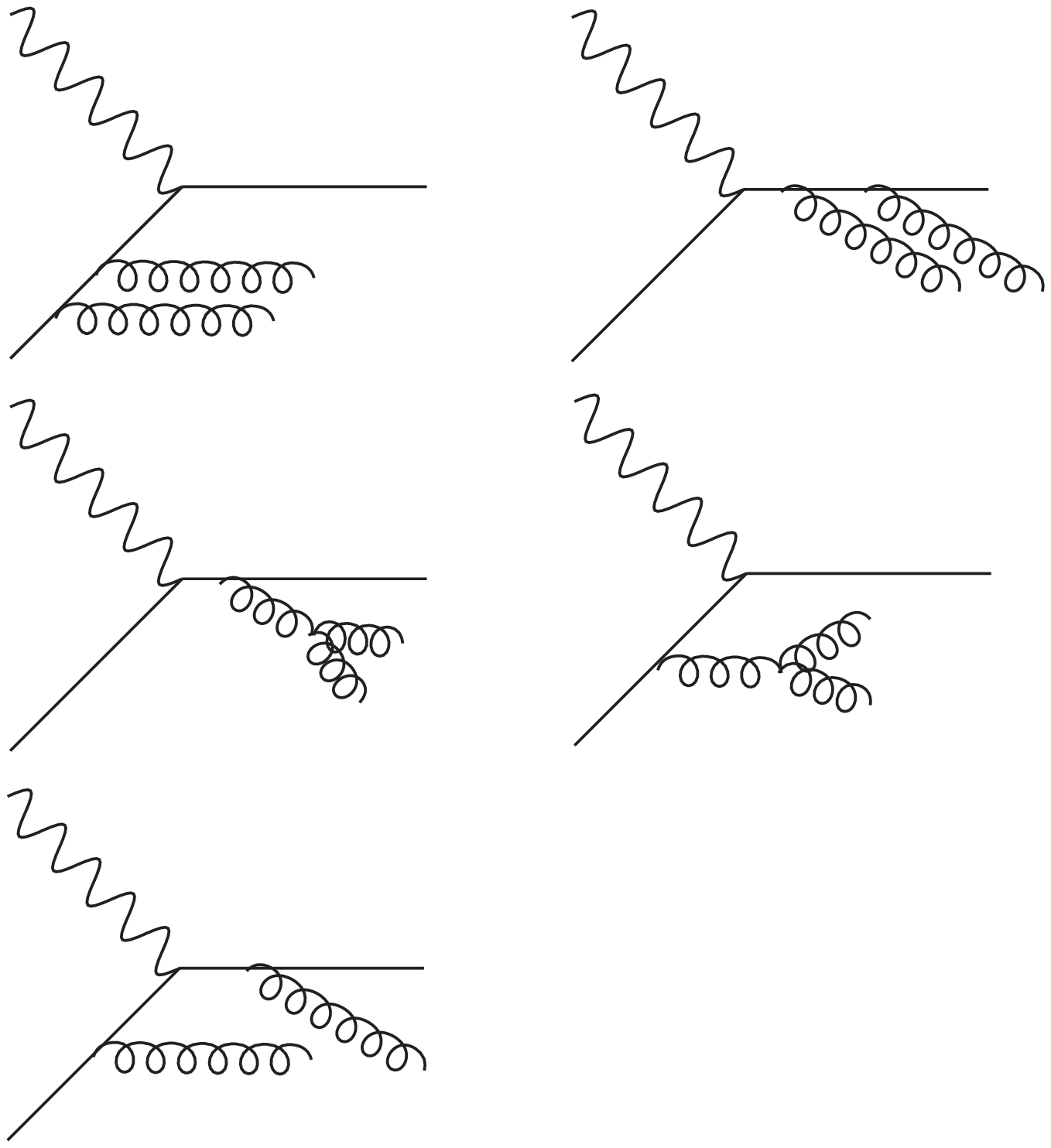}  
  \caption{${\cal O} (\alpha_S^2)$ - $\gamma^*q_i\rightarrow q_igg$.
Contributes to $\cqns{a}$ (and hence to $\cqs{a}$) but not to $\cqps{a}$.
This part is independent of $n_f$.
 }
  \label{fig:q2qgg_2}
 \end{center}
\end{figure} 
\begin{figure}[t]
 \begin{center} 
  \includegraphics[scale=0.23]{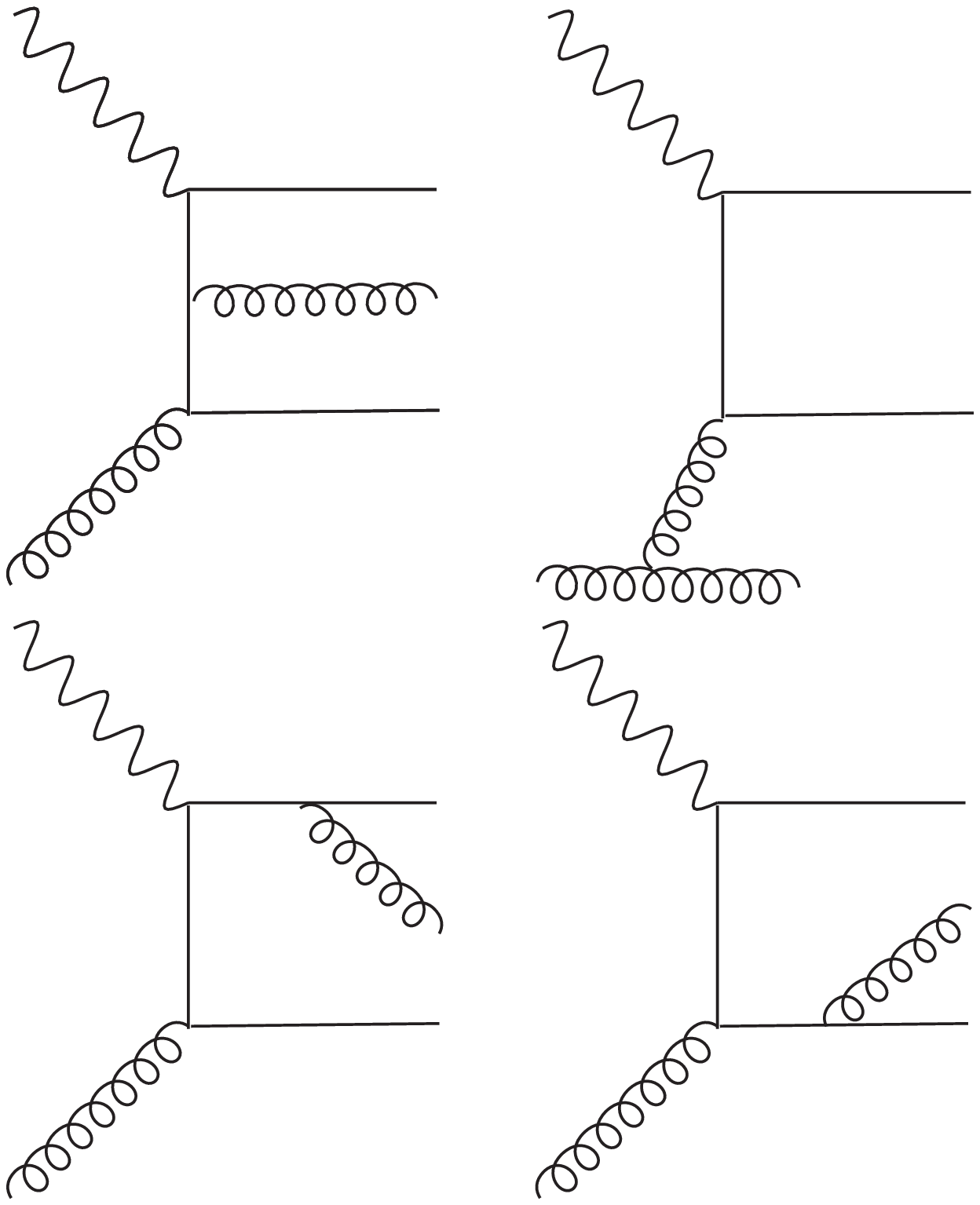} 
  \caption{${\cal O} (\alpha_S^2)$ - $\gamma^*g\rightarrow q_j\bar q_j g$. }
  \label{fig:g2qqbg_2}
 \end{center}
\end{figure}
%
\begin{figure}
\subfloat[Contribution proportional to $n_f$ for $\cqps{a}$.
\label{fig:q2qq'qb'_2_a}]{
\includegraphics[scale=0.23]{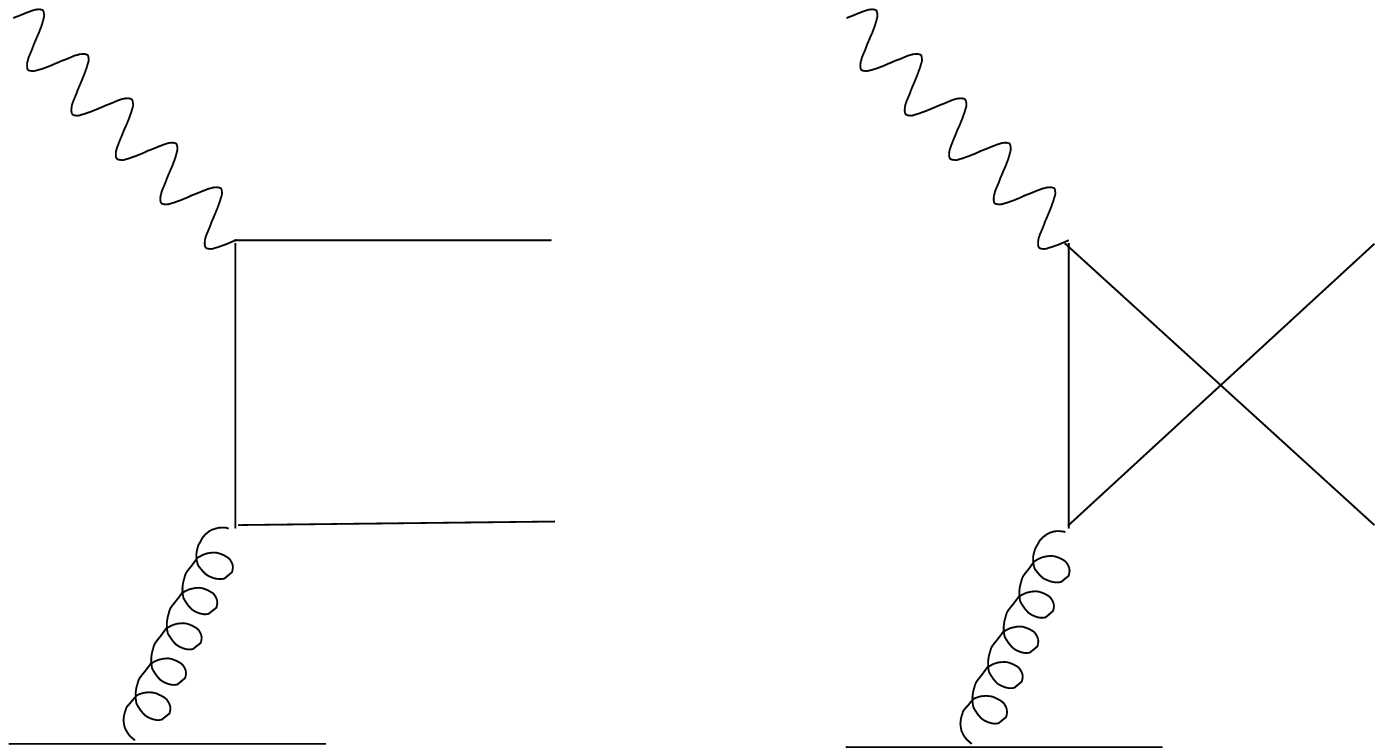}
}
\\
\subfloat[Contribution proportional to $n_f$ for $\cqns{a}$.
\label{fig:q2qq'qb'_2_b}]{
\includegraphics[scale=0.23]{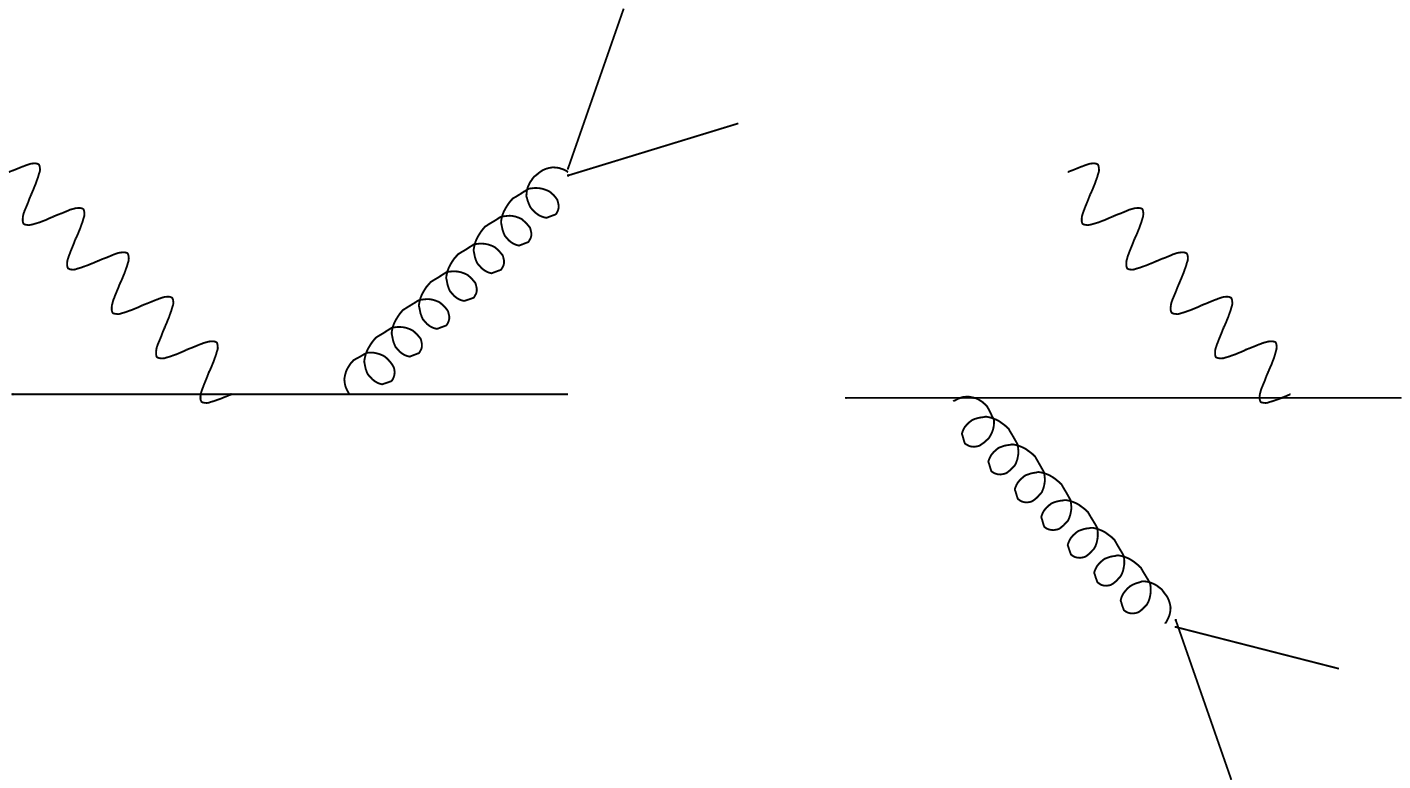}
}
\caption{${\cal O} (\alpha_S^2)$ - $\gamma^*q_i\rightarrow q_i q_j\bar{q_j}$.}
\label{fig:q2qq'qb'_2}
\end{figure}
\begin{figure}[t]
 \begin{center} 
  \includegraphics[scale=0.23]{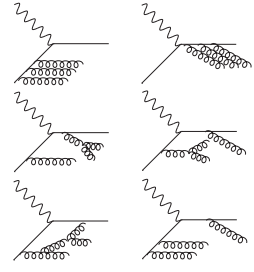}  
  \caption{${\cal O} (\alpha_S^3)$ - $\gamma^*q_i\rightarrow q_i ggg$.
Contribution to $\cqns{a}$ not proportional to $n_f$.
 }
  \label{fig:q2qggg_3}
 \end{center}
\end{figure} 
\begin{figure}[t]
 \begin{center} 
  \includegraphics[scale=0.23]{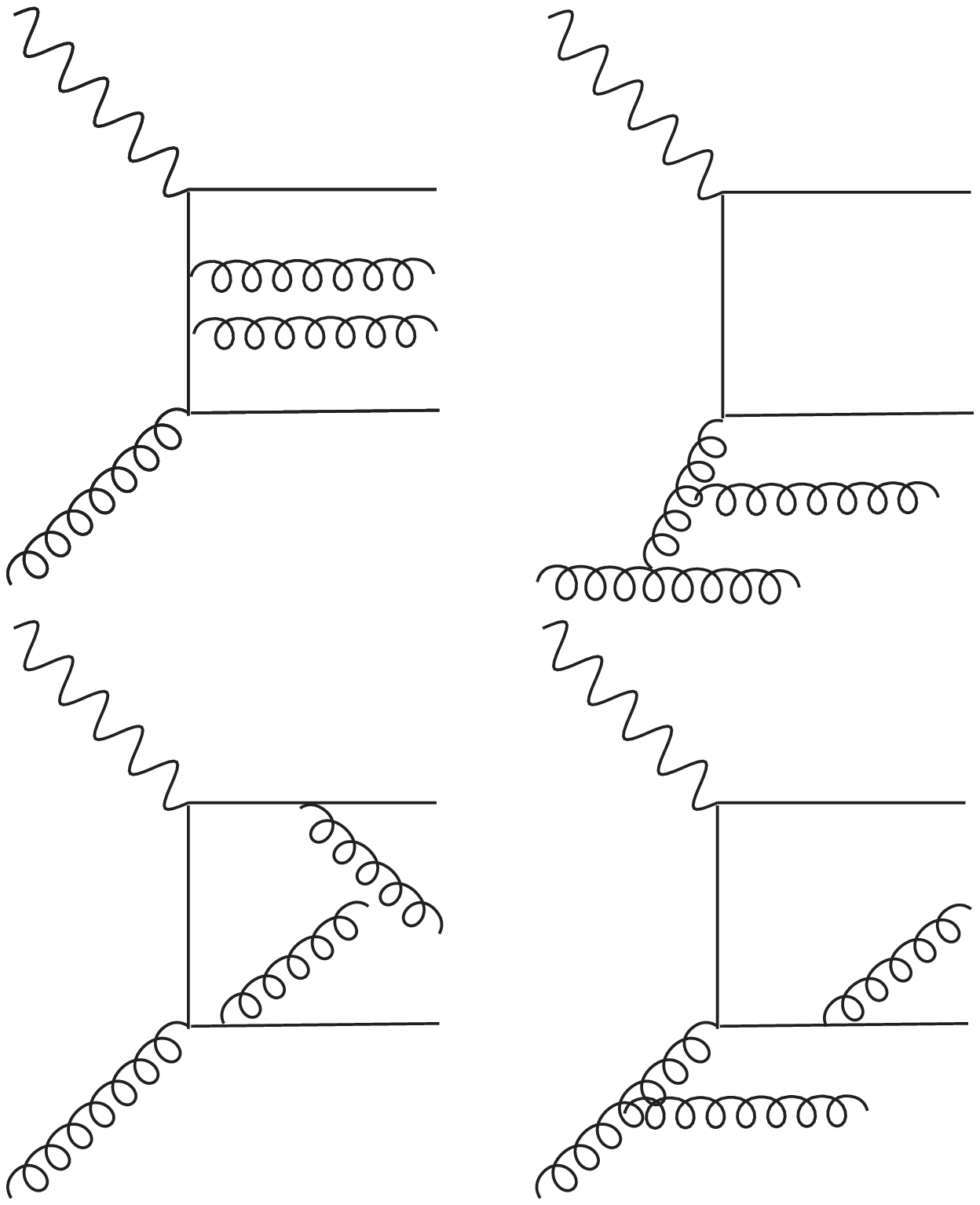} 
  \caption{${\cal O} (\alpha_S^3)$ - $\gamma^*g\rightarrow q_j\bar q_j gg$. }
  \label{fig:g2qqbgg_3}
 \end{center}
\end{figure}
%
\begin{figure}[t]
\subfloat[Contribution proportional to $n_f$ for $\cqps{a}$.
\label{fig:q2qq'qb'g_3_a}]{
\includegraphics[scale=0.23]{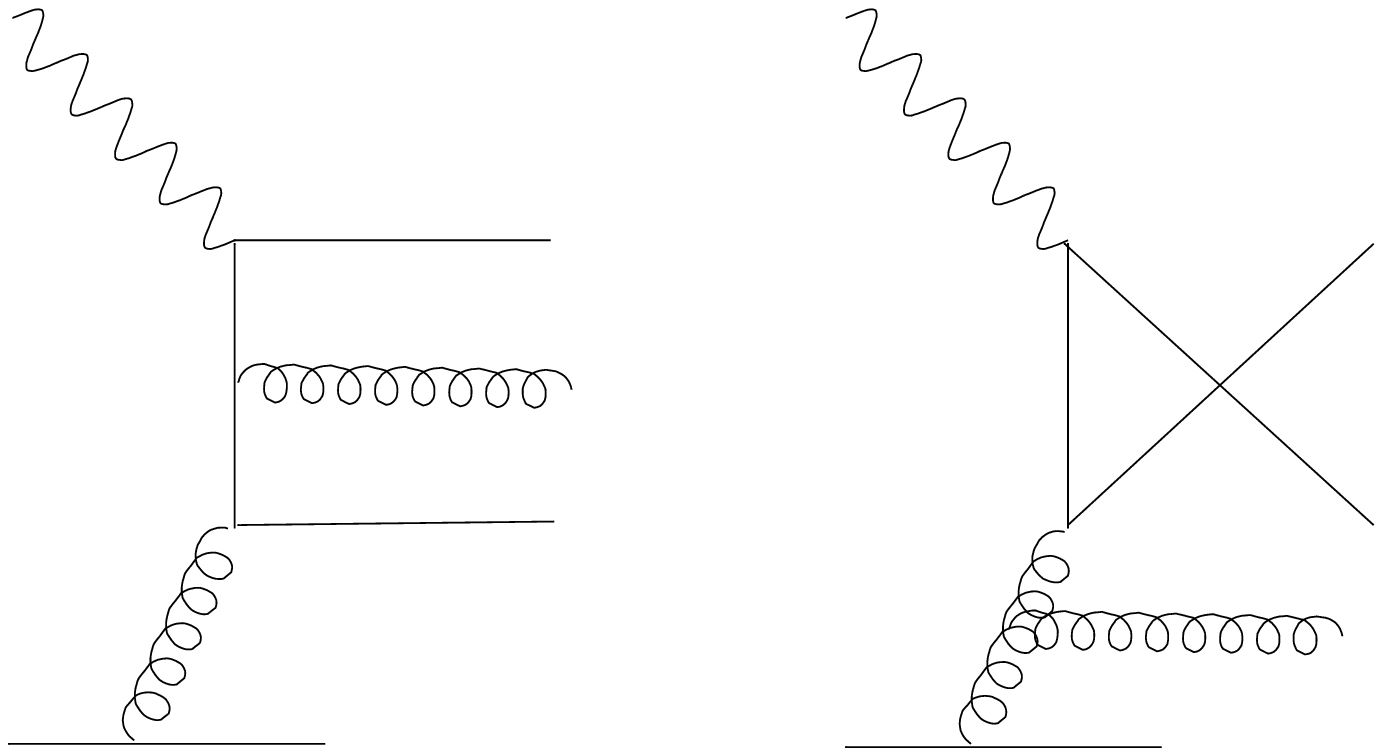}
}
\\
\subfloat[Contribution proportional to $n_f$ for $\cqns{a}$.
\label{fig:q2qq'qb'g_3_b}]{
\includegraphics[scale=0.23]{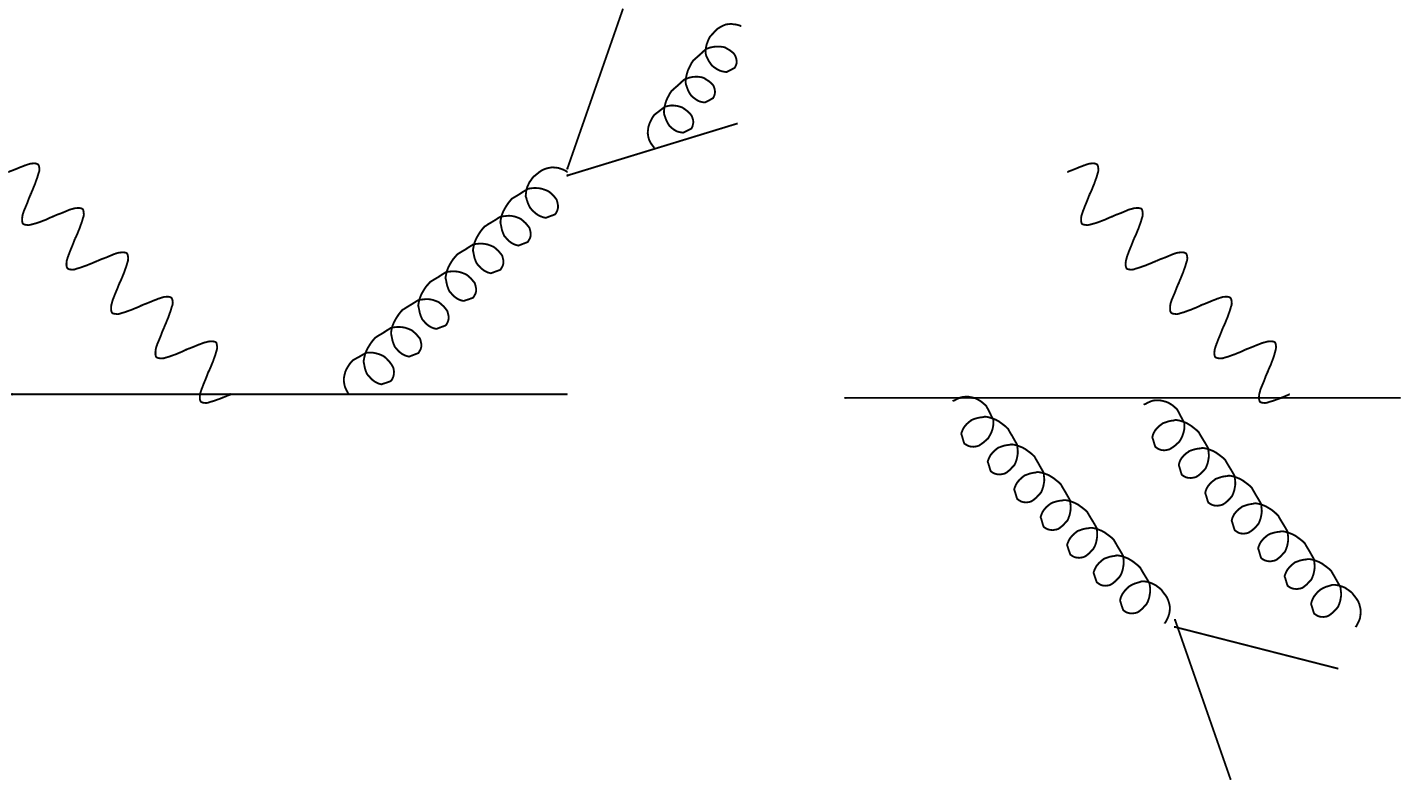}
}
\caption{${\cal O} (\alpha_S^3)$ - $\gamma^*q_i\rightarrow q_i q_j\bar{q_j}g$.}
\end{figure}
\begin{figure}[t]
 \begin{center} 
  \includegraphics[scale=0.23]{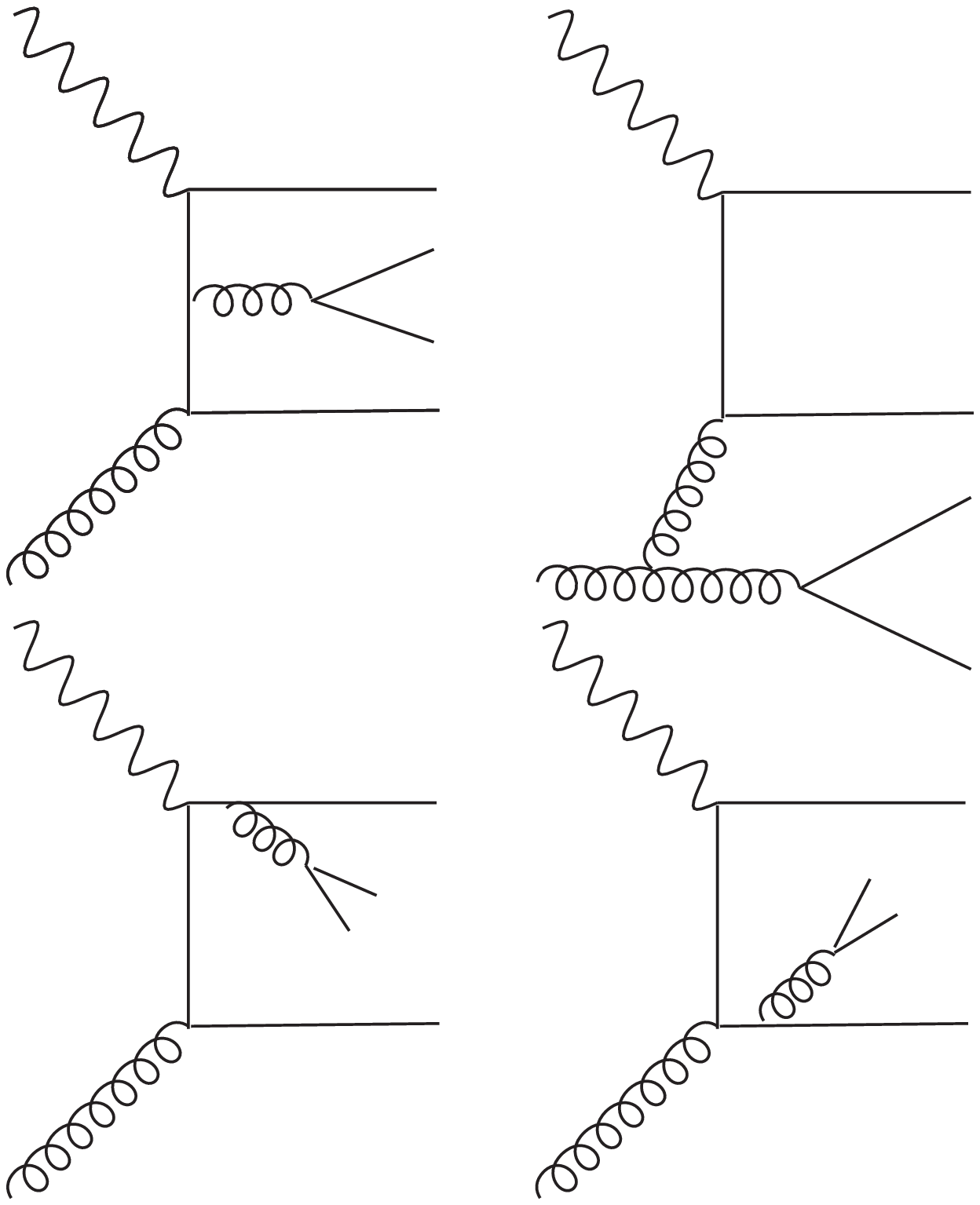}  
  \caption{${\cal O} (\alpha_S^3)$ - $\gamma^*g\rightarrow q_j\bar{q_j} q_k\bar{q_k}$.}
  \label{fig:g2qqbq'qb'_3}
 \end{center}
\end{figure} 


In this appendix we present the decomposition of the Wilson coefficients
used to implement the scheme. 
We will need to decompose the structure function $F$ in terms 
of the individual partonic contributions, 
\begin{equation}
F=\sum_{i=0}^5 \sum_{j=1}^6 F^{ij}
\label{F:decomp}
\end{equation}
where  the indices $i$ and $j$ represent initial and
final-state partons respectively (see captions of Figs.~\ref{fig:q2q_0}--\ref{fig:g2qqbq'qb'_3}). 
More specifically, $i=0$ denotes a gluon  
and $i,j = 1,2,3,\ldots$ denotes $u$, $d$, $s$, \ldots quarks and anti-quarks.
A top quark PDF ($i=6$) is not included in this study.

Let us consider the  heavy
quark structure functions  $F^c_{2,L}$ as
an example.
This is obtained by requiring that there is a charm in the initial
state while summing over the final-state flavors up to and including
charm in Eq.~\eqref{F:decomp},  or by requiring that there is a charm in
the final-state and summing over the initial flavors up to and
including charm. Thus, we obtain:
\begin{equation}
F^c=\sum_{i=0}^{3}F^{i4}+\sum_{j=1}^3F^{4j} + F^{44} \quad .
\label{F:decompFc}
\end{equation}
 The case where the initial and final-state are both
charm quarks ($F^{44}$)  has been written explicitly in the equation to avoid
double counting this contribution.\footnote{Note that in our
  decomposition, diagrams with a bottom quark in the initial or final
  state, contribute to the bottom structure function, even in the
  presence of a charm quark.}
The first sum in Eq.~\eqref{F:decompFc} includes cases, as in
Fig.~\ref{fig:q2qq'qb'_2}, where the incoming quark is a light quark while the 
charm quark is one of the quarks in the quark anti-quark pair.   

In order to obtain the required decomposition, there are some manipulations that need to be performed
to transform from the singlet ($s$), non-singlet ($ns$), and purely-singlet ($ps$) 
structure function combinations  found in the literature into individual partonic components. 

The general expression for the structure function is given by:
\begin{eqnarray}
x^{-1} F_{a} & = &
 q_{ns} \otimes \cqns{a}+\langle e^2\rangle
\left(\qs \otimes \cqs{a} 
+   g \otimes \cg{a}
\right) 
\nonumber \\
\label{F:totF}
\end{eqnarray}
where $a=\{2,L\}$, and 
\begin{eqnarray}
\qns &=& \sum_{i=1}^{n_f} (e_i^2 - \langle e^2\rangle)q_i^+ \nonumber \\
\quad \qs &=& \sum_{i=1}^{n_f} q_i^+, 
\qquad 
q_i^+ = q_i + \bar{q}_i 
\nonumber \\
\langle e^2\rangle &=& \langle e^2\rangle^{(n_f)} = \frac{1}{n_f} \sum_{i=1}^{n_f} e_i^2\qquad , 
\label{F:totFdef}
\end{eqnarray}
and $\cqns{a}$, $\cqs{a}$,  $\cg{a}$ are the Wilson coefficients.  
From Eq.~\eqref{F:totFdef}
one can extract the contribution from a single initial-state quark as: 
\begin{equation}
x^{-1} F_{a,q_i}  = q_i^+ \otimes \left[e_i^2\ \cqns{a} + \langle e^2\rangle \cqps{a}\right]
\label{F:Finitq}
\end{equation}
where $\cqps{a}$ is 
\begin{equation}
\cqps{a} = \cqs{a} - \cqns{a} \quad .
\end{equation}

To further decompose Eq.~\eqref{F:Finitq} into the different final-state contributions, we
examine the diagrams that contribute to the non-singlet and purely-singlet coefficients.
Diagrams 
in which the photon couples to the incoming quark contribute to \cqns{a}
(Figs.~\ref{fig:q2q_0},~\ref{fig:q2qg_1},~\ref{fig:q2qgg_2},~\ref{fig:q2qq'qb'_2_b}, etc.),
whereas the diagrams where 
the photon does not couple to the incoming quark contribute to \cqps{a};
these contributions appear for the first time at $\Ord(\alpha_S^2)$ in 
Figs.~\ref{fig:q2qq'qb'_2_a},~\ref{fig:q2qq'qb'g_3_a}. 
Separating out the final-state quark from Eq.~\eqref{F:Finitq} we obtain: 
\begin{eqnarray}
\label{F:Finitfinq}
x^{-1} F_{a}^{ij} &=& q_i^+ \otimes \bigg \{  
 e_i^2  \ \Big[\cqns{a}(n_f=0)\ \delta_{ij}
\nonumber \\
&+& 
 \cqns{a}(j) - \cqns{a}(j-1) \Big]  
\\
&+& \langle e^2\rangle^{(j)}  \cqps{a}(j)
- \langle e^2\rangle^{(j-1)}\ \cqps{a}(j-1) \bigg \} \, .
\nonumber
\end{eqnarray}
We have  introduced  $\delta_{ij}$ in the non-singlet contribution to 
account for contributions in which the photon couples
to the initial and final-state quark. 
When this is not the case, 
(i.e., in all purely-singlet contributions
and in non-singlet contributions such as the ones in
Fig.~\ref{fig:q2qq'qb'_2_b}),
the difference of the coefficient functions with $n_f=j$
and $n_f=j-1$ flavors is taken.

Some comments are in order:
\begin{itemize}
\item We have verified analytically and numerically that one recovers
Eq.\ (\ref{F:Finitq}) when summing over the final state quark partons
($j=1,\ldots,n_f$) in Eq.\ (\protect\ref{F:Finitfinq}).
\item The corresponding decomposition for the gluon-initiated subprocesses
is simpler than the one in Eq.\ (\protect\ref{F:Finitfinq}) 
since there are only purely-singlet contributions:
\begin{eqnarray}
x^{-1} F_{a}^{0j} &=& g \otimes \bigg \{  
\langle e^2\rangle^{(j)}  \cg{a}(j)
\nonumber \\
&-& \langle e^2\rangle^{(j-1)}\ \cg{a}(j-1) \bigg \} \, .
\label{F:Finitfing}
\end{eqnarray}
\item We remark that the decomposition in Eq. (\ref{F:Finitfinq}) also includes the 
contributions from virtual diagrams to the Wilson coefficients.
As has been discussed in the literature \protect\cite{Chuvakin:1999nx},
such a decomposition is ambiguous at $\Ord(\alpha_S^2)$ and beyond
due to the treatment of heavy quark loops contributing to the light quark structure functions.
However, numerically the ambiguous terms are small and it is standard to analyze the 
heavy quark structure functions $F_{2,L}^c$ and $F_{2,L}^b$ in addition to the inclusive structure 
functions $F_{2,L}$ without any further prescription.
 \end{itemize}

For the general neutral current case (including $Z$-boson exchange),
the electromagnetic couplings should be replaced by electroweak
couplings as follows:
\begin{equation}
e_i^2 \rightarrow a_{q_i}^+=e_i^2 - 2 e_i v_e v_q \chi_Z
+ (v_e^2 + a_e^2)(v_q^2 + a_q^2) \chi_Z^2
\end{equation}
where
\begin{equation}
v_f = T^3_f - 2 Q_f\ \sin^2 \theta_W  \, ,\qquad  a_f = T_3^f
\end{equation}
are the standard (axial-)vector couplings of the $Z$-boson to the leptons ($f=e$) and quarks ($f=q$).
Furthermore, $\chi_Z$ is the ratio of the $Z$-boson propagator with respect to the photon propagator
including additional coupling factors:
\begin{equation}
\chi_Z = \frac{G_F M_Z^2}{2 \sqrt{2} \, \pi \, \alpha_{em}}\  \frac{Q^2}{Q^2 + M_Z^2}\quad .
\end{equation}
Finally, the average squared charge is modified as
\begin{equation}
\langle e^2\rangle^{(n_f)}\rightarrow \aplus={1\over n_f}\sum_{i=1}^{n_f}a_{q_i}^+ \quad .
\end{equation}

\section{Matching Across Heavy Flavor Thresholds \label{subsec:matching}}

As we compute at higher orders, we find 
the matching conditions of the PDFs become discontinuous at 
${\cal O}(\alpha_{s}^{2})$ (NNLO),
and the matching of the $\msbar$ $\alpha_{s}(\mu)$ becomes discontinuous
at  ${\cal O}(\alpha_{s}^{3})$  (N$^{3}$LO).

While the discontinuities in the PDFs and $\alpha_{s}$ (which are
unphysical quantities) persist at all orders, physical observables
(such as cross sections and structure functions) will match across
thresholds up to the computed order of the perturbation theory; for
example, a physical observable in an $N$-flavor and an ($N+1$)-flavor
scheme will match up to higher order terms when computed to order
$\alpha_{s}^{M}$ in the perturbation expansion: 
$$
\sigma^{N}=\sigma^{N+1}+{\cal O}(\alpha_{s}^{M+1})\quad .
\label{eq:order}
$$
As it is not immediately obvious how the discontinuities cancel order-by-order,
we shall examine a NNLO numeric case, and also a simple analytic example.

\subsection{Discontinuities across the flavor transition.}

To illustrate the behavior of the discontinuities, we will work at
NNLO where the DGLAP evolution and the flavor-threshold boundary
conditions have been computed and implemented.\footnote{At present,
the full set of matching conditions and DGLAP kernels have not been
computed at N${}^3$LO.}
Since $\mu=m_{c}$ is often used for the initial evolution scale, we
will focus on the transition from $N_{F}=4$ to $N_{F}=5$ flavors at
$\mu=m_{b}$.

The matching conditions across flavor thresholds can 
be summarized as~\cite{Buza:1995ie}
\begin{equation}
f_{a}^{N+1}=A_{ab}\otimes f_{b}^{N}
\label{eq:matching}
\end{equation}
 where $f^{N}$ and $f^{N+1}$ are the PDFs for $N$ and $N+1$~flavors,
and $A_{ab}$ can be expanded perturbatively.
In the VFNS  for $\mu<m_{b}$,  the $b$-quark PDF is zero and the gluon PDF is finite and positive. 
Using Eq.~(\ref{eq:matching})  for $\mu > m_{b}$,  
we find the $b$-quark is negative for  $\mu\sim m_{b}$,  and it becomes more negative as we move to smaller $x$. 
In contrast, the gluon has a positive discontinuity 
as it must to ensure the momentum sum rule is satisfied.

\subsection{The $b$-quark flavor transition.}

\begin{figure*}[t]
\begin{center}
\subfloat[$F_2$ vs. $Q$ for $x=\{10^{-1},10^{-3},10^{-5}\}$ (left to right)
for different $n$ scalings.]
{\includegraphics[width=0.3\textwidth]{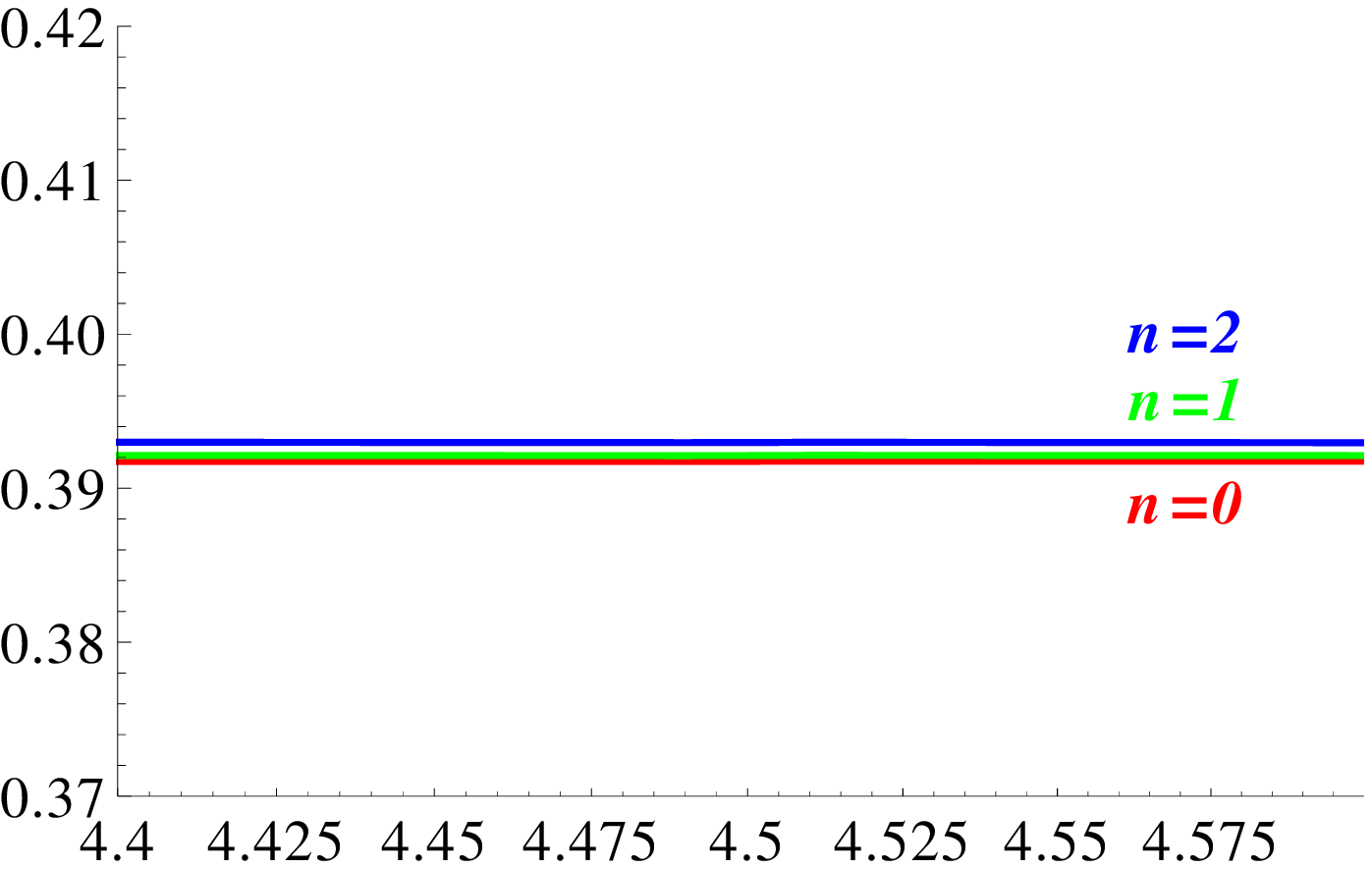} \hfil 
\includegraphics[width=0.3\textwidth]{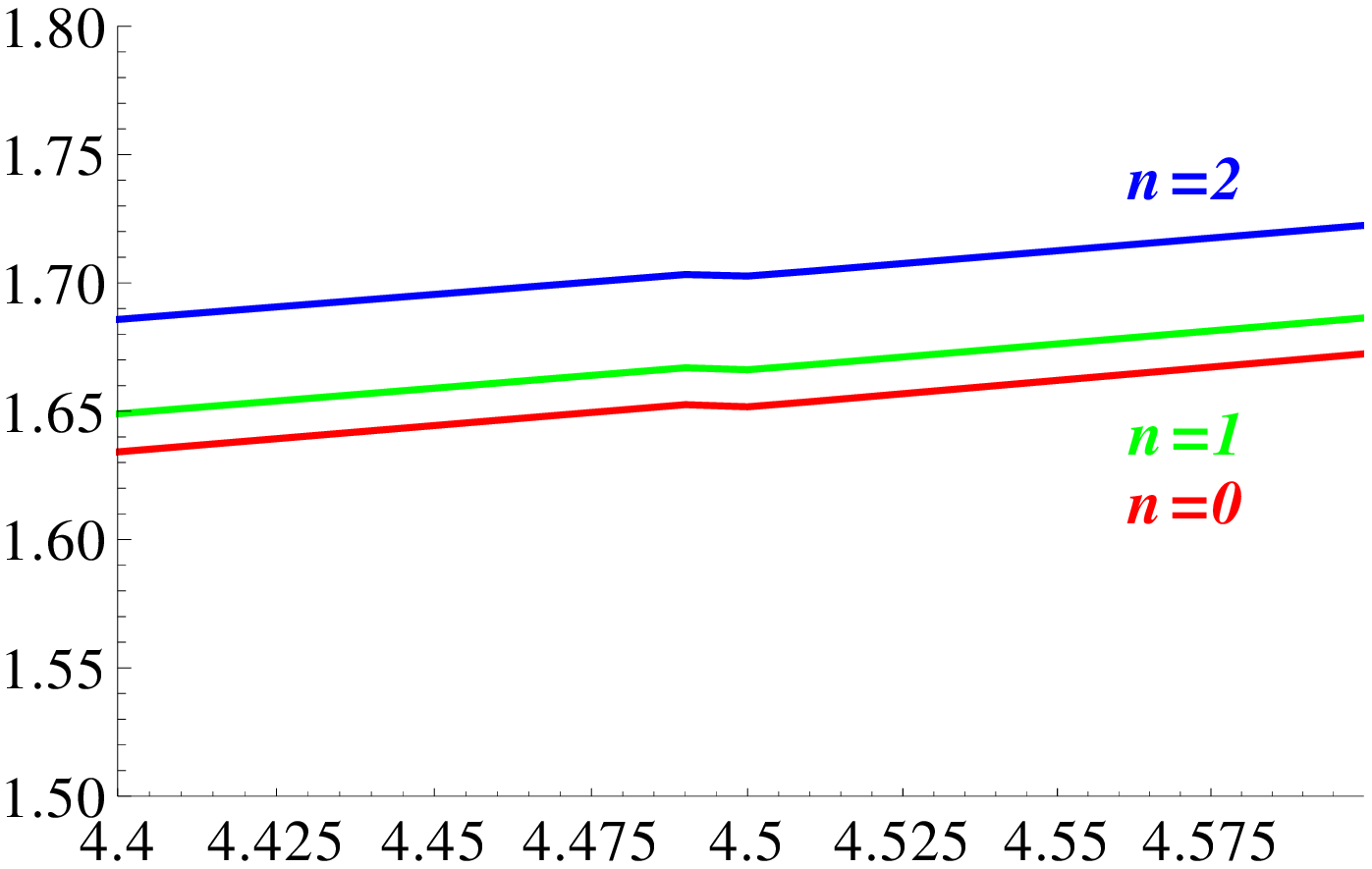} \hfil 
\includegraphics[width=0.3\textwidth]{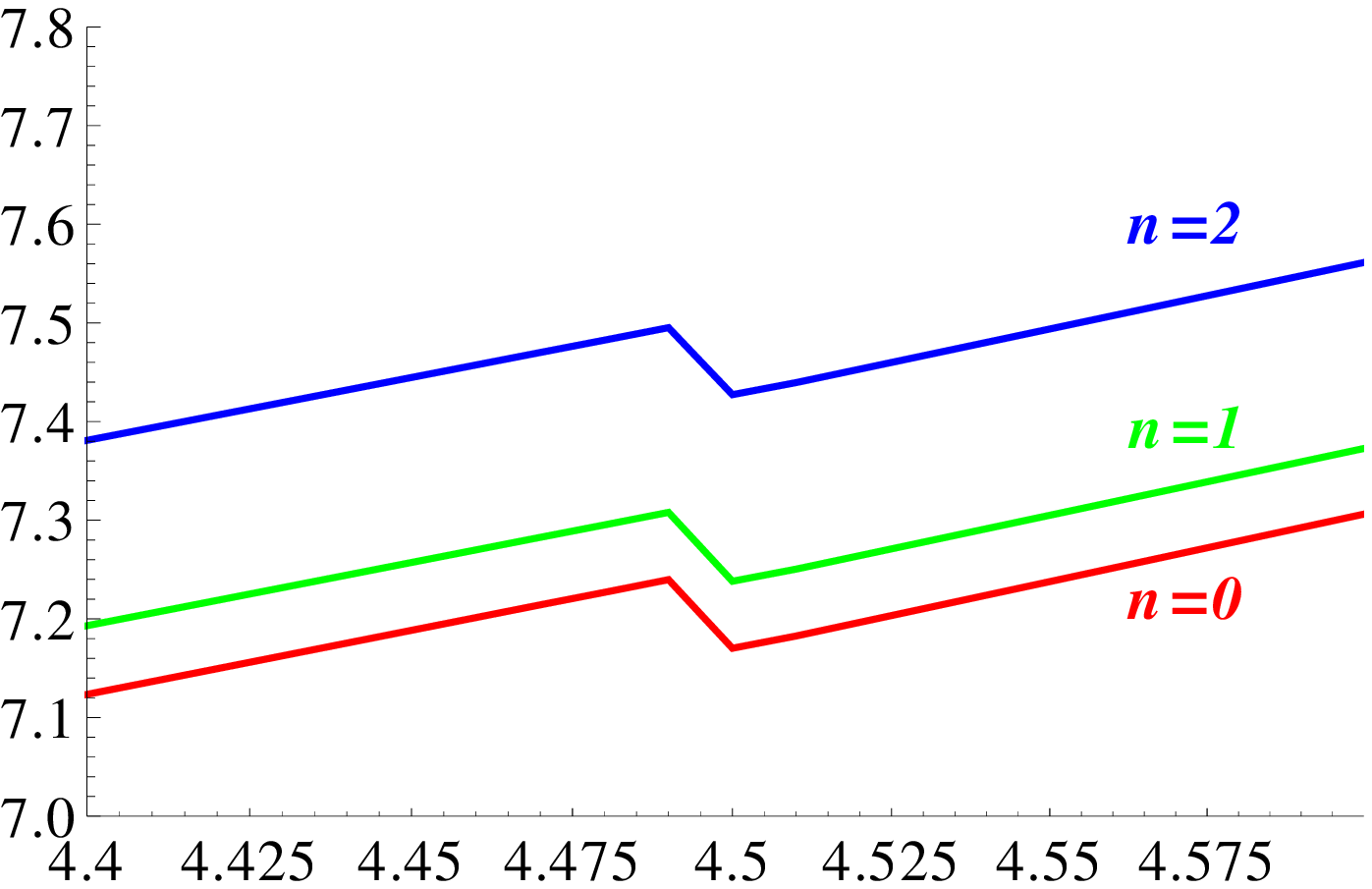}
\label{subfig:F2Ldisca}
}
\newline
\subfloat[$F_L$ vs. $Q$ for $x=\{10^{-1},10^{-3},10^{-5}\}$ (left to right)
for different $n$ scalings.
]
{
\includegraphics[width=0.3\textwidth]{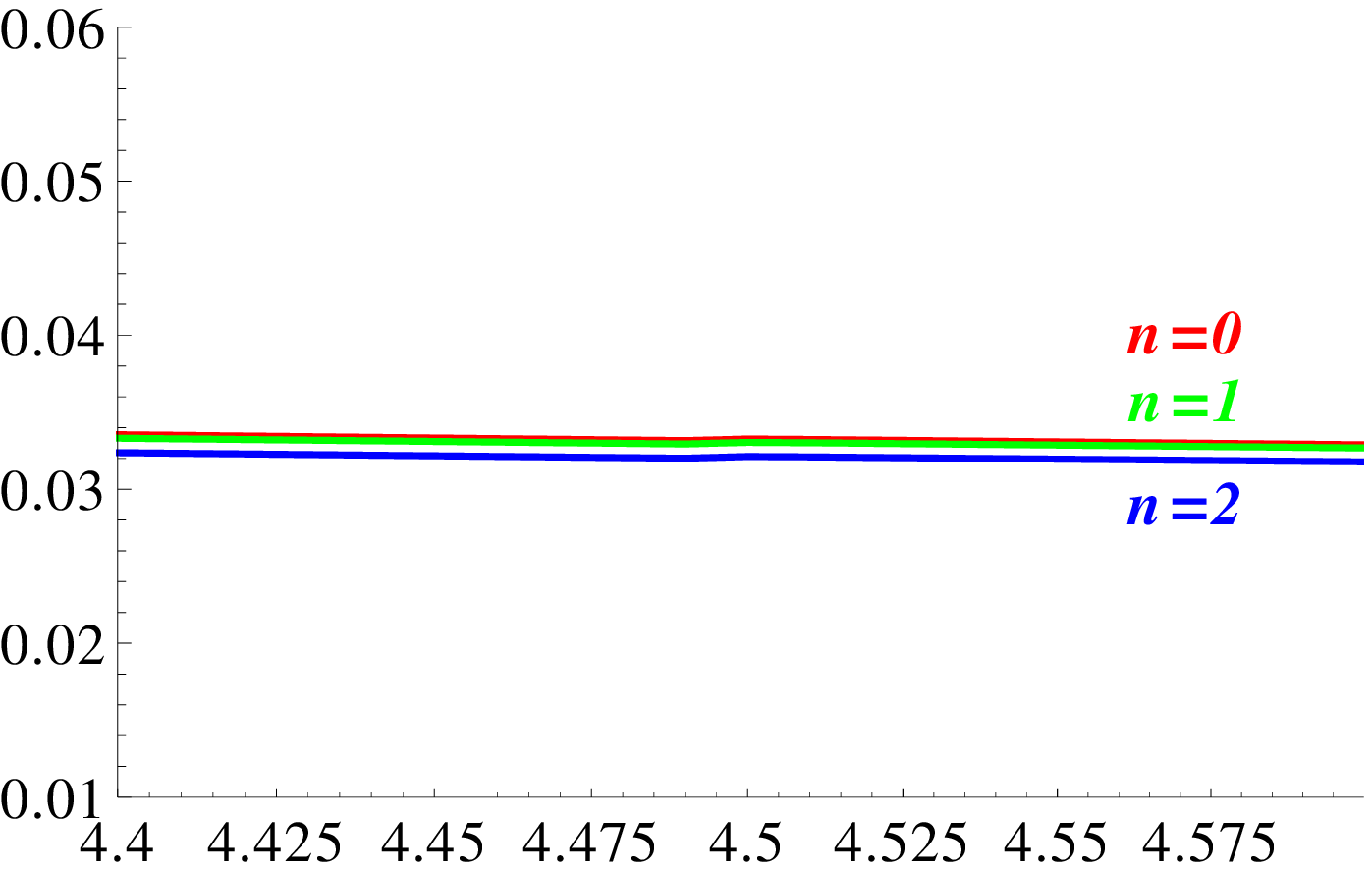} \hfil 
\includegraphics[width=0.3\textwidth]{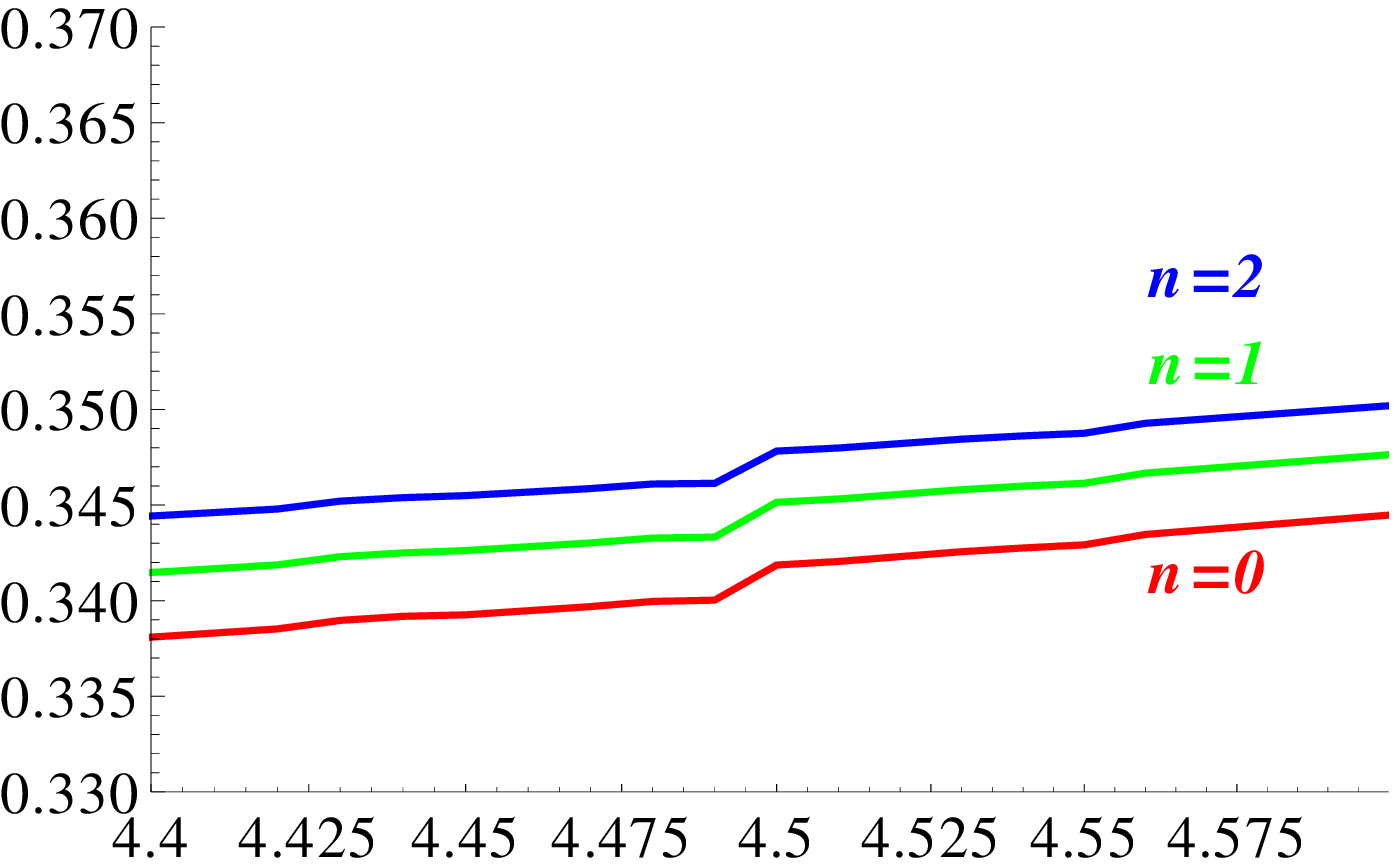} \hfil 
\includegraphics[width=0.3\textwidth]{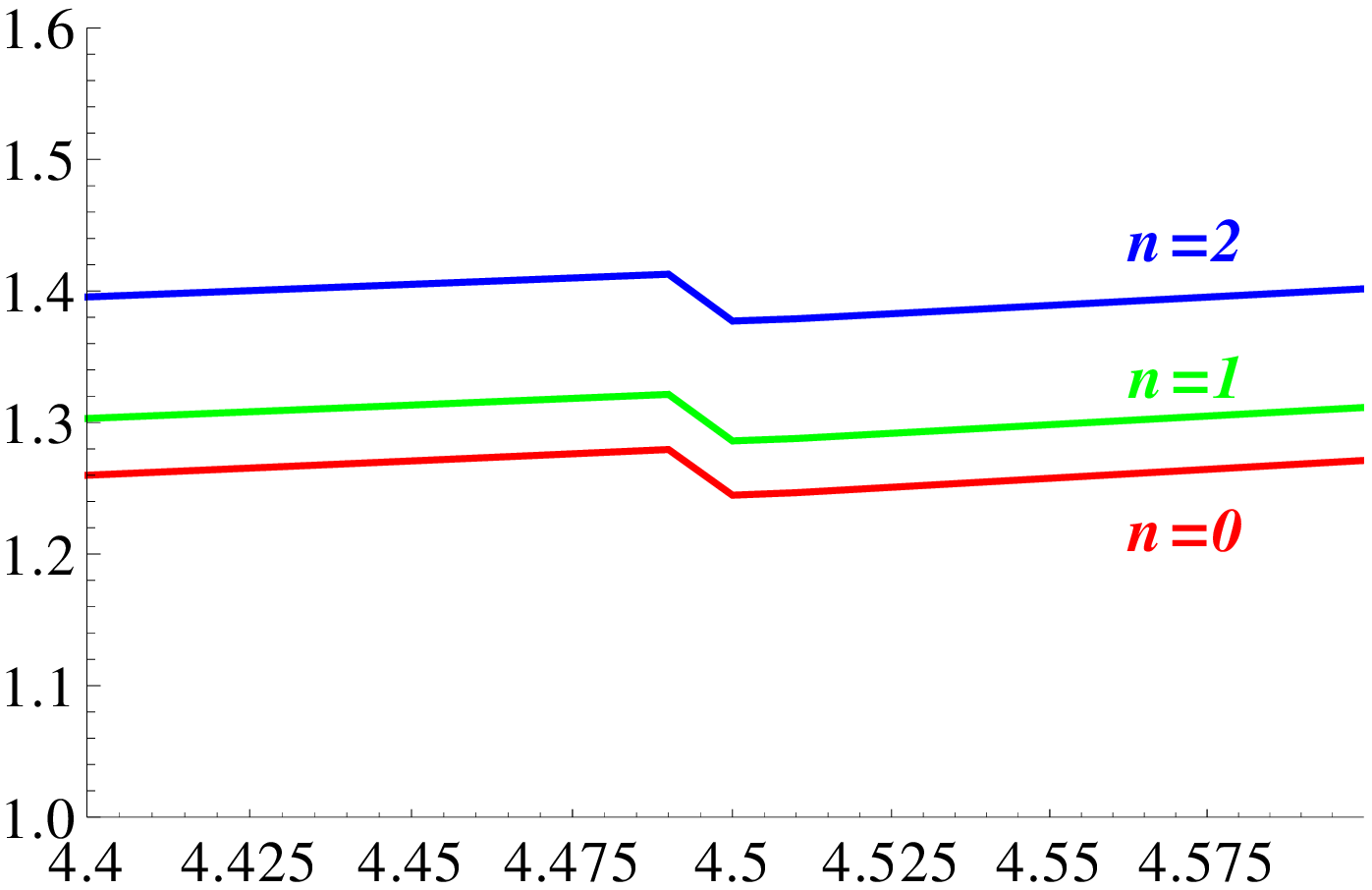}
\label{subfig:F2Ldiscb}
}
\caption{Discontinuity for $F_2$, $F_L$ at NNLO in the region of the bottom mass,
$m_b=4.5$~GeV.}
\label{fig:F2Ldisc}
\end{center}
\end{figure*}

Although these discontinuities are too small to be noticeable in the 
figures of Sec.IV, 
in Figure~\ref{fig:F2Ldisc}  
we have magnified the axes 
so the discontinuities are visible. 
Here, we display $F_{2}$ and $F_{L}$
for a selection of $x$-values.

The first  general feature we notice in Fig.~\ref{fig:F2Ldisc}  is that the 
size of the discontinuity generally grows as we go to smaller $x$  values. 
This is consistent with the fact that the  discontinuity computed 
by  Eq.~(\ref{eq:matching}) also grows for smaller~$x$. 
We display the results for a selection of $n$-scaling values;
note that the uncertainty arising from the discontinuity is typically
on the order of the difference due to the choice of scaling. 

Another feature  that is most evident for the series of $F_L$ plots 
(Fig.~\ref{subfig:F2Ldiscb}) is that the discontinuity 
can change sign for different $x$ values. 
This can happen because the mix of quark and gluon initiated terms 
is changing as a function of~$x$. 

This observation is key to understanding how the (unphysical) PDFs may
have a relatively large discontinuity, while the effect on the
physical quantities (such as $\sigma$ and $F_{2,L}$) is moderated.  
Because physical quantities will contain a sum of gluon and quark
initiated contributions, and because the discontinuity of the quark
and gluon PDFs have {\em opposite signs}, the discontinuities of the quark
and gluon PDFs can partially cancel so that the physical quantity may
have a reduced discontinuity.

This discontinuity, in part, reflects the theoretical uncertainty of the 
perturbation theory at a given order. As we compute the physical observables 
to higher and higher orders, this 
discontinuity will be reduced even though the discontinuity in 
the PDFs and $\alpha_s$ remain. 
We will demonstrate this mechanism in the following.

\subsection{A ``Toy'' Example at NLO''}

We now illustrate how the cancellation of the quark and gluon PDF 
discontinuities work
analytically using a ``toy'' calculation.

Expanding Eq.~(\ref{eq:matching})
 in the region of  $\mu=m_{b}$ we
have:
\begin{eqnarray}
f_{b}^{5}&=&
\left\{ 0+\frac{\alpha_{s}}{2\pi}P_{qg}\left(L+a_{qg}\right)+O(\alpha_{s}^{2})\right\} \otimes f_{g}^{4}
\nonumber\\
f_{g}^{5}&=&
\left\{ 1+\frac{\alpha_{s}}{2\pi}P_{gg}\left(L+a_{gg}\right)+O(\alpha_{s}^{2})\right\} \otimes f_{g}^{4}
\nonumber\\
\label{eq:expansion}
\end{eqnarray}
where 
$L=\ln(\mu^{2}/m_{b}^{2})$.
It happens that the constant terms $a_{ij}$ in Eq.~(\ref{eq:expansion}) are zero
at ${\cal O}(\alpha_{s}^{1})$ in the $\msbar$  scheme; this is not due to any underlying
symmetry, and in fact at ${\cal O}(\alpha_{s}^{2})$ these terms are non-zero. 
Because  $a_{ij}$  are zero, if we perform the matching
at $\mu=m_{b}$, we find that the gluon PDF is continuous 
$f_{b}^{5}(x,m_{b})=f_{g}^{4}(x,m_{b})$, and the bottom PDF starts
from zero $f_{b}^{5}(x,m_{b})=0$.

\subsubsection{If $a_{ij}$ was non-zero at ${\cal O}(\alpha_{s}^{1})$}

To illustrate how the discontinuities cancel in the ACOT renormalization
scheme, we will suppose 
(for this ``toy'' calculation) 
that the constant terms ($a_{ij}$) 
in the matching conditions are non-vanishing at order $\alpha_{s}^{1}$;
thus, the gluon and bottom PDFs will now have ${\cal O}(\alpha_{s}^{1})$
discontinuities, {\em but} the physical observables computed with different
$N_{F}$ values will still match up to ${\cal O}(\alpha_{s}^{2})$.

In the ACOT scheme, the total cross section can be decomposed as:
$\sigma_{TOT}=\sigma_{LO}+\sigma_{NLO}-\sigma_{SUB}$, 
where $\sigma_{LO}$ represents $\gamma b\to b$, 
$\sigma_{NLO}$ represents $\gamma g\to b\bar{b}$,
and $\sigma_{SUB}$ represents the $(g\to b)\otimes(\gamma b\to b)$  ``subtraction'' 
contribution.\footnote{Note, 
we will focus on the gluon-initiated terms, but the demonstration
for the quark-initiated pieces is analogous.} 
We will now perturbatively compute $\sigma_{TOT}$ in the region $\mu\sim m_{b}$
for both $N_{F}=4$ and $N_{F}=5$.

\subsubsection{ACOT for $N_{F}=4$}

For $\mu<m_{b}$, we have $N_{F}=4$ and $f_{b}=0$; thus, $\sigma_{LO}$
and $\sigma_{SUB}$ vanish, and we have:
\[
\sigma_{TOT}^{N_{F}=4}=\sigma_{NLO}=C^{1}\otimes f_{g}^{4}+O(\alpha_{s}^{2})
\]
where $C^{1}$ represents the ${\cal O}(\alpha_{s}^{1})$ process
$\gamma g\to b\bar{b}$.

\subsubsection{ACOT for $N_{F}=5$}

For $\mu>m_{b}$, we have $N_{F}=5$ and $f_{b}\not=0$. For the contributions
we have: 
\begin{eqnarray*}
\sigma_{LO}&=&
C^{0}\otimes f_{b}^{5}\simeq C^{0}\otimes\left\{ 0+ \frac{\alpha_{s}}{2\pi}P_{qg}\left(L+a_{qg}\right)\right\} \otimes f_{g}^{4}
\nonumber\\
\sigma_{NLO}&=&
C^{1}\otimes f_{g}^{5}\simeq C^{1}\otimes\left\{ 1+\frac{\alpha_{s}}{2\pi}P_{gg}\left(L+a_{gg}\right)\right\} \otimes f_{g}^{4}
\nonumber\\
\sigma_{SUB} & = & 
C^{0}\otimes\widetilde{f}_{g\to q}\otimes f_{g}^{5}\simeq C^{0}\otimes\left\{ \frac{\alpha_{s}}{2\pi}P_{qg}\left(L+a_{qg}\right)\right\} 
\nonumber \\
&\otimes&
\left\{ 1+\frac{\alpha_{s}}{2\pi}P_{gg}\left(L+a_{gg}\right)\right\} \otimes f_{g}^{4}
\nonumber\\
\end{eqnarray*}
Keeping terms to ${\cal O}(\alpha_{s}^{1})$ we have: 
\[
\sigma_{TOT}^{N_{F}=5}=\sigma_{LO}+\sigma_{NLO}-\sigma_{SUB}=C^{1}\otimes f_{g}^{4}+O(\alpha_{s}^{2})
\]
Notice that the discontinuity introduced by $a_{qg}$ in the PDFs
is canceled by $a_{qg}$ from the SUB contribution.%
\footnote{The explicit form of the ACOT subtraction is defined in
Sec.IV.C ({\it cf.}, Eq.~(36)) of Ref.~\cite{Collins:1998rz}.
For an example of the cancellation between $\sigma_{LO}$ and
$\sigma_{SUB}$ in a more general context,
see Ref.~\cite{Olness:1997yn}.
}

\subsubsection{Comparison of $N_{F}=5$ and $N_{F}=4$}

Comparing the  $N_{F}=5$ and $N_{F}=4$ results, we find
\[
\sigma_{TOT}^{N_{F}=5}=\sigma_{TOT}^{N_{F}=4}+O(\alpha_{s}^{2})
\]
so that the total physical results 
match up the order of the perturbation theory.

In the above illustration, we have retained the log terms ($L$);
the cancellation of the logs is ensured in a well defined renormalization
scheme, and the $a_{ij}$ constant terms get carried along with the logs and
will thus cancel order by order. 

Therefore, the discontinuity of the physical quantities ($\sigma$, $F_{2,L}$)
reflects the perturbative uncertainty, and this will be systematically
reduced at higher orders.

\section*{Acknowledgment}

We thank 
M.~Botje, 
A.~M.~Cooper-Sarkar, 
A.~Glazov,
C.~Keppel, 
J.~G.~Morf\'in, 
P.~Nadolsky, 
M.~Guzzi,
J.~F.~Owens,
V.~A.~Radescu,
and
A.~Vogt
for discussions.

F.I.O., I.S., and J.Y.Y.\ acknowledge the hospitality of 
CERN, DESY, Fermilab, and Les Houches where a portion of this work
was performed. This work was partially supported by the U.S.\ Department
of Energy under grant DE-FG02-04ER41299, and the Lightner-Sams Foundation.
F.I.O thanks the Galileo
Galilei Institute for Theoretical Physics for their hospitality
and the INFN for partial support during the
completion of this work.
The research of T.S. is supported by a fellowship from the
Th\'eorie LHC France initiative funded by the CNRS/IN2P3.
This work has been supported by  {\it Projet international de cooperation
scientifique} PICS05854 between France and the USA.
The work of J.~Y.~Yu was supported by the Deutsche Forschungsgemeinschaft
(DFG) through grant No.~YU~118/1-1.

\bibliographystyle{apsrev}
\bibliography{acot_nnlo}

\begin{thebibliography}{42}
\expandafter\ifx\csname natexlab\endcsname\relax\def\natexlab#1{#1}\fi
\expandafter\ifx\csname bibnamefont\endcsname\relax
  \def\bibnamefont#1{#1}\fi
\expandafter\ifx\csname bibfnamefont\endcsname\relax
  \def\bibfnamefont#1{#1}\fi
\expandafter\ifx\csname citenamefont\endcsname\relax
  \def\citenamefont#1{#1}\fi
\expandafter\ifx\csname url\endcsname\relax
  \def\url#1{\texttt{#1}}\fi
\expandafter\ifx\csname urlprefix\endcsname\relax\def\urlprefix{URL }\fi
\providecommand{\bibinfo}[2]{#2}
\providecommand{\eprint}[2][]{\url{#2}}

\bibitem[{\citenamefont{Schienbein et~al.}(2008)\citenamefont{Schienbein,
  Radescu, Zeller, Christy, Keppel et~al.}}]{Schienbein:2007gr}
\bibinfo{author}{\bibfnamefont{I.}~\bibnamefont{Schienbein}},
  \bibinfo{author}{\bibfnamefont{V.~A.} \bibnamefont{Radescu}},
  \bibinfo{author}{\bibfnamefont{G.}~\bibnamefont{Zeller}},
  \bibinfo{author}{\bibfnamefont{M.}~\bibnamefont{Christy}},
  \bibinfo{author}{\bibfnamefont{C.}~\bibnamefont{Keppel}},
  \bibnamefont{et~al.}, \bibinfo{journal}{J.Phys.G}
  \textbf{\bibinfo{volume}{G35}}, \bibinfo{pages}{053101}
  (\bibinfo{year}{2008}), \eprint{0709.1775}.

\bibitem[{\citenamefont{{H1 and ZEUS Collaborations}}(2010)}]{hera:FL}
\bibinfo{author}{\bibnamefont{{H1 and ZEUS Collaborations}}}
  (\bibinfo{year}{2010}), \eprint{Reports Nos. H1prelim-10-044 and
  ZEUS-prel-10-008 (unpublished),
  \url{http://www-h1.desy.de/psfiles/confpap/DIS2010/H1prelim-10-044.pdf}}.

\bibitem[{\citenamefont{Aaron et~al.}(2011)}]{Collaboration:2010ry}
\bibinfo{author}{\bibfnamefont{F.~D.} \bibnamefont{Aaron}}
  \bibnamefont{et~al.}, \bibinfo{journal}{Eur. Phys. J.}
  \textbf{\bibinfo{volume}{C71}}, \bibinfo{pages}{1579} (\bibinfo{year}{2011}),
  \eprint{1012.4355}.

\bibitem[{\citenamefont{Aivazis
  et~al.}(1994{\natexlab{a}})\citenamefont{Aivazis, Collins, Olness, and
  Tung}}]{Aivazis:1993pi}
\bibinfo{author}{\bibfnamefont{M.~A.~G.} \bibnamefont{Aivazis}},
  \bibinfo{author}{\bibfnamefont{J.~C.} \bibnamefont{Collins}},
  \bibinfo{author}{\bibfnamefont{F.~I.} \bibnamefont{Olness}},
  \bibnamefont{and} \bibinfo{author}{\bibfnamefont{W.~K.} \bibnamefont{Tung}},
  \bibinfo{journal}{Phys. Rev.} \textbf{\bibinfo{volume}{D50}},
  \bibinfo{pages}{3102} (\bibinfo{year}{1994}{\natexlab{a}}),
  \eprint{hep-ph/9312319}.

\bibitem[{\citenamefont{Collins}(1998)}]{Collins:1998rz}
\bibinfo{author}{\bibfnamefont{J.~C.} \bibnamefont{Collins}},
  \bibinfo{journal}{Phys. Rev.} \textbf{\bibinfo{volume}{D58}},
  \bibinfo{pages}{094002} (\bibinfo{year}{1998}), \eprint{hep-ph/9806259}.

\bibitem[{\citenamefont{Kretzer and Schienbein}(1998)}]{Kretzer:1998ju}
\bibinfo{author}{\bibfnamefont{S.}~\bibnamefont{Kretzer}} \bibnamefont{and}
  \bibinfo{author}{\bibfnamefont{I.}~\bibnamefont{Schienbein}},
  \bibinfo{journal}{Phys. Rev.} \textbf{\bibinfo{volume}{D58}},
  \bibinfo{pages}{094035} (\bibinfo{year}{1998}), \eprint{hep-ph/9805233}.

\bibitem[{\citenamefont{Thorne and
  Roberts}(1998{\natexlab{a}})}]{Thorne:1997uu}
\bibinfo{author}{\bibfnamefont{R.}~\bibnamefont{Thorne}} \bibnamefont{and}
  \bibinfo{author}{\bibfnamefont{R.}~\bibnamefont{Roberts}},
  \bibinfo{journal}{Phys.Lett.} \textbf{\bibinfo{volume}{B421}},
  \bibinfo{pages}{303} (\bibinfo{year}{1998}{\natexlab{a}}),
  \eprint{hep-ph/9711223}.

\bibitem[{\citenamefont{Kr{\"a}mer et~al.}(2000)\citenamefont{Kr{\"a}mer,
  Olness, and Soper}}]{Kramer:2000hn}
\bibinfo{author}{\bibfnamefont{M.}~\bibnamefont{Kr{\"a}mer}},
  \bibinfo{author}{\bibfnamefont{F.~I.} \bibnamefont{Olness}},
  \bibnamefont{and} \bibinfo{author}{\bibfnamefont{D.~E.} \bibnamefont{Soper}},
  \bibinfo{journal}{Phys. Rev.} \textbf{\bibinfo{volume}{D62}},
  \bibinfo{pages}{096007} (\bibinfo{year}{2000}), \eprint{hep-ph/0003035}.

\bibitem[{\citenamefont{Barnett}(1976)}]{Barnett:1976ak}
\bibinfo{author}{\bibfnamefont{R.~M.} \bibnamefont{Barnett}},
  \bibinfo{journal}{Phys. Rev. Lett.} \textbf{\bibinfo{volume}{36}},
  \bibinfo{pages}{1163} (\bibinfo{year}{1976}).

\bibitem[{\citenamefont{Tung et~al.}(2002)\citenamefont{Tung, Kretzer, and
  Schmidt}}]{Tung:2001mv}
\bibinfo{author}{\bibfnamefont{W.~K.} \bibnamefont{Tung}},
  \bibinfo{author}{\bibfnamefont{S.}~\bibnamefont{Kretzer}}, \bibnamefont{and}
  \bibinfo{author}{\bibfnamefont{C.}~\bibnamefont{Schmidt}},
  \bibinfo{journal}{J. Phys.} \textbf{\bibinfo{volume}{G28}},
  \bibinfo{pages}{983} (\bibinfo{year}{2002}), \eprint{hep-ph/0110247}.

\bibitem[{\citenamefont{Kretzer et~al.}(2004)\citenamefont{Kretzer, Lai,
  Olness, and Tung}}]{Kretzer:2003it}
\bibinfo{author}{\bibfnamefont{S.}~\bibnamefont{Kretzer}},
  \bibinfo{author}{\bibfnamefont{H.}~\bibnamefont{Lai}},
  \bibinfo{author}{\bibfnamefont{F.}~\bibnamefont{Olness}}, \bibnamefont{and}
  \bibinfo{author}{\bibfnamefont{W.}~\bibnamefont{Tung}},
  \bibinfo{journal}{Phys.Rev.} \textbf{\bibinfo{volume}{D69}},
  \bibinfo{pages}{114005} (\bibinfo{year}{2004}), \eprint{hep-ph/0307022}.

\bibitem[{\citenamefont{Guzzi et~al.}(2011{\natexlab{a}})\citenamefont{Guzzi,
  Nadolsky, Lai, and Yuan}}]{Guzzi:2011ew}
\bibinfo{author}{\bibfnamefont{M.}~\bibnamefont{Guzzi}},
  \bibinfo{author}{\bibfnamefont{P.~M.} \bibnamefont{Nadolsky}},
  \bibinfo{author}{\bibfnamefont{H.-L.} \bibnamefont{Lai}}, \bibnamefont{and}
  \bibinfo{author}{\bibfnamefont{C.-P.} \bibnamefont{Yuan}}
  (\bibinfo{year}{2011}{\natexlab{a}}), \eprint{1108.5112}.

\bibitem[{\citenamefont{Laenen et~al.}(1993)\citenamefont{Laenen, Riemersma,
  Smith, and van Neerven}}]{Laenen:1992zk}
\bibinfo{author}{\bibfnamefont{E.}~\bibnamefont{Laenen}},
  \bibinfo{author}{\bibfnamefont{S.}~\bibnamefont{Riemersma}},
  \bibinfo{author}{\bibfnamefont{J.}~\bibnamefont{Smith}}, \bibnamefont{and}
  \bibinfo{author}{\bibfnamefont{W.~L.} \bibnamefont{van Neerven}},
  \bibinfo{journal}{Nucl. Phys.} \textbf{\bibinfo{volume}{B392}},
  \bibinfo{pages}{162} (\bibinfo{year}{1993}).

\bibitem[{\citenamefont{Guzzi et~al.}(2011{\natexlab{b}})\citenamefont{Guzzi,
  Nadolsky, Lai, and Yuan}}]{Guzzi:2011dk}
\bibinfo{author}{\bibfnamefont{M.}~\bibnamefont{Guzzi}},
  \bibinfo{author}{\bibfnamefont{P.~M.} \bibnamefont{Nadolsky}},
  \bibinfo{author}{\bibfnamefont{H.-L.} \bibnamefont{Lai}}, \bibnamefont{and}
  \bibinfo{author}{\bibfnamefont{C.~P.} \bibnamefont{Yuan}}
  (\bibinfo{year}{2011}{\natexlab{b}}), \eprint{1108.4008}.

\bibitem[{\citenamefont{Thorne and
  Roberts}(1998{\natexlab{b}})}]{Thorne:1997ga}
\bibinfo{author}{\bibfnamefont{R.}~\bibnamefont{Thorne}} \bibnamefont{and}
  \bibinfo{author}{\bibfnamefont{R.}~\bibnamefont{Roberts}},
  \bibinfo{journal}{Phys.Rev.} \textbf{\bibinfo{volume}{D57}},
  \bibinfo{pages}{6871} (\bibinfo{year}{1998}{\natexlab{b}}),
  \eprint{hep-ph/9709442}.

\bibitem[{\citenamefont{Thorne}(2006)}]{Thorne:2006qt}
\bibinfo{author}{\bibfnamefont{R.}~\bibnamefont{Thorne}},
  \bibinfo{journal}{Phys.Rev.} \textbf{\bibinfo{volume}{D73}},
  \bibinfo{pages}{054019} (\bibinfo{year}{2006}), \eprint{hep-ph/0601245}.

\bibitem[{\citenamefont{Cacciari et~al.}(1998)\citenamefont{Cacciari, Greco,
  and Nason}}]{Cacciari:1998it}
\bibinfo{author}{\bibfnamefont{M.}~\bibnamefont{Cacciari}},
  \bibinfo{author}{\bibfnamefont{M.}~\bibnamefont{Greco}}, \bibnamefont{and}
  \bibinfo{author}{\bibfnamefont{P.}~\bibnamefont{Nason}},
  \bibinfo{journal}{JHEP} \textbf{\bibinfo{volume}{9805}}, \bibinfo{pages}{007}
  (\bibinfo{year}{1998}), \eprint{hep-ph/9803400}.

\bibitem[{\citenamefont{Forte et~al.}(2010)\citenamefont{Forte, Laenen, Nason,
  and Rojo}}]{Forte:2010ta}
\bibinfo{author}{\bibfnamefont{S.}~\bibnamefont{Forte}},
  \bibinfo{author}{\bibfnamefont{E.}~\bibnamefont{Laenen}},
  \bibinfo{author}{\bibfnamefont{P.}~\bibnamefont{Nason}}, \bibnamefont{and}
  \bibinfo{author}{\bibfnamefont{J.}~\bibnamefont{Rojo}},
  \bibinfo{journal}{Nucl.Phys.} \textbf{\bibinfo{volume}{B834}},
  \bibinfo{pages}{116} (\bibinfo{year}{2010}), \eprint{1001.2312}.

\bibitem[{\citenamefont{Ball et~al.}(2011)\citenamefont{Ball, Bertone, Cerutti,
  Del~Debbio, Forte et~al.}}]{Ball:2011mu}
\bibinfo{author}{\bibfnamefont{R.~D.} \bibnamefont{Ball}},
  \bibinfo{author}{\bibfnamefont{V.}~\bibnamefont{Bertone}},
  \bibinfo{author}{\bibfnamefont{F.}~\bibnamefont{Cerutti}},
  \bibinfo{author}{\bibfnamefont{L.}~\bibnamefont{Del~Debbio}},
  \bibinfo{author}{\bibfnamefont{S.}~\bibnamefont{Forte}},
  \bibnamefont{et~al.}, \bibinfo{journal}{Nucl.Phys.}
  \textbf{\bibinfo{volume}{B849}}, \bibinfo{pages}{296} (\bibinfo{year}{2011}),
  \eprint{1101.1300}.

\bibitem[{\citenamefont{Andersen et~al.}(2010)}]{Binoth:2010ra}
\bibinfo{author}{\bibfnamefont{J.}~\bibnamefont{Andersen}} \bibnamefont{et~al.}
  (\bibinfo{collaboration}{SM and NLO Multileg Working Group}), pp.
  \bibinfo{pages}{21--189} (\bibinfo{year}{2010}), \eprint{1003.1241}.

\bibitem[{\citenamefont{Gottschalk}(1981)}]{PhysRevD.23.56}
\bibinfo{author}{\bibfnamefont{T.}~\bibnamefont{Gottschalk}},
  \bibinfo{journal}{Phys. Rev. D} \textbf{\bibinfo{volume}{23}},
  \bibinfo{pages}{56} (\bibinfo{year}{1981}).

\bibitem[{\citenamefont{Gluck et~al.}(1996)\citenamefont{Gluck, Kretzer, and
  Reya}}]{Gluck:1996ve}
\bibinfo{author}{\bibfnamefont{M.}~\bibnamefont{Gluck}},
  \bibinfo{author}{\bibfnamefont{S.}~\bibnamefont{Kretzer}}, \bibnamefont{and}
  \bibinfo{author}{\bibfnamefont{E.}~\bibnamefont{Reya}},
  \bibinfo{journal}{Phys.Lett.} \textbf{\bibinfo{volume}{B380}},
  \bibinfo{pages}{171} (\bibinfo{year}{1996}), \eprint{hep-ph/9603304}.

\bibitem[{\citenamefont{Blumlein et~al.}(2011)\citenamefont{Blumlein,
  Hasselhuhn, Kovacikova, and Moch}}]{Blumlein:2011zu}
\bibinfo{author}{\bibfnamefont{J.}~\bibnamefont{Blumlein}},
  \bibinfo{author}{\bibfnamefont{A.}~\bibnamefont{Hasselhuhn}},
  \bibinfo{author}{\bibfnamefont{P.}~\bibnamefont{Kovacikova}},
  \bibnamefont{and} \bibinfo{author}{\bibfnamefont{S.}~\bibnamefont{Moch}},
  \bibinfo{journal}{Phys.Lett.} \textbf{\bibinfo{volume}{B700}},
  \bibinfo{pages}{294} (\bibinfo{year}{2011}), \eprint{1104.3449}.

\bibitem[{\citenamefont{Buza and van Neerven}(1997)}]{Buza:1997mg}
\bibinfo{author}{\bibfnamefont{M.}~\bibnamefont{Buza}} \bibnamefont{and}
  \bibinfo{author}{\bibfnamefont{W.}~\bibnamefont{van Neerven}},
  \bibinfo{journal}{Nucl.Phys.} \textbf{\bibinfo{volume}{B500}},
  \bibinfo{pages}{301} (\bibinfo{year}{1997}), \eprint{hep-ph/9702242}.

\bibitem[{\citenamefont{Furmanski and Petronzio}(1982)}]{Furmanski:1981cw}
\bibinfo{author}{\bibfnamefont{W.}~\bibnamefont{Furmanski}} \bibnamefont{and}
  \bibinfo{author}{\bibfnamefont{R.}~\bibnamefont{Petronzio}},
  \bibinfo{journal}{Zeit. Phys.} \textbf{\bibinfo{volume}{C11}},
  \bibinfo{pages}{293} (\bibinfo{year}{1982}).

\bibitem[{\citenamefont{Bardeen et~al.}(1978)\citenamefont{Bardeen, Buras,
  Duke, and Muta}}]{Bardeen:1978yd}
\bibinfo{author}{\bibfnamefont{W.~A.} \bibnamefont{Bardeen}},
  \bibinfo{author}{\bibfnamefont{A.~J.} \bibnamefont{Buras}},
  \bibinfo{author}{\bibfnamefont{D.~W.} \bibnamefont{Duke}}, \bibnamefont{and}
  \bibinfo{author}{\bibfnamefont{T.}~\bibnamefont{Muta}},
  \bibinfo{journal}{Phys. Rev.} \textbf{\bibinfo{volume}{D18}},
  \bibinfo{pages}{3998} (\bibinfo{year}{1978}).

\bibitem[{\citenamefont{Altarelli et~al.}(1978)\citenamefont{Altarelli, Ellis,
  and Martinelli}}]{Altarelli:1978id}
\bibinfo{author}{\bibfnamefont{G.}~\bibnamefont{Altarelli}},
  \bibinfo{author}{\bibfnamefont{R.~K.} \bibnamefont{Ellis}}, \bibnamefont{and}
  \bibinfo{author}{\bibfnamefont{G.}~\bibnamefont{Martinelli}},
  \bibinfo{journal}{Nucl. Phys.} \textbf{\bibinfo{volume}{B143}},
  \bibinfo{pages}{521} (\bibinfo{year}{1978}).

\bibitem[{\citenamefont{van Neerven and Zijlstra}(1991)}]{vanNeerven:1991nn}
\bibinfo{author}{\bibfnamefont{W.~L.} \bibnamefont{van Neerven}}
  \bibnamefont{and} \bibinfo{author}{\bibfnamefont{E.~B.}
  \bibnamefont{Zijlstra}}, \bibinfo{journal}{Phys. Lett.}
  \textbf{\bibinfo{volume}{B272}}, \bibinfo{pages}{127} (\bibinfo{year}{1991}).

\bibitem[{\citenamefont{Zijlstra and van Neerven}(1991)}]{Zijlstra:1991qc}
\bibinfo{author}{\bibfnamefont{E.~B.} \bibnamefont{Zijlstra}} \bibnamefont{and}
  \bibinfo{author}{\bibfnamefont{W.~L.} \bibnamefont{van Neerven}},
  \bibinfo{journal}{Phys. Lett.} \textbf{\bibinfo{volume}{B273}},
  \bibinfo{pages}{476} (\bibinfo{year}{1991}).

\bibitem[{\citenamefont{Zijlstra and van Neerven}(1992)}]{Zijlstra:1992qd}
\bibinfo{author}{\bibfnamefont{E.~B.} \bibnamefont{Zijlstra}} \bibnamefont{and}
  \bibinfo{author}{\bibfnamefont{W.~L.} \bibnamefont{van Neerven}},
  \bibinfo{journal}{Nucl. Phys.} \textbf{\bibinfo{volume}{B383}},
  \bibinfo{pages}{525} (\bibinfo{year}{1992}).

\bibitem[{\citenamefont{Vermaseren et~al.}(2005)\citenamefont{Vermaseren, Vogt,
  and Moch}}]{Vermaseren:2005qc}
\bibinfo{author}{\bibfnamefont{J.~A.~M.} \bibnamefont{Vermaseren}},
  \bibinfo{author}{\bibfnamefont{A.}~\bibnamefont{Vogt}}, \bibnamefont{and}
  \bibinfo{author}{\bibfnamefont{S.}~\bibnamefont{Moch}},
  \bibinfo{journal}{Nucl. Phys.} \textbf{\bibinfo{volume}{B724}},
  \bibinfo{pages}{3} (\bibinfo{year}{2005}), \eprint{hep-ph/0504242}.

\bibitem[{\citenamefont{van Neerven and
  Vogt}(2000{\natexlab{a}})}]{vanNeerven:1999ca}
\bibinfo{author}{\bibfnamefont{W.~L.} \bibnamefont{van Neerven}}
  \bibnamefont{and} \bibinfo{author}{\bibfnamefont{A.}~\bibnamefont{Vogt}},
  \bibinfo{journal}{Nucl. Phys.} \textbf{\bibinfo{volume}{B568}},
  \bibinfo{pages}{263} (\bibinfo{year}{2000}{\natexlab{a}}),
  \eprint{hep-ph/9907472}.

\bibitem[{\citenamefont{van Neerven and
  Vogt}(2000{\natexlab{b}})}]{vanNeerven:2000uj}
\bibinfo{author}{\bibfnamefont{W.~L.} \bibnamefont{van Neerven}}
  \bibnamefont{and} \bibinfo{author}{\bibfnamefont{A.}~\bibnamefont{Vogt}},
  \bibinfo{journal}{Nucl. Phys.} \textbf{\bibinfo{volume}{B588}},
  \bibinfo{pages}{345} (\bibinfo{year}{2000}{\natexlab{b}}),
  \eprint{hep-ph/0006154}.

\bibitem[{\citenamefont{Moch et~al.}(2002)\citenamefont{Moch, Vermaseren, and
  Vogt}}]{Moch:2002sn}
\bibinfo{author}{\bibfnamefont{S.}~\bibnamefont{Moch}},
  \bibinfo{author}{\bibfnamefont{J.~A.~M.} \bibnamefont{Vermaseren}},
  \bibnamefont{and} \bibinfo{author}{\bibfnamefont{A.}~\bibnamefont{Vogt}},
  \bibinfo{journal}{Nucl. Phys.} \textbf{\bibinfo{volume}{B646}},
  \bibinfo{pages}{181} (\bibinfo{year}{2002}), \eprint{hep-ph/0209100}.

\bibitem[{\citenamefont{Sanchez~Guillen
  et~al.}(1991)\citenamefont{Sanchez~Guillen, Miramontes, Miramontes, Parente,
  and Sampayo}}]{SanchezGuillen:1990iq}
\bibinfo{author}{\bibfnamefont{J.}~\bibnamefont{Sanchez~Guillen}},
  \bibinfo{author}{\bibfnamefont{J.}~\bibnamefont{Miramontes}},
  \bibinfo{author}{\bibfnamefont{M.}~\bibnamefont{Miramontes}},
  \bibinfo{author}{\bibfnamefont{G.}~\bibnamefont{Parente}}, \bibnamefont{and}
  \bibinfo{author}{\bibfnamefont{O.~A.} \bibnamefont{Sampayo}},
  \bibinfo{journal}{Nucl. Phys.} \textbf{\bibinfo{volume}{B353}},
  \bibinfo{pages}{337} (\bibinfo{year}{1991}).

\bibitem[{\citenamefont{Moch et~al.}(2005)\citenamefont{Moch, Vermaseren, and
  Vogt}}]{Moch:2004xu}
\bibinfo{author}{\bibfnamefont{S.}~\bibnamefont{Moch}},
  \bibinfo{author}{\bibfnamefont{J.~A.~M.} \bibnamefont{Vermaseren}},
  \bibnamefont{and} \bibinfo{author}{\bibfnamefont{A.}~\bibnamefont{Vogt}},
  \bibinfo{journal}{Phys. Lett.} \textbf{\bibinfo{volume}{B606}},
  \bibinfo{pages}{123} (\bibinfo{year}{2005}), \eprint{hep-ph/0411112}.

\bibitem[{\citenamefont{Aivazis
  et~al.}(1994{\natexlab{b}})\citenamefont{Aivazis, Olness, and
  Tung}}]{Aivazis:1993kh}
\bibinfo{author}{\bibfnamefont{M.~A.~G.} \bibnamefont{Aivazis}},
  \bibinfo{author}{\bibfnamefont{F.~I.} \bibnamefont{Olness}},
  \bibnamefont{and} \bibinfo{author}{\bibfnamefont{W.~K.} \bibnamefont{Tung}},
  \bibinfo{journal}{Phys. Rev.} \textbf{\bibinfo{volume}{D50}},
  \bibinfo{pages}{3085} (\bibinfo{year}{1994}{\natexlab{b}}),
  \eprint{hep-ph/9312318}.

\bibitem[{\citenamefont{Botje}(2011)}]{Botje:2010ay}
\bibinfo{author}{\bibfnamefont{M.}~\bibnamefont{Botje}},
  \bibinfo{journal}{Comput. Phys. Commun.} \textbf{\bibinfo{volume}{182}},
  \bibinfo{pages}{490} (\bibinfo{year}{2011}), \eprint{1005.1481}.

\bibitem[{\citenamefont{Giele et~al.}(2002)\citenamefont{Giele, Glover,
  Hinchliffe, Huston, Laenen et~al.}}]{Giele:2002hx}
\bibinfo{author}{\bibfnamefont{W.}~\bibnamefont{Giele}},
  \bibinfo{author}{\bibfnamefont{E.}~\bibnamefont{Glover}},
  \bibinfo{author}{\bibfnamefont{I.}~\bibnamefont{Hinchliffe}},
  \bibinfo{author}{\bibfnamefont{J.}~\bibnamefont{Huston}},
  \bibinfo{author}{\bibfnamefont{E.}~\bibnamefont{Laenen}},
  \bibnamefont{et~al.}, pp. \bibinfo{pages}{275--426} (\bibinfo{year}{2002}),
  \eprint{hep-ph/0204316}.

\bibitem[{\citenamefont{Chuvakin et~al.}(2000)\citenamefont{Chuvakin, Smith,
  and van Neerven}}]{Chuvakin:1999nx}
\bibinfo{author}{\bibfnamefont{A.}~\bibnamefont{Chuvakin}},
  \bibinfo{author}{\bibfnamefont{J.}~\bibnamefont{Smith}}, \bibnamefont{and}
  \bibinfo{author}{\bibfnamefont{W.~L.} \bibnamefont{van Neerven}},
  \bibinfo{journal}{Phys. Rev.} \textbf{\bibinfo{volume}{D61}},
  \bibinfo{pages}{096004} (\bibinfo{year}{2000}), \eprint{hep-ph/9910250}.

\bibitem[{\citenamefont{Buza et~al.}(1996)\citenamefont{Buza, Matiounine,
  Smith, Migneron, and van Neerven}}]{Buza:1995ie}
\bibinfo{author}{\bibfnamefont{M.}~\bibnamefont{Buza}},
  \bibinfo{author}{\bibfnamefont{Y.}~\bibnamefont{Matiounine}},
  \bibinfo{author}{\bibfnamefont{J.}~\bibnamefont{Smith}},
  \bibinfo{author}{\bibfnamefont{R.}~\bibnamefont{Migneron}}, \bibnamefont{and}
  \bibinfo{author}{\bibfnamefont{W.}~\bibnamefont{van Neerven}},
  \bibinfo{journal}{Nucl.Phys.} \textbf{\bibinfo{volume}{B472}},
  \bibinfo{pages}{611} (\bibinfo{year}{1996}), \eprint{hep-ph/9601302}.

\bibitem[{\citenamefont{Olness and Scalise}(1998)}]{Olness:1997yn}
\bibinfo{author}{\bibfnamefont{F.~I.} \bibnamefont{Olness}} \bibnamefont{and}
  \bibinfo{author}{\bibfnamefont{R.~J.} \bibnamefont{Scalise}},
  \bibinfo{journal}{Phys.Rev.} \textbf{\bibinfo{volume}{D57}},
  \bibinfo{pages}{241} (\bibinfo{year}{1998}), \eprint{hep-ph/9707459}.

\end{thebibliography}

\end{document}